\documentclass[aps,prd,preprint,showpacs,floatfix,preprintnumbers,nofootinbib,superscriptaddress,showkeys,epsfig]{revtex4}
\usepackage[utf8]{inputenc}
\usepackage[usenames,dvipsnames,svgnames,x11names]{xcolor}
\usepackage{fancybox}
\usepackage{hhline}
\usepackage{dcolumn}
\usepackage{textcomp}
\usepackage{epsfig,graphics,graphicx}
\usepackage{amsfonts,amssymb,amsmath}
\usepackage{pifont}
\usepackage{bm}
\usepackage{longtable} 
\usepackage{appendix}
\usepackage{lscape}
\usepackage[mathscr]{euscript}
\usepackage{mathrsfs}
\usepackage{multirow}
\usepackage{rotating}
\usepackage{color}   %May be necessary if you want to color links
\usepackage[dvipsnames]{xcolor}

\newcommand{\beq}{\begin{equation}}
\newcommand{\eeq}{\end{equation}}
\newcommand{\bea}{\begin{eqnarray}}
\newcommand{\eea}{\end{eqnarray}}

\newcommand{\ord}[1]{{\cal{O}}\left( #1 \right)}
\newcommand{\B}{{\bf B}}
\newcommand{\Bdag}{{\bf B^\dagger}}

\DeclareFontFamily{OT1}{pzc}{}
\DeclareFontShape{OT1}{pzc}{m}{it}%
              {<-> s * [0.900] pzcmi7t}{}
\DeclareMathAlphabet{\mathpzc}{OT1}{pzc}%
                                 {m}{it}
\DeclareMathAlphabet{\mathcalligra}{T1}{calligra}{m}{n}
\usepackage{hyperref}
\hypersetup{pdfauthor={me}, colorlinks=true, citecolor=blue, urlcolor=blue, linkcolor=black}
\begin{document}
\preprint{\vbox{\hbox{ JLAB-THY-14-1910} }}

\title{ 
The baryon vector current in the combined chiral and $1/N_c$ expansions
}

\author{
Rub\'en Flores-Mendieta\footnote{On sabbatical leave from Instituto de F{\'\i}sica, Universidad Aut\'onoma de San Luis Potos{\'\i}, San Luis Potos{\'\i}, S.L.P.\ M\'exico}
}
\affiliation{
Theory Center, Thomas Jefferson National Accelerator Facility, Newport News, Virginia 23606, USA
}

\author{
Jos\'e L.\ Goity
}
\affiliation{
Theory Center, Thomas Jefferson National Accelerator Facility, Newport News, Virginia 23606, USA
}
\affiliation{
Department of Physics, Hampton University, Hampton, Virginia 23668, USA
}

\date{\today}

\begin{abstract}
The baryon vector current is computed at one-loop order in large-$N_c$ baryon chiral perturbation theory, where $N_c$ is the number of colors. Loop graphs with octet and decuplet intermediate states are systematically incorporated into the analysis and the effects of the decuplet-octet mass difference and $SU(3)$ flavor symmetry breaking are accounted for. There are large-$N_c$ cancellations between different one-loop graphs as a consequence of the large-$ N_c$ spin-flavor symmetry of QCD baryons. The results  are compared against the available experimental data through several fits in order to extract information about the unknown parameters. The large-$N_c$ baryon chiral perturbation theory predictions are in very good agreement both with the expectations from the $1/N_c$ expansion and with the experimental data. The effect of $SU(3)$ flavor symmetry breaking  for the $|\Delta S|=1$ vector current form factors $f_1(0)$  results in a reduction  by a few percent with respect to the corresponding $SU(3)$ symmetric values.
\end{abstract}


\maketitle

\section{\label{sec:intro}Introduction}

Baryon semileptonic decays (BSD) have served as key source of information and tests of the weak interactions, and through the strictures of $SU(3)$ and chiral symmetries also of the strong interactions. Super-allowed nuclear $\beta$ decay provides the most accurate determination of the Cabibbo angle, and hyperon semileptonic decays (HSD) provide key information on chiral $SU(3)\times SU(3)$ symmetry and its breaking by the quark masses, and also give access to independent determinations of the CKM matrix element $|V_{us}|$. 

BSD, denoted here by $B_1(p_1)\to B_2(p_2) + e^-(p_\ell) + \overline{\nu}_e(p_\nu)$, are described by the effective Hamiltonian
\begin{equation}
H_W = \frac{G_F}{\sqrt{2}} L^\alpha J_\alpha + \mathrm{H.c.}, \label{eq:ec2}
\end{equation}
where $L_\alpha$ and $J_\alpha$ are the leptonic and hadronic weak currents, respectively, which possess the $V-A$ structure of the weak interactions, and $G_F$ is the Fermi constant. The leptonic current is given by
\begin{eqnarray}
L^\alpha = \overline \psi_e \gamma^\alpha (1 - \gamma_5) \psi_{\nu_e} + \overline \psi_\mu \gamma^\alpha (1 - \gamma_5)
\psi_{\nu_\mu}, \label{eq:ec3}
\end{eqnarray}
and the hadronic current is $J_\alpha = V_\alpha - A_\alpha$, where
\begin{equation}
V_\alpha = V_{ud} \;\overline u \gamma_\alpha d + V_{us} \;\overline u \gamma_\alpha s,
\end{equation}
and
\begin{equation}
A_\alpha = V_{ud} \;\overline u \gamma_\alpha \gamma_5 d + V_{us}\; \overline u \gamma_\alpha \gamma_5 s.
\end{equation}
$V_\alpha$ and $A_\alpha$ are the weak vector and axial-vector currents, respectively, and $V_{ud}$ and $V_{us}$ are elements of the CKM matrix. The matrix elements of $J_\alpha$ between spin-$1/2$ baryon states  have the most general forms:
\begin{eqnarray}
\langle B_2 |V_\alpha|B_1 \rangle & = & V_{\rm CKM} \, \overline u_{B_2} (p_2) \left[ f_1(q^2) \gamma_\alpha + \frac{f_2(q^2)}{M_{B_1}} \sigma_{\alpha\beta}q^\beta + \frac{f_3(q^2)}{M_{B_1}} q_\alpha \right] u_{B_1}(p_1), \label{eq:ec5}
\end{eqnarray}
and
\begin{eqnarray}
\langle B_2 |A_\alpha|B_1 \rangle & = & V_{\rm CKM} \, \overline u_{B_2} (p_2) \left[g_1(q^2) \gamma_\alpha + \frac{g_2(q^2)}{M_{B_1}} \sigma_{\alpha\beta}q^\beta + \frac{g_3(q^2)}{M_{B_1}} q_\alpha \right] \gamma_5 u_{B_1}(p_1), \label{eq:ec6}
\end{eqnarray}
where $q \equiv p_1 - p_2$ is the four-momentum transfer, $u_{B_1}$ and $\overline{u}_{B_2}$ are the Dirac spinors of the decaying and emitted baryons, respectively, and $V_{\rm CKM}$ stands for $V_{ud}$ or $V_{us}$, as the case may be. Here the metric and $\gamma$-matrix conventions of Ref.~\cite{Garcia:1985xz} are used.

The matrix elements (\ref{eq:ec5}) and (\ref{eq:ec6}) are characterized by three form factors each, $f_i(q^2)$ and $g_i(q^2)$, respectively, where the weak decays probe their charged components. Additional information is of course obtained from the EM current, which is not discussed here. As a shorthand notation, $f_i\equiv f_i(0)$ and $g_i\equiv g_i(0)$ will be used hereafter. For the leading form factors,  $f_1(0)=g_V$ and $g_1(0)=g_A$ are also used. The latter couplings are related by Cabibbo's theory, with the further generalization to six quarks by Kobayashi and Maskawa.

At the present level of experimental accuracy on BSD, only the form factors $f_1(q^2)$ and $f_2(q^2)$ of the vector current and $g_1(q^2)$ and $g_2(q^2)$ of the axial vector current are involved in electron modes, whereas the $f_3(q^2)$ and $g_3(q^2)$ contributions can be neglected because of the small factor  $m_e^2$  that comes along with them. At a more detailed level, the $q^2-$dependence of the leading form factors can be parametrized in a dipole form whereas the $q^2-$dependence of $f_2$ and $g_2$ can be neglected due to the $q$ factor already present in the matrix elements (\ref{eq:ec5}) and (\ref{eq:ec6}).

In  the limit of exact flavor $SU(3)$ symmetry  $f_1$ and $f_2$ are predicted in terms of the EM form factors of $p$ and $n$ via $SU(3)$ transformations. The $g_2$ form factor for diagonal matrix elements of hermitian currents vanishes by hermiticity and time-reversal invariance. Therefore, $SU(3)$ symmetry yields $g_2=0$ in the symmetry limit. Finally, $g_1$ is given in terms of  the familiar   couplings $F$ and $D$.

The decay widths driven by vector and axial vector currents do not interfere, thus, $\Gamma=\Gamma_V+\Gamma_A$. The determination of $|V_{us}|$ and the mentioned form factors can be extracted from the total decay rate $R$, and, to a high degree of precision, $R$ must include radiative corrections. The actual expression for $R$ reads,
\begin{equation}
R = R^0\left(1+\frac{\alpha}{\pi}\Phi\right), \label{eq:decayrate}
\end{equation}
where $R^0$ is the uncorrected decay rate and model-independent radiative corrections are encoded in the term $\frac{\alpha}{\pi}\Phi$ \cite{Garcia:1985xz}. $R^0$ is a quadratic function of the form factors and can be written in the most general form as\footnote{Strictly speaking, the model-dependence of radiative corrections can be absorbed into the leading form factors $f_1$ and $g_1$ \cite{Garcia:1985xz} so Eq.~(\ref{eq:decayrate}) should be written in terms of $f_1^\prime$ and $g_1^\prime$. Actually, these primed form factors are the ones accessible to experiment.}
\begin{equation}
R^0 = |V_{\textrm{CKM}}|^2 \left( \sum_{i\leq j=1}^6 a_{ij}^R \; f_i f_j + \sum_{i\leq j=1}^6 b_{ij}^R \;(f_i \lambda_{f_j} + f_j \lambda_{f_i}) \right), \label{eq:rnum}
\end{equation}
where the dipole parametrizations assumed for all form factors introduce six slope parameters $\lambda_{f_i}$. For the sake of shortening Eq.~(\ref{eq:rnum}), $g_1=f_4$, $g_2 =f_5$, $g_3=f_6$, $\lambda_{g_1}=\lambda_{f_4}$, $\lambda_{g_2}=\lambda_{f_5}$, and $\lambda_{g_3}=\lambda_{f_6}$ have been momentarily redefined. The analytic expressions for $R^0$ in HSD can be found in Ref.~\cite{FloresMendieta:2004sk}. The short distance contributions of radiative corrections, given by the factor $S_{ew}$, can be accounted for in the usual way by defining an effective weak coupling constant.

The $|\Delta S|=1$ form factors $f_1$ satisfy the Ademollo-Gatto (AG) theorem, which states that the $SU(3)$ symmetry breaking (SB) corrections to their $SU(3)$ limit values are proportional to $(m_s-\hat{m})^2$. One must note that this does not mean the corrections are $\ord{p^4}$ in the chiral expansion. As it happens with {$K_{\ell 3}$} decays \cite{Leutwyler:1984je,Gasser:1984ux}, the dominant such corrections are non-analytic in quark masses and stem from the chiral loop contributions.  Those corrections, if expanded in $(m_s-\hat{m})$ will behave as the AG theorem requires but with small denominators proportional to quark masses, and therefore the non-analytic corrections are $\ord{p^2}$. The analytic contributions are of course $\ord{p^4}$ and beyond the accuracy of the calculation in this work. Therefore, the dominant $SU(3)$ SB corrections to $f_1$ calculated here are ultraviolet finite and well defined. 

In this work, the formalism of the $1/N_c$ expansion combined with HBChPT is used to calculate the one-loop corrections to the baryon vector currents. The approach has been successfully applied to compute flavor-$\mathbf{27}$ baryon mass splittings \cite{Jenkins:1995gc}, baryon axial-vector couplings \cite{FloresMendieta:2006ei,FloresMendieta:2012dn} and baryon magnetic moments \cite{FloresMendieta:2009rq,Ahuatzin:2010ef}, as well as to the study of lattice QCD results for baryon masses and axial couplings \cite{CalleCordon:2012xz,Cordon:2014sda}. Here its applicability is extended to the analysis of one-loop corrections to the baryon vector current operator.

Consistency with the $1/N_c$ expansion requires that the baryon decuplet be also included with specific couplings. Here it is shown how to carry out the calculation following the strictures of the $1/N_c$ expansion, which imposes relations between the various couplings involved. The present work will give the $SU(3)$ SB corrections to the vector current at the leading order of the breaking, i.e.\ $\ord{p^2}$, and represents an important step towards a more accurate calculation where the first sub-leading $SU(3)$ SB effects are also included. Thus the approximations involved, which will be discussed in more detail later, are the following: (i) The $SU(3)$ breaking mass splittings in the baryon propagators involved in the loop are disregarded; it will be shown that such effects are of sub-leading order in the chiral expansion. (ii) The calculation involves the mass splittings between octet and decuplet baryons; in the present work the $SU(3)$ SB in those splittings are ignored as per (i). The $SU(3)$ SB corrections to (i) and (ii) will be studied in detail in future work as they will contribute to sub-leading $SU(3)$ SB effects. (iii) The one-loop correction, as discussed below, is proportional to $A^{ia}\otimes A^{ib}$, where $A^{ia}$ is the axial vector current operator. The $1/N_c$ expansion of $A^{ia}$ is truncated at the physical value $N_c=3$, so in the correction there appear up to six-body operators, which are suppressed by $1/N_c^4$ factors. Working out to this order is two-fold. First, the operator reductions are doable; secondly, the complete expressions will allow a rigorous comparison with chiral perturbation theory results order by order. Knowing that the chiral and $1/N_c$ expansions do not commute, a more rigorous expansion scheme can be implemented, such as the low scale or $\xi$ expansion discussed recently in \cite{CalleCordon:2012xz}, at the cost of substantially lengthier calculations. That scheme will be implemented in the future work as well. The present work will serve as a reference mark for the effects of those improvements.

In order to set the stage, at this point it is convenient to outline the expansions involved in the relevant form factors in Eqs.~(\ref{eq:ec5}) and (\ref{eq:ec6}). In the rest frame of the decaying baryon, the dominant contribution to the matrix elements of the vector current is the corresponding charge term given by $f_1$, which is $\ord{p^0 N_c^0}$. The sub-leading terms involve (i) the recoil piece of the convection current, which is $\ord{q/M_B}$, where $q$ is the momentum transfer through the current which is $q\sim M_{B_2}-M_{B_1}\sim m_s=\ord{p^2}$, thus the recoil term is $\ord{p^2/N_c}$, (ii) the weak magnetism terms from the term proportional to $f_2$ and from the spin component proportional to $f_1$, are respectively $\ord{q N_c/\Lambda_{QCD}}$ and $\ord{q/M_B}$, and thus $\ord{p^2 N_c}$ and $\ord{p^2/N_c}$ respectively, (iii) the term proportional to $f_3$ vanishes in the $SU(3)$ symmetry limit, and is therefore proportional to $(m_s-\hat{m}) q_\mu=\ord{p^4}$. A similar discussion can be done for the axial vector current, where (i) the term proportional to $g_1$ gives matrix elements $\ord{p^0 N_c}$ for the spatial components of the current and $\ord{q/M_B}=\ord{p^2/N_c}$ for the time component, (ii) the term proportional to $g_3$ is highly suppressed as $\ord{q^2/M_B}=\ord{p^4/N_c}$, and (iii) $g_2$ vanishes in the limit of $SU(3)$ symmetry.

The earliest attempts at computing corrections to the leading vector form factors in HSD beyond tree level in chiral perturbation theory can be traced back to the works by Krause \cite{Krause:1990xc} and Anderson and Luty \cite{Anderson:1993as}. More recent analyses can be found in the works by Villadoro \cite{Villadoro:2006nj}, Lacour, Kubis and Meissner \cite{Lacour:2007wm} and Geng, Martin-Camalich and Vicente-Vacas \cite{Geng:2009ik}. Reference \cite{Krause:1990xc} presents the calculation in BChPT to order $\mathcal{O}(p^2)$ with only octet baryons in the loop. Reference \cite{Anderson:1993as} goes beyond and partially computes corrections to order $\mathcal{O}(p^3)$ in HBChPT, also without considering decuplet baryons in the loops. Reference \cite{Villadoro:2006nj} performs the analysis also in HBChPT to $\mathcal{O}(p^3)$ and includes $\mathcal{O}(1/M_0)$ corrections. This analysis accounts for decuplet dynamical degrees of freedom in the loops. References \cite{Lacour:2007wm} and \cite{Geng:2009ik} perform the analyses in covariant BChPT to order $\mathcal{O}(p^4)$, except for the fact that the latter does include dynamical octet and decuplet contributions. Out of the above analyses, only the latter finds positive $SU(3)$ SB corrections. This result is compatible with the SB pattern found in the context of the $1/N_c$ expansion alone of Ref.~\cite{FloresMendieta:2004sk}. The present analysis may provide some insight into this  issue.

This article is organized as follows. In Sec.~\ref{sec:bchpt} some general aspects of baryon chiral perturbation theory in the $1/N_c$ expansion are provided. In Sec.~\ref{sec:tree} the tree-level contribution of the baryon vector current is dealt with as a prelude to discuss in Sec.~\ref{sec:loop} the one-loop correction, where each Feynman diagram is  individually discussed in detail. In Sec.~\ref{sec:numbers} a numerical analysis is performed to compare the resultant theoretical expressions against the experimental information through several different  least-squared fits. In Sec.~\ref{sec:summary} the summary and  concluding remarks are given. This work is complemented by three appendices. In Appendix \ref{app:integrals} all the analytical results of the loop integrals that appear in the calculation are provided. In Appendix \ref{app:reduc1} the baryon operator reductions performed are listed; this way in Appendix \ref{app:coeff} some useful formulas are given in a compact form. 

\section{\label{sec:bchpt}Baryon chiral perturbation theory in the $1/N_c$ expansion}

The $1/N_c$ expansion for baryons has been discussed in detail in Refs.~\cite{Dashen:1993as,Dashen:1994qi,Jenkins:1995gc}, thus this section only provides a brief summary introducing notations and conventions. In the large-$N_c$ limit, the lowest-lying baryons are given by the completely symmetric spin-flavor representation of $N_c$ quarks ${SU}(2N_f)$ \cite{Dashen:1993as,Gervais:1984rc}. Under ${SU}(2)\times{SU}(N_f)$, this representation decomposes into a tower of baryon flavor representations with spins $J=1/2,3/2,\ldots,N_c/2$, where the states with vanishing strangeness satisfy $I=J$. This tower is degenerate in the large-$N_c$ limit, and the hyperfine mass splittings $\Delta$ between states with spin $J$ of order $\ord{N_c^0}$ are $ \ord{1/N_c}$. In general, corrections to the large-$N_c$ limit of observables are expressed in terms of $1/N_c$ suppressed operators \cite{Dashen:1993as}, which leads to the $1/N_c$ expansion of QCD. Note however that there are also non-analytic dependencies on the ratios $m_\pi/\Delta$ which are not captured by the expansion in operators, but which emerge from the finite pieces of loop corrections in the chiral expansion, as discussed below.

When a QCD operator is considered, for the purpose of its matrix elements between the ground state spin-flavor multiplet of baryon states, it can be represented by a series of effective operators organized in a power series in $1/N_c$. The $1/N_c$ expansion of a QCD $m$-body quark operator acting can then be expressed as follows \cite{Dashen:1994qi}
\begin{equation}
\mathcal{O}_{\textrm{QCD}}^{\textrm{$m$-body}} = \sum_{n=m}^{N_c}\sum_{i=1}^{i_n} c^i_n \frac{1}{N_c^{n-m}} \mathcal{O}^i_n\, ,
\end{equation}
where the $\mathcal{O}^i_n$ constitute a complete set of linearly independent effective $n$-body operators. These operators are represented by products of $n$ spin-flavor generators $J^i$, $T^a$ and $G^{ia}$, and the $c^i_n(1/N_c)$ are
unknown coefficients which have an expansion, possibly non-analytic due to loop effects,  in $1/N_c$ beginning at order unity. These effective coefficients are determined by the QCD dynamics, and are obtainable through phenomenological analysis or in certain cases also lattice QCD.

Among the most relevant QCD operators studied in the $1/N_c$ expansion are the Hamiltonian (baryon masses) \cite{Dashen:1994qi,Jenkins:1995td}, axial \cite{Dashen:1993ac,Dai:1995zg,FloresMendieta:2006ei,FloresMendieta:2012dn,CalleCordon:2012xz} and vector \cite{FloresMendieta:1998ii} currents  and magnetic moments \cite{Dai:1995zg,FloresMendieta:2009rq,Ahuatzin:2010ef}.

The expansion for the baryon mass operator is given by \cite{Dashen:1994qi}
\begin{equation}
\mathcal{M} = m_{0}^{0,\bf{1}} N_c \openone + \sum_{n=1}^{N_c-1} \frac{m_{n}^{0,\bf{1}}}{N_c^{2n-1}}J^{2n} + SU(3) ~{\rm breaking~operators}\, , \label{eq:massop}
\end{equation}
where the coefficients $m_{n}^{0,\bf{1}}$ are order $\mathcal{O}(\Lambda_{QCD})$. The first term in Eq.~(\ref{eq:massop}) represents the overall spin-independent mass of the baryon spin-flavor multiplet and the remaining spin-dependent terms constitute $\mathcal{M}_{\textrm{HF}}$, where HF stands for hyperfine. The $SU(3)$ breaking pieces are omitted here as they are not needed in the present work; they have been given in Ref. \cite{Jenkins:1995td}. 

In the limit of exact $SU(3)$ flavor symmetry, the $1/N_c$ expansion of the baryon axial vector current, can be written as \cite{Dashen:1994qi}
\begin{equation}
A^{ia} = a_1 G^{ia} + \sum_{n=2}^{N_c} b_n \frac{1}{N_c^{n-1}} \mathcal{D}_n^{ia} + \sum_{n=3}^{N_c} c_n \frac{1}{N_c^{n-1}} \mathcal{O}_n^{ia}, \label{eq:akc}
\end{equation}
where the coefficients $a_1$, $b_n$ and $c_n$ are of order unity and the leading operators that come along with them read
\begin{eqnarray}
\mathcal{D}_2^{ia} & = & J^iT^a, \label{eq:d2kc} \\
\mathcal{O}_2^{ia} & = & 0, \\
\mathcal{D}_3^{ia} & = & \{J^i,\{J^j,G^{ja}\}\}, \label{eq:d3kc} \\
\mathcal{O}_3^{ia} & = & \{J^2,G^{ia}\} - \frac12 \{J^i,\{J^j,G^{ja}\}\}. \label{eq:o3kc}
\end{eqnarray}
Higher order operators are constructed from the previous ones by anticommuting them with $J^2$. The operators $\mathcal{D}_n^{ia}$ and $\mathcal{O}_n^{ia}$ have non-vanishing matrix elements only between states of equal and different spin, respectively, so they are referred to as diagonal and off-diagonal operators. The axial currents enter in the present calculation via the pseudoscalar-baryon couplings in the one-loop diagrams, and up to the considered  chiral order of the calculation there is no need to include the $SU(3)$ SB corrections to them. For details on those effects, see \cite{FloresMendieta:2012dn} and references therein.

An interesting feature of the large-$N_c$ counting scheme is the determination of the $N_c$ dependence of the matrix elements of the generators $J^i$, $T^a$ and $G^{ia}$. The baryon matrix elements of $J^i$ for the low-lying baryons in the $SU(6)$ representation are of order unity. The $N_c$ dependence of the matrix elements of $T^a$ and $G^{ia}$ is by far more subtle because it depends on the component $a$ and on the initial and final baryon states. Specifically, for baryons with strangeness $\ord{N_c^0}$ the matrix elements of $T^a$ $(a=1,2,3)$ and $G^{i8}$ are $\ord{N_c^0}$; the matrix elements of $T^a$ and $G^{ia}$ $(a=4,5,6,7)$ are $\ord{\sqrt{N_c}}$; and the matrix elements of $T^8$ and $G^{ia}$ $(a=1,2,3)$ are $\ord{N_c}$ \cite{Dashen:1994qi}. For concreteness, the naive estimate that matrix elements of $T^a$ and $G^{ia}$ are both $\ord{N_c}$, which is the largest they can be, will be implemented here. This estimate is legitimate provided the analysis is restricted to the lowest-lying baryon states, namely, those states that make up the $\mathbf{56}$ dimensional representation of $SU(6)$.

The scaling of the baryon masses proportional to $N_c$ implies that an expansion in $1/N_c$ naturally leads to a formulation of the effective theory in the framework of heavy baryon chiral perturbation theory (HBChPT) \cite{Jenkins:1990jv}. In addition, and as mentioned earlier, the $SU(2N_f)$ dynamical spin-flavor symmetry in large-$N_c$ requires that the ground state baryons appear in a multiplet of such symmetry, namely the totally symmetric one with $N_c$ boxes in the Young tableaux. The chiral Lagrangian can be then constructed to satisfy the strictures of chiral symmetry and spin-flavor symmetry, with the breaking of these symmetries expanded in a Taylor series in quark masses and $1/N_c$ respectively \cite{Jenkins:1995gc}. 

In the baryon rest frame, the combined HBChPT and $1/N_c$ expansion effective Lagrangian at lowest order is given by 
\cite{Jenkins:1995gc,FloresMendieta:2000mz,CalleCordon:2012xz}:{\color{black} ~}
\beq
{\cal{L}}_\B^{(1)} = 
\Bdag \left(
i D_0 + \mathring{g}_A u^{ia} G^{ia}
 - \frac{m_2^{ 0,\mathbf{1}}}{N_c}-\frac{C_{HF}}{N_c}{  J^2 }
 -\frac{c_1}{2} N_c\; \chi_+
\right)
\B,
\label{eq:Lagrangian-LO}
\eeq
where $\B$ is the symmetric spin-flavor baryon multiplet with states  
$J=1/2,\cdots,N_c/2$, and  $G^{ia}$ 
are the spin-flavor generators of $SU(6)$  with matrix elements are
$\ord{N_c}$, where  $i$ are spatial indices and $a$ are $SU(3)$ flavor indices. 
The Goldstone boson pseudoscalar octet $\pi^a$ resides in the unitary matrix
\begin{equation}
u \equiv \exp \left( \frac{i\pi^a\lambda^a}{2F_0} \right),
\end{equation}
where $F_0$ is the pion decay constant in the chiral limit,  which for  the purpose of the present work can be taken to be  $F_0=F_\pi=93\,\textrm{MeV}/c^2$. The chiral operators in the Lagrangian are
\begin{equation}
u^\mu=i(u^\dag(\partial^\mu-i(v^\mu+a^\mu))u - u(\partial^\mu-i(v^\mu-a^\mu))u^\dag))=-\frac{1}{F_0} \partial^i \pi^a\lambda^a+\cdots,
\end{equation}
which gives $u^{ia}=(1/2)\, \mathrm{Tr} \, (\lambda^a u^i)$, and the covariant derivative $D_\mu=\partial_\mu-i \Gamma_\mu$ with
\begin{equation}
\Gamma_\mu = \frac{i}{2} (u^\dag(\partial_\mu-i (v_0+a_0))u + u(\partial_\mu-i(v_\mu-a_\mu))u^\dag).
\end{equation}
$v_\mu$ are the sources coupling to the vector currents, namely $v_\mu=v_\mu^a T^a/2$, and similarly $a_\mu$ are sources coupling to the axial vector currents, and the quark masses  reside in $\chi_+$. The low energy constants $m_2^{0,\mathbf{1}}$, $\mathring{g}_A$, $C_{HF}$, and $c_1$ are $\ord{N_c^0}$. As defined here, the axial coupling $ \mathring{g}_A$ is related to the one of the nucleon at $N_c=3$ by $ \mathring{g}_A=\frac{6}{5}\, g_A$, where $g_A=1.27$ is the well known nucleon axial coupling. At this order the single meson-baryon couplings are all fixed by $ \mathring{g}_A$, which are also entirely determined by the corresponding axial couplings as there is an underlying exact Goldberger-Treiman relation. The commonly used axial vector couplings are then given by $F=\mathring{g}_A/3$, $D=\mathring{g}_A/2$, ${\cal{C}}=-\mathring{g}_A$ and ${\cal{H}}=-3\mathring{g}_A/2$. Deviations from these values are due to effects $\ord{1/N_c}$.

The vector current is affected by the $SU(3)$ SB effects at higher orders in the chiral expansion. The effects stemming from tree contributions appear in the chiral Lagrangian at $\ord{p^3}$ for the magnetic components and for the corresponding charges, which are of the main interest in this work, at $\ord{p^5}$, which is beyond the order needed in this work. Thus, for the present calculations  only the above displayed Lagrangian is needed, to which the terms that correspond to $1/N_c$ corrections will be added. In particular higher order in $1/N_c$ corrections to the pseudoscalar-baryon couplings, i.e. the $F$, $D$, ${\cal{ C}}$  and ${\cal{ H}}$ couplings, through the corresponding corrections to the axial currents will be included. This will serve the purpose of determining  how important such corrections are for the weak decays as well as their impact on the strong decays, which are also included in the fits.

\section{\label{sec:tree}The baryon vector current at tree level}

At $q^2=0$ the baryon matrix elements for the vector current are given by the associated charge or $SU(3)$ generator. Therefore, the $1/N_c$ expansion of $V^{0c}$ reduces to \cite{FloresMendieta:1998ii}. 
\begin{equation}
V^{0a} = T^a, \label{eq:vcsu3}
\end{equation}
which is valid to all orders in the $1/N_c$ expansion. Due to the AG theorem, tree-level corrections to $V^{0a}$ first appear to $\ord{p^4}$, which is far beyond the order considered here.

The matrix elements of $V^{0a}$ between $SU(6)$ baryon states yield the actual values of the vector form factors at zero momentum transfer in the limit of exact $SU(3)$ symmetry, as they are introduced in the semileptonic decays of baryons. These matrix elements are listed in the first row of Table \ref{t:mtx1} for five $|\Delta S|=1$ processes of interest. These particular form factors  will be referred to as $f_1^{SU(3)}$. In passing, notice that $A^{ia}$ and $T^a$ are $\ord{N_c}$ according to the naive power counting discussed above.

\section{\label{sec:loop}One-loop corrections to the baryon vector current}

$SU(3)$ flavor SB will be considered in the exact isospin limit. As mentioned earlier,  the leading $SU(3)$ flavor SB corrections to the vector currents occur at one-loop order in the chiral expansion. Previous works focused on computing one-loop corrections to other baryon static properties \cite{FloresMendieta:2006ei,FloresMendieta:2009rq,FloresMendieta:2012dn,Ahuatzin:2010ef} will provide some feedback, so a close parallelism with them will be kept. Also, results of those works are used in the global analysis involving both weak and strong decays in Sec.~\ref{sec:numbers}.

\begin{figure}[ht]
%\scalebox{1.05}{\includegraphics{Diagrams.eps}}
\scalebox{1.05}{\includegraphics{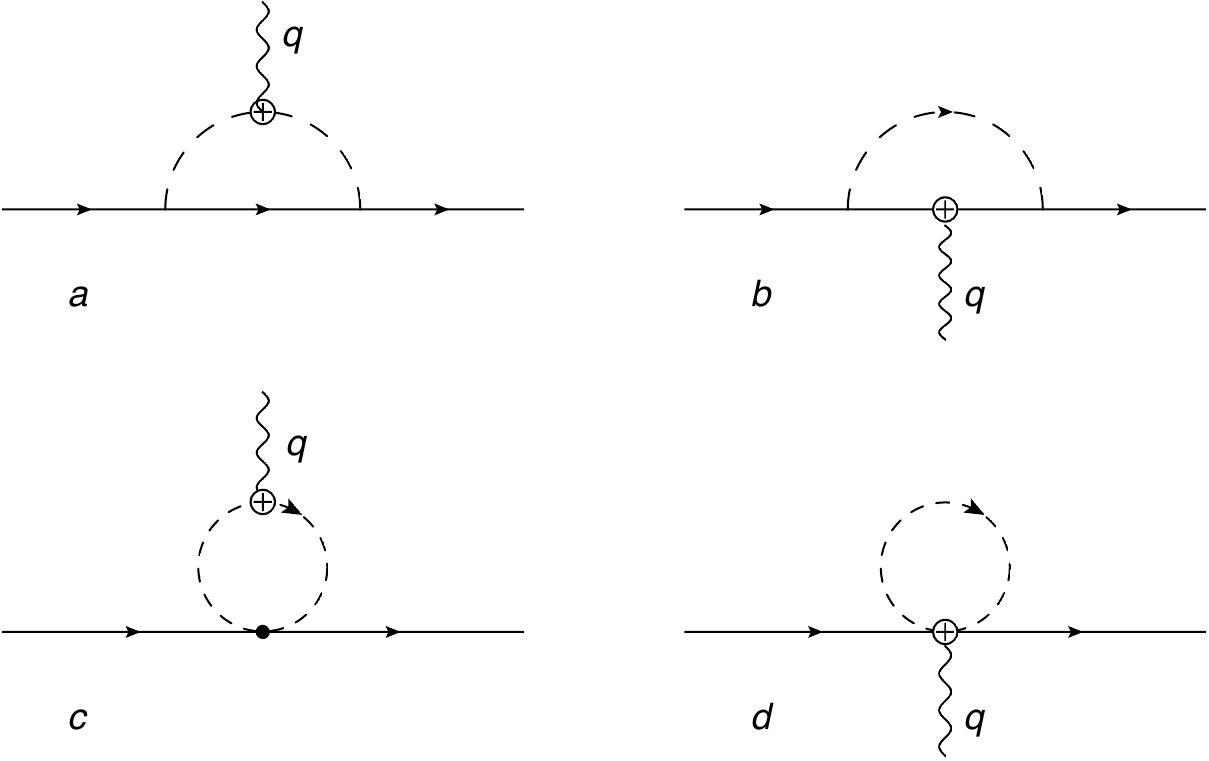}}
\caption{\label{fig:vcloop}Feynman diagrams which yield one-loop corrections to the baryon vector current. Dashed lines and solid lines denote mesons and baryons, respectively. The inner solid lines in (a) and (b) can also denote decuplet baryons. Although the wavefunction renormalization graphs are not displayed, they nevertheless have been included in the analysis.}
\end{figure}

The one-loop corrections to the baryon vector current operator are displayed in Fig.~\ref{fig:vcloop}. All these graphs can be written as the product of a baryon operator times a flavor tensor which can be written in terms of the integrals over the loops. Let us recall that the pion-baryon vertex is proportional to $g_A/F_\pi$; in the large-$N_c$ limit, $g_A\propto N_c$ and $F_\pi\propto \sqrt{N_c}$, so the pion-baryon vertex scales as $\sqrt{N_c}$. Although the $N_c$ dependence of each diagram can be deduced straightforwardly from the naive $N_c$ counting rule, the group theoretical structure for $N_c=3$ will be rigorously computed here. As for the loop integrals, they have a non-analytic dependence on $m_q$. The appropriate combination of diagrams, however, yields corrections that respect the  AG  theorem. The overall one-loop correction is thus $\ord{(m_s-\hat{m})^2}$ when expanded in a Taylor series in the mass difference, as mentioned in the introduction.

At this point it is convenient to spell out the general chiral and $1/N_c$ power counting which allows one to simplify the one-loop calculation. Since the transitions involved are only those with initial and final baryons in the octet, the energy transfer through the current $q^0\sim {M}_{\mathcal{B}}-{M}_{\mathcal{B}^\prime}$ which is a quantity of $\ord{p^2}$ in the chiral expansion. On the other hand the decuplet--octet HF mass splittings $\Delta$ have a piece $\ord{1/N_c}$ plus an $SU(3)$ SB contribution $\ord{p^2}$. If one works in the linked power counting where $1/N_c=\ord{p}$ \cite{CalleCordon:2012xz}, one concludes that the heavy baryon propagator can be Taylor expanded in the $SU(3)$ breaking mass shifts. Also the loop contributions can be expanded in powers of $q^0$. This points to the fact that the dominant $SU(3)$ SB effects on the one-loop corrections stem from the masses of the $\pi$, $K$ and $\eta$ mesons involved, with the $SU(3)$ SB effects in the baryon masses playing a sub-leading role, appearing with an additional suppression factor $\ord{(m_s-\hat{m})/\Lambda_\chi}$.

The starting point in the  analysis is the fact that $SU(3)$ flavor SB transforms as a flavor octet. The $SU(3)$ SB correction to the baryon vector current is then obtained from the tensor product of the vector current itself and the perturbation, which both transform as $(0,\mathbf{8})$. Let us also keep in mind that the tensor product of two octet representations can be separated into an antisymmetric and a symmetric product, $(\mathbf{8}\times \mathbf{8})_A$ and $(\mathbf{8}\times \mathbf{8})_S$, respectively, which can be written as \cite{Dashen:1994qi}
\begin{subequations}
\begin{eqnarray}
&  & (\mathbf{8}\times \mathbf{8})_A = \mathbf{8}+\mathbf{10}+\overline{\mathbf{10}}, \label{eq:oant} \\
&  & (\mathbf{8}\times \mathbf{8})_S = \mathbf{1}+\mathbf{8}+\mathbf{27}. \label{eq:osym}
\end{eqnarray}
\end{subequations}
The one-loop SB corrections to the baryon vector current will therefore fall in the ${SU(2)}\times{SU(3)}$ representations $(0,\mathbf{1})$, $(0,\mathbf{8})$, $(0,\mathbf{8})$, $(0,\mathbf{10}+\overline{\mathbf{10}})$, and $(0,\mathbf{27})$. Let us proceed to analyze each one of them separately.

\subsection{Figure \ref{fig:vcloop}(a)}

The one-loop contribution to the baryon vector current arising from the Feynman diagram of Fig.~\ref{fig:vcloop}(a) can be written as
\begin{equation}
\delta V_{\textrm{(a)}}^c = \sum_{\textsf{j}} A^{ia} \mathcal{P}_{\textsf{j}} A^{ib} P^{abc}(\Delta_{\textsf{j}}). \label{eq:loop1a}
\end{equation}
Here $A^{ia}$ and $A^{jb}$ are used at the meson-baryon vertices; $\mathcal{P}_{\mathsf{j}}$ is the baryon projector for spin $J=\mathsf{j}$ \cite{Jenkins:1995gc}
\begin{equation}
\frac{i\mathcal{P}_{\mathsf{j}}}{k^0-\Delta_{\mathsf{j}}}, \label{eq:barprop}
\end{equation}
which satisfies by definition
\begin{subequations}
\label{eq:spr}
\begin{eqnarray}
&  & \mathcal{P}_{\mathsf{j}}^2=\mathcal{P}_{\mathsf{j}}, \\
&  & \mathcal{P}_{\mathsf{j}} \mathcal{P}_{\mathsf{j^\prime}} = 0, \qquad \mathsf{j} \neq \mathsf{j^\prime},
\end{eqnarray}
\end{subequations}
and $\Delta_{\mathsf{j}}$ stands for the difference of the hyperfine mass splittings between the intermediate baryon with spin $J=\mathsf{j}$ and the external baryon, namely,
\begin{equation}
\Delta_{\mathsf{j}} = \mathcal{M}_{\textrm{HF}}|_{J^2=\mathsf{j}(\mathsf{j}+1)}-\mathcal{M}_{\textrm{HF}}|_{J^2=\mathsf{j}_{\textrm{ext}}(\mathsf{j}_{\textrm{ext}}+1)}.
\end{equation}
Notice that as only octet to octet weak transitions are of interest, the external baryons have $J=1/2$. In Eq.~(\ref{eq:vc1a}) the sum over spin $\mathsf{j}$ has been explicitly indicated whereas the sums over repeated spin and flavor indices are understood. In this work  $\mathsf{j}=1/2$

The general expressions for $\mathcal{P}_{\mathsf{j}}$ and $\Delta_{\mathsf{j}}$ have been introduced in Ref.~\cite{Jenkins:1995gc}. For the lowest-lying baryons,
\begin{subequations}
\label{eq:projnc3}
\begin{eqnarray}
\mathcal{P}_\frac12 & = & -\frac13 \left(J^2-\frac{15}{4}\right), \\
\mathcal{P}_\frac32 & = & \frac13 \left(J^2-\frac{3}{4}\right),
\end{eqnarray}
\end{subequations}
along with
\begin{subequations}
\begin{equation}
\Delta_\frac12 = \left\{
\begin{array}{ll}
\displaystyle 0, & \mathsf{j}_{\textrm{ext}}=\frac12, \\[2mm]
\displaystyle -\Delta, & \mathsf{j}_{\textrm{ext}}=\frac32,
\end{array}
\right.
\end{equation}
\begin{equation}
\Delta_\frac32 = \left\{
\begin{array}{ll}
\displaystyle \Delta, & \mathsf{j}_{\textrm{ext}}=\frac12, \\[2mm]
\displaystyle 0, & \mathsf{j}_{\textrm{ext}}=\frac32, \\[2mm]
\end{array}
\right.
\end{equation}
\end{subequations}
and
\begin{equation}
\Delta = \frac{3}{N_c}m_{2}^{0,\mathbf{1}}, \label{eq:mssspl}
\end{equation}
where $m_{2}^{0,\mathbf{1}}$ is the leading coefficient of the $1/N_c$ expansion of the baryon mass operator (\ref{eq:massop}). It is important to remark that expressions (\ref{eq:projnc3})--(\ref{eq:mssspl}) have been truncated at the physical value $N_c=3$.

On the other hand, $P^{abc}(\Delta_{\textsf{j}})$ is an antisymmetric tensor which can be expressed as
\begin{equation}
P^{abc}(\Delta_{\textsf{j}}) = A_{\mathbf{8}}(\Delta_{\textsf{j}}) if^{acb} + A_{\mathbf{10}+\overline{\mathbf{10}}}(\Delta_{\textsf{j}}) i(f^{aec}d^{be8} - f^{bec}d^{ae8} - f^{abe}d^{ec8}), \label{eq:pabc}
\end{equation}
where $f^{acb}$ and $f^{aec}d^{be8}-f^{bec}d^{ae8}-f^{abe}d^{ec8}$ break $SU(3)$as $\mathbf{8}$ and $\mathbf{10}+\overline{\mathbf{10}}$, respectively. The integral over the loop, $I_a(m_1,m_2,\Delta_{\textsf{j}},\mu;0)$, is contained in the tensor $P^{abc}(\Delta_{\textsf{j}})$ through
\begin{subequations}
\label{eq:ais}
\begin{eqnarray}
A_{\mathbf{8}}(\Delta_{\textsf{j}}) & = & \frac12 [ I_a(m_\pi,m_K,\Delta_{\textsf{j}},\mu;0) + I_a(m_K,m_\eta,\Delta_{\textsf{j}},\mu;0) ], \\
A_{\mathbf{10}+\overline{\mathbf{10}}}(\Delta_{\textsf{j}}) & = & -\frac{\sqrt{3}}{2}[ I_a(m_\pi,m_K,\Delta_{\textsf{j}},\mu;0) - I_a(m_K,m_\eta,\Delta_{\textsf{j}},\mu;0) ].
\end{eqnarray}
\end{subequations}
The explicit expression for $I_a(m_\pi,m_K,\Delta_{\textsf{j}},\mu;0)$ is given in Eq.~(\ref{eq:ia})

Thus, the full contribution to the baryon vector current operator from Fig.~\ref{fig:vcloop}(a) can be cast into the form
\begin{equation}
\delta V_{\textrm(a)}^c = \mathcal{P}_\frac12 A^{ia} \mathcal{P}_\frac12 A^{ib} \mathcal{P}_\frac12 P^{abc}(0) + \mathcal{P}_\frac12 A^{ia} \mathcal{P}_\frac32 A^{ib} \mathcal{P}_\frac12 P^{abc}(\Delta). \label{eq:vc1a}
\end{equation}

Naively, it could be expected $\delta V_{\textrm(a)}^c$ to be $\ord{N_c}$: two factors of the pion-baryon vertex $g_A/F_\pi$ would yield a factor $N_c$. However, the operator $A^{ia}\mathcal{P}_{\mathsf{j}}A^{jb}$ can be decomposed as $\alpha A^{ia}A^{ib} + \beta A^{ia}J^2 A^{ib}$, where $\alpha$ and $\beta$ are some coefficients. Next, $f^{acb}A^{ia}A^{ib}$ can be rewritten as $(1/2)f^{acb}\{A^{ia},A^{ib}\}+(1/2)f^{acb}[A^{ia},A^{ib}]$; the anticommutator vanishes whereas the commutator of an $n$-body operator with and $m$-body operator is an $(n+m-1)$-operator. Therefore, $f^{acb}A^{ia}A^{ib}$ is $\ord{N_c}$. For $f^{acb}A^{ia}J^2A^{ib}$ the relation
\begin{eqnarray}
A^{ia}J^2A^{ib} & = & \frac12\{J^2,A^{ia}A^{ib}\} + \frac14 [[A^{ia},J^2],A^{ib}] + \frac14 [A^{ia},[J^2,A^{ib}]] + \frac14 \{[A^{ia},J^2],A^{ib}\} \nonumber \\
&  & \mbox{} + \frac14 \{A^{ia},[J^2,A^{ib}]\},
\end{eqnarray}
can be used to verify that $f^{acb}A^{ia}J^2A^{ib}$ is also $\ord{N_c}$. In consequence, $\delta V_{\textrm(a)}^a$ is $\ord{N_c^0}$, or equivalently, $1/N_c$ times the tree level value, which is $\ord{N_c}$. In actual calculations, there will appear up to eight-body operators in the operator products on the right-hand side of Eq.~(\ref{eq:vc1a}) if the $1/N_c$ expansion of $A^{ia}$ is truncated at the physical value $N_c=3$. Because the operator basis is complete \cite{Dashen:1994qi}, the reduction, although long and tedious, is doable.

The way these operator reductions are performed can be better seen through a sample calculation. For the $if^{acb}A^{ia}A^{ib}$ piece, using the form of $A^{ia}$ of (\ref{eq:akc}) truncated at $N_c=3$, one finds,
\begin{equation}
if^{acb}A^{ia}A^{ib} = a_1^2 if^{acb}G^{ia}G^{ib} + \frac{1}{N_c}a_1b_2 if^{acb}G^{ia}\mathcal{D}_2^{ib} + \ldots + \frac{1}{N_c^4}c_3^2 if^{acb}\mathcal{O}_3^{ia}\mathcal{O}_3^{ib},
\end{equation}
where only some contributions are displayed for simplicity. Computing the leading order piece is straightforward by using the $SU(6)$ commutation relations \cite{Dashen:1994qi}, namely,
\begin{equation}
if^{acb}G^{ia}G^{ib} = \frac{i}{2} f^{acb}[G^{ia},G^{ib}] = \frac{i}{2} f^{acb} \left( \frac{i}{4} \delta^{ii} f^{abe} T^e \right) = \frac38 N_f T^c.
\end{equation}
The computation of all subleading pieces (at the order worked here) is possible by systematically using the $SU(6)$ commutation relations along with some operator identities. The full reductions are listed in Appendix \ref{app:reduc1} for the sake of completeness. The $N_c$ dependence is explicitly kept.

Gathering together partial results, the various contributions from Fig.~\ref{fig:vcloop}(a) can be organized as
\begin{equation}
if^{acb}A^{ia}A^{ib} = \sum_{n=1}^7 a_{n}^{\mathbf{8}} S_n^c,
\end{equation}
and
\begin{equation}
if^{acb}A^{ia} J^2 A^{ib} = \sum_{n=1}^7 \overline{a}_{n}^{\mathbf{8}} S_n^c,
\end{equation}
for the octet contribution, and
\begin{equation}
i(f^{aec}d^{be8}-f^{bec}d^{ae8}-f^{abe}d^{ec8}) A^{ia}A^{ib} = \sum_{n=1}^{13} b_{n}^{\mathbf{10}+\overline{\mathbf{10}}} O_n^c,
\end{equation}
and
\begin{equation}
i(f^{aec}d^{be8}-f^{bec}d^{ae8}-f^{abe}d^{ec8}) A^{ia} J^2 A^{ib} = \sum_{n=1}^{13} \overline{b}_{n}^{\mathbf{10}+\overline{\mathbf{10}}} O_n^c,
\end{equation}
for the $\mathbf{10}+\overline{\mathbf{10}}$ contribution. The coefficients $a_{n}^{\mathbf{8}}$, $\overline{a}_{n}^{\mathbf{8}}$, $b_{n}^{\mathbf{10}+\overline{\mathbf{10}}}$ and $\overline{b}_{n}^{\mathbf{10}+\overline{\mathbf{10}}}$ are listed in full in Appendix \ref{app:coeff}. The corresponding operator bases are:
\begin{eqnarray}
\begin{array}{lll}
S_1^c = T^c, &
S_2^c = \{J^r, G^{rc}\}, & 
S_3^c = \{J^2, T^c\}, \\
S_4^c = \{J^2, \{J^r, G^{rc}\}\}, & 
S_5^c = \{J^2, \{J^2, T^c\}\}, &
S_6^c = \{J^2, \{J^2, \{J^r, G^{rc}\}\}\}, \\
S_7^c = \{J^2, \{J^2, \{J^2, T^c\}\}\}, &  & \label{eq:1op}
\end{array}
\end{eqnarray}
and
\begin{eqnarray}
\begin{array}{lll}
O_{1}^c = d^{c8e} T^e, &
O_{2}^c =d^{c8e} \{J^r,G^{re}\} , &
O_{3}^c =d^{c8e} \{J^2,T^e\}, \\
O_{4}^c = \{T^c,\{J^r,G^{r8}\}\}, &
O_{5}^c = \{T^8,\{J^r,G^{rc}\}\}, &
O_{6}^c = d^{c8e} \{J^2,\{J^r,G^{re}\}\}, \\
O_{7}^c = d^{c8e} \{J^2,\{J^2,T^e\}\}, &
O_{8}^c = \{J^2,\{T^c,\{J^r,G^{r8}\}\}\}, &
O_{9}^c = \{J^2,\{T^8,\{J^r,G^{rc}\}\}\}, \\
O_{10}^c = d^{c8e} \{J^2,\{J^2,\{J^r,G^{re}\}\}\}, &
O_{11}^c = d^{c8e} \{J^2,\{J^2,\{J^2,T^e\}\}\}, &
O_{12}^c = \{J^2,\{J^2,\{T^c,\{J^r,G^{r8}\}\}\}\}, \\
O_{13}^c = \{J^2,\{J^2,\{T^8,\{J^r,G^{rc}\}\}\}\}. &  & \label{eq:8op}
\end{array}
\end{eqnarray}
The matrix elements of the operators $S_n$ and $O_n$ between baryon octet states are listed in tables \ref{t:mtx1} and \ref{t:mtx8} for completeness.

All the pieces of the one-loop contribution (\ref{eq:vc1a}) for the process $\Lambda \to p$ can be put together to illustrate how the approach works for concreteness. In terms of the operator coefficients introduced in Eq.~(\ref{eq:akc}), at $N_c=3$ one gets
\begin{eqnarray}
\left[ \frac{f_1^{\mathrm{(a)}}}{f_1^{SU(3)}} \right]_{\Lambda p} & = & \left[ \frac{17}{16} a_1^2 + \frac38 a_1 b_2 + \frac{17}{24} a_1 b_3 + \frac{1}{16} b_2^2 + \frac18 b_2 b_3 + \frac{17}{144} b_3^2 \right] I_a(m_\pi,m_K,0,\mu;0) \nonumber \\
&  & \mbox{} + \left[ \frac{9}{16} a_1^2 + \frac38 a_1 b_2 + \frac38 a_1 b_3 + \frac{1}{16} b_2^2 + \frac18 b_2 b_3 + \frac{1}{16} b_3^2 \right] I_a(m_\eta,m_K,0,\mu;0) \nonumber \\
&  & \mbox{} + \left[ -\frac12 a_1^2 - \frac12 a_1 c_3 - \frac18 c_3^2 \right] I_a(m_\pi,m_K,\Delta,\mu;0). \label{eq:mtxva}
\end{eqnarray}
Similar expressions can be found for the rest of the processes of interest. In order to display the leading $N_c$ dependence of each term, in this expression and similar ones that will follow, one simply replaces $b_n\to (3/N_c)^{n-1} b_n$ and similarly for $c_n$. 

\begin{table*}
\caption{\label{t:mtx1}Matrix elements of baryon singlet operators.}
\begin{ruledtabular}
\begin{tabular}{rrrrrr}
& $\Lambda p $ & $\Sigma^-n$ & $\Xi^-\Lambda$ & $\Xi^-\Sigma^0$ & $\Xi^0\Sigma^+$ \\
\hline
$\langle S_{1}^c \rangle$ & $-\sqrt{\frac32}$              & $-1$             & $\sqrt{\frac32}$              & $\frac{1}{\sqrt{2}}$    & $1$            \\
$\langle S_{2}^c \rangle$ & $-\frac32 \sqrt{\frac32}$      & $\frac12$        & $\frac12 \sqrt{\frac32}$      & $\frac{5}{2 \sqrt{2}}$  & $\frac52$      \\
$\langle S_{3}^c \rangle$ & $-\frac32 \sqrt{\frac32}$      & $-\frac32$       & $\frac32 \sqrt{\frac32}$      & $\frac{3}{2 \sqrt{2}}$  & $\frac32$      \\
$\langle S_{4}^c \rangle$ & $-\frac94 \sqrt{\frac32}$      & $\frac34$        & $\frac34 \sqrt{\frac32}$      & $\frac{15}{4 \sqrt{2}}$ & $\frac{15}{4}$ \\
$\langle S_{5}^c \rangle$ & $-\frac94 \sqrt{\frac32}$      & $-\frac94$       & $\frac94 \sqrt{\frac32}$      & $\frac{9}{4 \sqrt{2}}$  & $\frac94$      \\
$\langle S_{6}^c \rangle$ & $-\frac{27}{8} \sqrt{\frac32}$ & $\frac98$        & $\frac98 \sqrt{\frac32}$      & $\frac{45}{8 \sqrt{2}}$ & $\frac{45}{8}$ \\
$\langle S_{7}^c \rangle$ & $-\frac{27}{8} \sqrt{\frac32}$ & $-\frac{27}{8}$  & $\frac{27}{8} \sqrt{\frac32}$ & $\frac{27}{8 \sqrt{2}}$ & $\frac{27}{8}$ \\
\end{tabular}
\end{ruledtabular}
\end{table*}

\begin{table*}
\caption{\label{t:mtx8}Matrix elements of baryon octet operators.}
\begin{ruledtabular}
\begin{tabular}{lrrrrr}
 & $\Lambda p$ & $\Sigma^-n$ & $\Xi^-\Lambda$ & $\Xi^-\Sigma^0$ & $\Xi^0\Sigma^+$ \\
\hline
$\langle O_{1}^c  \rangle$ & $\frac{1}{2 \sqrt{2}}$    & $\frac{1}{2 \sqrt{3}}$    & $-\frac{1}{2 \sqrt{2}}$    & $-\frac{1}{2 \sqrt{6}}$        & $-\frac{1}{2 \sqrt{3}}$   \\
$\langle O_{2}^c  \rangle$ & $\frac{3}{4 \sqrt{2}}$    & $-\frac{1}{4 \sqrt{3}}$   & $-\frac{1}{4 \sqrt{2}}$    & $-\frac{5}{4 \sqrt{6}}$        & $-\frac{5}{4 \sqrt{3}}$   \\
$\langle O_{3}^c  \rangle$ & $\frac{3}{4 \sqrt{2}}$    & $\frac{\sqrt{3}}{4}$      & $-\frac{3}{4 \sqrt{2}}$    & $-\frac14\sqrt{\frac32}$       & $-\frac{\sqrt{3}}{4}$     \\
$\langle O_{4}^c  \rangle$ & $\frac{3}{4 \sqrt{2}}$    & $-\frac{3 \sqrt{3}}{4}$   & $-\frac{15}{4 \sqrt{2}}$   & $-\frac14\sqrt{\frac32}$       & $-\frac{\sqrt{3}}{4}$     \\
$\langle O_{5}^c  \rangle$ & $-\frac{9}{4 \sqrt{2}}$   & $\frac{\sqrt{3}}{4}$      & $-\frac{3}{4 \sqrt{2}}$    & $-\frac54\sqrt{\frac32}$       & $-\frac{5 \sqrt{3}}{4}$   \\
$\langle O_{6}^c  \rangle$ & $\frac{9}{8 \sqrt{2}}$    & $-\frac{\sqrt{3}}{8}$     & $-\frac{3}{8 \sqrt{2}}$    & $-\frac58\sqrt{\frac32}$       & $-\frac{5 \sqrt{3}}{8}$   \\
$\langle O_{7}^c  \rangle$ & $\frac{9}{8 \sqrt{2}}$    & $\frac{3 \sqrt{3}}{8}$    & $-\frac{9}{8 \sqrt{2}}$    & $-\frac38\sqrt{\frac32}$       & $-\frac{3 \sqrt{3}}{8}$   \\
$\langle O_{8}^c  \rangle$ & $\frac{9}{8 \sqrt{2}}$    & $-\frac{9 \sqrt{3}}{8}$   & $-\frac{45}{8 \sqrt{2}}$   & $-\frac38\sqrt{\frac32}$       & $-\frac{3 \sqrt{3}}{8}$   \\
$\langle O_{9}^c  \rangle$ & $-\frac{27}{8 \sqrt{2}}$  & $\frac{3 \sqrt{3}}{8}$    & $-\frac{9}{8 \sqrt{2}}$    & $-\frac{15}{8}\sqrt{\frac32}$  & $-\frac{15 \sqrt{3}}{8}$  \\
$\langle O_{10}^c \rangle$ & $\frac{27}{16 \sqrt{2}}$  & $-\frac{3 \sqrt{3}}{16}$  & $-\frac{9}{16 \sqrt{2}}$   & $-\frac{15}{16}\sqrt{\frac32}$ & $-\frac{15 \sqrt{3}}{16}$ \\
$\langle O_{11}^c \rangle$ & $\frac{27}{16 \sqrt{2}}$  & $\frac{9 \sqrt{3}}{16}$   & $-\frac{27}{16 \sqrt{2}}$  & $-\frac{9}{16}\sqrt{\frac32}$  & $-\frac{9 \sqrt{3}}{16}$  \\
$\langle O_{12}^c \rangle$ & $\frac{27}{16 \sqrt{2}}$  & $-\frac{27 \sqrt{3}}{16}$ & $-\frac{135}{16 \sqrt{2}}$ & $-\frac{9}{16}\sqrt{\frac32}$  & $-\frac{9 \sqrt{3}}{16}$  \\
$\langle O_{13}^c \rangle$ & $-\frac{81}{16 \sqrt{2}}$ & $\frac{9 \sqrt{3}}{16}$   & $-\frac{27}{16 \sqrt{2}}$  & $-\frac{45}{16}\sqrt{\frac32}$ & $-\frac{45\sqrt{3}}{16}$  \\
\end{tabular}
\end{ruledtabular}
\end{table*}

\subsection{Figure \ref{fig:vcloop}(b)}

The correction to the baryon vector current arising from Fig.~\ref{fig:vcloop}(b), along with the corresponding wave function renormalization graphs not displayed but nevertheless accounted for in the analysis, can be written as [cf.\ Eq.~(14) of Ref.~\cite{FloresMendieta:2012dn}]
\begin{eqnarray}
\delta V_{\textrm{(b)}}^c & = & \frac12 [A^{ja},[A^{jb},V^c]] Q_{(1)}^{ab} - \frac12 \{A^{ja},[V^c,[\mathcal{M},A^{jb}]]\} Q_{(2)}^{ab} \nonumber \\
&  & \mbox{} + \frac16 \left([A^{ja},[[\mathcal{M},[\mathcal{M},A^{jb}]],V^c]] - \frac12 [[\mathcal{M},A^{ja}],[[\mathcal{M},A^{jb}],V^c]]\right) Q_{(3)}^{ab} + \ldots\;\, \; , \nonumber \\
\label{eq:vc1b}
\end{eqnarray}
where $A^{ja}$ and $A^{jb}$ represent the meson-baryon vertices, $V^c$ denotes the insertion of the baryon vector current operator and $\mathcal{M}$ is the baryon mass operator. $Q_{(n)}^{ab}$ is a symmetric tensor which encodes the loop integral; it decomposes into flavor singlet, flavor $\mathbf{8}$ and flavor $\mathbf{27}$ representations as \cite{Jenkins:1995gc}
\begin{eqnarray}
Q_{(n)}^{ab} = I_{b,\mathbf{1}}^{(n)} \delta^{ab} + I_{b,\mathbf{8}}^{(n)} d^{ab8} + I_{b,\mathbf{27}}^{(n)} \left[ \delta^{a8} \delta^{b8} - \frac18 \delta^{ab} - \frac35 d^{ab8} d^{888}\right], \label{eq:pisym}
\end{eqnarray}
where 
\begin{subequations}
\begin{eqnarray}
I_{b,\mathbf{1}}^{(n)} & = & \frac18 \left[3I_b^{(n)}(m_\pi,0,\mu) + 4I_b^{(n)}(m_K,0,\mu) + I_b^{(n)}(m_\eta,0,\mu) \right], \label{eq:F1} \\
I_{b,\mathbf{8}}^{(n)} & = & \frac{2\sqrt 3}{5} \left[\frac32 I_b^{(n)}(m_\pi,0,\mu) - I_b^{(n)}(m_K,0,\mu) - \frac12 I_b^{(n)}(m_\eta,0,\mu) \right], \label{eq:F8} \\
I_{b,\mathbf{27}}^{(n)} & = & \frac13 I_b^{(n)}(m_\pi,0,\mu) - \frac43 I_b^{(n)}(m_K,0,\mu) + I_b^{(n)}(m_\eta,0,\mu). \label{eq:F27}
\end{eqnarray}
\end{subequations}

Here $I_b^{(n)}(m,0,\mu)$ represents the degeneracy limit $\Delta \to 0$ of the general function $I_b^{(n)}(m,\Delta,\mu)$, defined as \cite{FloresMendieta:2000mz}
\begin{equation}
I_b^{(n)}(m,\Delta,\mu) \equiv \frac{\partial^n I_b(m,\Delta,\mu)}{\partial \Delta^n}, \label{eq:fn}
\end{equation}
where the function $I_b(m,\Delta,\mu)$ is given in Eq.~(\ref{eq:ib}).

The expansion contained in Eq.~(\ref{eq:vc1b}) was derived for the baryon axial vector current in Ref.~\cite{FloresMendieta:2000mz}; here that result is extended to the baryon vector current taking advantage of the fact that both currents transform as flavor octets so one can reach the very same conclusions in the discussion presented in Ref.~\cite{FloresMendieta:2000mz}. Naively, one would expect the double commutator alone in (\ref{eq:vc1b}) to be $\ord{N_c^3}$: one factor of $N_c$ from each baryon current. However, there are large-$N_c$ cancellations between the Feynman diagrams of Fig.~\ref{fig:vcloop}(b) provided that all baryon states in a complete multiplet of the large-$N_c$ $SU(6)$ spin-flavor symmetry are included in the sum over intermediate states and that the axial coupling ratios predicted by this spin-flavor symmetry are used. Thus it can be proved that the double commutator in (\ref{eq:vc1b}) is at most $\ord{N_c}$. The same behavior is observed in the second contribution in (\ref{eq:vc1b}), so it can be concluded that $\delta V_{\textrm{(b)}}^c$ is $\ord{N_c^0}$ and is of the \textit{same order} as $\delta V_{\textrm{(a)}}^c$.

The final form of $\delta V_{\textrm{(b)}}^c$ can be organized as
\begin{eqnarray}
\delta V_{\textrm{(b)}}^c & = & \sum_{n=1}^7 \left(c_{n}^{\mathbf{1}} S_n^c I_{b,\mathbf{1}}^{(1)} + d_n^{\mathbf{1}} S_n^c I_{b,\mathbf{1}}^{(2)} S_n^c + e_n^{\mathbf{1}} S_n^c I_{b,\mathbf{1}}^{(3)} \right) + \sum_{n=1}^{13} \left(c_{n}^{\mathbf{8}} O_n^c I_{b,\mathbf{8}}^{(1)} + d_n^{\mathbf{8}} O_n^c I_{b,\mathbf{8}}^{(2)} + e_n^{\mathbf{8}} O_n^c I_{b,\mathbf{8}}^{(3)} \right) \nonumber \\
&  & \mbox{} + \sum_{n=1}^9 \left(c_{n}^{\mathbf{27}} T_n^c I_{b,\mathbf{27}}^{(1)} + d_n^{\mathbf{27}} T_n^c I_{b,\mathbf{27}}^{(2)} + e_n^{\mathbf{27}} T_n^c I_{b,\mathbf{27}}^{(3)} \right) + \ldots,
\end{eqnarray}
where the coefficients $c_{n}^{\mathbf{r}}$, $d_{n}^{\mathbf{r}}$ and $e_{n}^{\mathbf{r}}$ and given in Appendix \ref{app:coeff}. While the singlet and octet operator bases are listed in Eqs.~(\ref{eq:1op}) and (\ref{eq:8op}), respectively, the $\mathbf{27}$ operator basis is
\begin{eqnarray}
\begin{array}{ll}
T_{1}^c = f^{a8e} f^{8eg} T^g, &
T_{2}^c = f^{a8e} f^{8eg} \{J^r,G^{rg}\}, \\
T_{3}^c = f^{a8e} f^{8eg} \{J^2,T^g\}, &
T_{4}^c = \epsilon^{ijk} f^{a8e} \{G^{ke},\{J^i,G^{j8}\}\}, \\
T_{5}^c = f^{a8e} f^{8eg} \{J^2,\{J^r,G^{rg}\}\}, &
T_{6}^c = f^{a8e} f^{8eg} \{J^2,\{J^2,T^g\}\}, \\
T_{7}^c = \epsilon^{ijk} f^{a8e} \{J^2,\{G^{ke},\{J^i,G^{j8}\}\}\}, &
T_{8}^c = f^{a8e} f^{8eg} \{J^2,\{J^2,\{J^2,T^g\}\}\}, \\
T_{9}^c = \epsilon^{ijk} f^{a8e} \{J^2,\{J^2,\{G^{ke},\{J^i,G^{j8}\}\}\}\}. & \label{eq:27op}
\end{array}
\end{eqnarray}
The corresponding matrix elements are given in Table \ref{t:mtx27}. The singlet and octet pieces should be subtracted off the $\mathbf{27}$ piece to have a truly $\mathbf{27}$ contribution.

\begin{table*}
\caption{\label{t:mtx27}Matrix elements of octet operators.}
\begin{ruledtabular}
\begin{tabular}{lrrrrr}
 & $\Lambda p$ & $\Sigma^-n$ & $\Xi^-\Lambda$ & $\Xi^-\Sigma^0$ & $\Xi^0\Sigma^+$ \\
\hline
$\langle T_{1}^c \rangle$ & $-\frac34\sqrt{\frac32}$ & $-\frac{3}{4}$ & $\frac34\sqrt{\frac32}$ & $\frac{3}{4 \sqrt{2}}$ & $\frac{3}{4}$ \\
$\langle T_{2}^c \rangle$ & $-\frac98\sqrt{\frac32}$ & $\frac38$ & $\frac38\sqrt{\frac32}$ & $\frac{15}{8 \sqrt{2}}$ & $\frac{15}{8}$ \\
$\langle T_{3}^c \rangle$ & $-\frac98\sqrt{\frac32}$ & $-\frac98$ & $\frac98\sqrt{\frac32}$ & $\frac{9}{8 \sqrt{2}}$ & $\frac{9}{8}$ \\
$\langle T_{4}^c \rangle$ & $0$ & $-\frac32$ & $\frac32\sqrt{\frac32}$ & $-\frac{3}{\sqrt{2}}$ & $-3$ \\
$\langle T_{5}^c \rangle$ & $-\frac{27}{16}\sqrt{\frac32}$ & $\frac{9}{16}$ & $\frac{9}{16}\sqrt{\frac32}$ & $\frac{45}{16 \sqrt{2}}$ & $\frac{45}{16}$ \\
$\langle T_{6}^c \rangle$ & $-\frac{27}{16}\sqrt{\frac32}$ & $-\frac{27}{16}$ & $\frac{27}{16}\sqrt{\frac32}$ & $\frac{27}{16 \sqrt{2}}$ & $\frac{27}{16}$ \\
$\langle T_{7}^c \rangle$ & $0$ & $-\frac94$ & $\frac94\sqrt{\frac32}$ & $-\frac{9}{2 \sqrt{2}}$ & $-\frac{9}{2}$ \\
$\langle T_{8}^c \rangle$ & $-\frac{81}{32}\sqrt{\frac32}$ & $-\frac{81}{32}$ & $\frac{81}{32}\sqrt{\frac32}$ & $\frac{81}{32 \sqrt{2}}$ & $\frac{81}{32}$ \\
$\langle T_{9}^c \rangle$ & $0$ & $-\frac{27}{8}$ & $\frac{27}{8}\sqrt{\frac32}$ & $-\frac{27}{4 \sqrt{2}}$ & $-\frac{27}{4}$ \\
\end{tabular}
\end{ruledtabular}
\end{table*}

The contribution of $\langle \delta V_{\mathrm{(b)}}^c \rangle$ to $f_1$ can be readily computed. Keeping the $\Lambda \to p$ process as an example, the contribution reads
\begin{eqnarray}
&  & \left[ \frac{f_1^{\textrm{(b)}}}{f_1^{SU(3)}} \right]_{\Lambda p} \nonumber \\
&  & \mbox{} = \left[\frac{9}{32} a_1^2 + \frac{3}{16} a_1 b_2 + \frac{17}{48} a_1 b_3 - \frac14 a_1 c_3 + \frac{1}{32} b_2^2 + \frac{1}{16} b_2 b_3 + \frac{17}{288} b_3^2 - \frac{1}{16} c_3^2 \right] I_b^{(1)}(m_\pi,0,\mu) \nonumber \\
&  & \mbox{\hglue0.5truecm} + \left[ \frac{9}{16} a_1^2 + \frac38 a_1 b_2 + \frac{13}{24} a_1 b_3 - \frac14 a_1 c_3 + \frac{1}{16} b_2^2 + \frac18 b_2 b_3 + \frac{13}{144} b_3^2 - \frac{1}{16} c_3^2 \right] I_b^{(1)}(m_K,0,\mu) \nonumber \\
&  & \mbox{\hglue0.5truecm} + \left[ \frac{9}{32} a_1^2 + \frac{3}{16} a_1 b_2 + \frac{3}{16} a_1 b_3 + \frac{1}{32} b_2^2 + \frac{1}{16} b_2 b_3 + \frac{1}{32} b_3^2 \right] I_b^{(1)}(m_\eta,0,\mu) \nonumber \\
&  & \mbox{\hglue0.5truecm} + \left[ -\frac34 a_1^2 - \frac34 a_1 c_3 - \frac{3}{16} c_3^2 \right] \frac{\Delta}{3} I_b^{(2)}(m_\pi,0,\mu) + \left[ -\frac34 a_1^2 - \frac34 a_1 c_3 - \frac{3}{16} c_3^2 \right] \frac{\Delta}{3} I_b^{(2)}(m_K,0,\mu) \nonumber \\
&  & \mbox{\hglue0.5truecm} + \left[ -\frac98 a_1^2 - \frac98 a_1 c_3 - \frac{9}{32} c_3^2 \right] \frac{\Delta^2}{9} I_b^{(3)}(m_\pi,0,\mu) + \left[ -\frac98 a_1^2 - \frac98 a_1 c_3 - \frac{9}{32} c_3^2 \right] \frac{\Delta^2}{9} I_b^{(3)}(m_K,0,\mu) + \ldots \nonumber \\
\label{eq:mtxvb}
\end{eqnarray}

Equations (\ref{eq:mtxva}) and (\ref{eq:mtxvb}) are now added together to get
\begin{eqnarray}
\left[ \frac{f_1^{\textrm{(a)}}+f_1^{\textrm{(b)}}}{f_1^{SU(3)}} \right]_{\Lambda p} & = & \left[ \frac{17}{32} a_1^2 + \frac{3}{16} a_1 b_2 + \frac{17}{48} a_1 b_3 + \frac{1}{32} b_2^2 + \frac{1}{16} b_2 b_3 + \frac{17}{288} b_3^2 \right] H(m_\pi,m_K) \nonumber \\
&  & \mbox{} + \left[ \frac{9}{32} a_1^2 + \frac{3}{16} a_1 b_2 + \frac{3}{16} a_1 b_3 + \frac{1}{32} b_2^2 + \frac{1}{16} b_2 b_3 + \frac{1}{32} b_3^2 \right] H(m_K,m_\eta) \nonumber \\
&  & \mbox{} + \left[ -\frac14 a_1^2 - \frac14 a_1 c_3 - \frac{1}{16} c_3^2 \right] K(m_\pi,m_\eta,\Delta), \label{eq:vcab}
\end{eqnarray}
where
\begin{equation}
H(m_1,m_2) \equiv 2I_a(m_1,m_2,0,\mu;0) + I_b^{(1)}(m_1,0,\mu) + I_b^{(1)}(m_2,0,\mu),
\end{equation}
and
\begin{eqnarray}
K(m_1,m_2,\Delta) & \equiv & 2I_a(m_1,m_2,\Delta,\mu;0) + I_b^{(1)}(m_1,0,\mu) + I_b^{(1)}(m_2,0,\mu) \nonumber \\
&  & \mbox{} + \left[ I_b^{(2)}(m_1,0,\mu) + I_b^{(2)}(m_2,0,\mu) \right] \Delta \nonumber \\
&  & \mbox{} + \left[ I_b^{(3)}(m_1,0,\mu) + I_b^{(3)}(m_2,0,\mu) \right] \frac{\Delta^2}{2} + \ldots \nonumber \\
& = & 2I_a(m_1,m_2,\Delta,\mu;0) + I_b^{(1)}(m_1,\Delta,\mu) + I_b^{(1)}(m_2,\Delta,\mu).
\end{eqnarray}
The final form of $K(m_1,m_2,\Delta)$ recovers the full form of the function $I_b^{(1)}(m_1,\Delta,\mu)$, which was originally expanded in a power series in $\Delta$ in Eq.~(\ref{eq:vc1b}). This is a remarkable result.

On the other hand, the explicit form of the function $H(m_1,m_2)$ becomes
\begin{equation}
H(m_1,m_2) = \frac{1}{16\pi^2f^2} \left[-\frac12(m_1^2+m_2^2) + \frac{m_1^2m_2^2}{m_1^2-m_2^2} \ln{\frac{m_1^2}{m_2^2}} \right],
\end{equation}
which is ultraviolet finite. The function $K(m_1,m_2,\Delta)$ can be easily constructed from $I_a(m_1,m_2,\Delta,\mu;0)$ and $I_b^{(1)}(m,\Delta,\mu)$ given in Eqs.~(\ref{eq:ia}) and (\ref{eq:ib}), respectively; the explicit expression will not be provided here. However, some important properties of this function are
\begin{enumerate}
\item $\displaystyle \lim_{\Delta\to 0} K(m_1,m_2,\Delta) = H(m_1,m_2),$
\item $\displaystyle \lim_{\Delta \to \infty} K(m_1,m_2,\Delta) = 0.$
\end{enumerate}
Property (2) above has some interesting physical implications. The present calculation exploits the near degeneracy between octet and decuplet baryons. For instance, in the loop integral $I_b(m,\Delta,\mu)$, Eq.~(\ref{eq:ib}), the full functional dependence on the ratio $m_\pi/\Delta$ has been retained. This ratio does not have to be small necessarily because the conditions for HBChPT to be valid are $m_\pi \ll \Lambda_\chi$ and $\Delta \ll \Lambda_\chi$. In the chiral limit $\Delta \gg m_\pi$ so the decuplet cannot contribute to the non-analytical corrections for octet processes since these corrections come from infrared divergences. The decuplet thus decouples in the large-$N_c$ limit and property (2) holds.

A further aim of the approach can be achieved by rewriting the results in terms of the $SU(3)$ invariant couplings $D$, $F$ and $\mathcal{C}$ introduced in HBChPT \cite{Jenkins:1990jv,Jenkins:1991ne}. These couplings are related to the $1/N_c$ coefficients $a_1$, $b_2$, $b_3$, and $c_3$ as
\begin{subequations}
\label{eq:rel1}
\begin{eqnarray}
&  & D = \frac12 a_1 + \frac16 b_3, \\
&  & F = \frac13 a_1 + \frac16 b_2 + \frac19 b_3, \\
&  & \mathcal{C} = - a_1 - \frac12 c_3, \\
&  & \mathcal{H} = -\frac32 a_1 - \frac32 b_2 - \frac52 b_3.
\end{eqnarray}
\end{subequations}
In the large-$N_c$ limit the standard $SU(6)$ ratios $D:F:C:H=1:\frac{2}{3}:-2:-3$ result. In the canonical example worked out so far, substituting Eqs.~(\ref{eq:rel1}) into Eq.~(\ref{eq:vcab}) yields
\bea
\left[ \frac{f_1^{\textrm{(a)}}+f_1^{\textrm{(b)}}}{f_1^{SU(3)}} \right]_{\Lambda p} & = & \frac18 (9D^2 + 6 D F + 9F^2) H(m_\pi,m_K) \nonumber \\
&   + & \frac18 (D^2 + 6 D F + 9F^2) H(m_K,m_\eta) - \frac14 \mathcal{C}^2 K(m_\pi,m_K,\Delta),
\eea
which exactly matches the ones obtained within (H)BChPT: When the decuplet fields are not explicitly retained in the effective theory but integrated out, this result agrees with those presented in Refs.~\cite{Krause:1990xc, Anderson:1993as,Lacour:2007wm}. When the decuplet fields are retained, there is a full agreement with the ones presented in Ref.~\cite{Villadoro:2006nj} (in that reference $f_\pi=131$ MeV is used). Moreover, it can be shown that
\begin{equation}
K(m_p,m_q,\Delta) = \frac43 \left(G_{pq}-\frac38 H_{pq} \right),
\end{equation}
where the functions $H_{pq}$ and $G_{pq}$ are given in Eqs.~(22) and (31) of that reference, respectively.

Note that the coupling $\mathcal{H}$ does not appear in the corrections to the vector currents, but it does in the corrections to the axial currents. Its determination is addressed in the analysis below.

\subsection{Figure \ref{fig:vcloop}(c)}

The tadpole diagrams of Figs.~\ref{fig:vcloop}(c) and \ref{fig:vcloop}(d) can be easily computed within the combined approach. These diagrams do not depend on the coefficients of the $1/N_c$ expansion of $A^{ia}$. 

The loop graph \ref{fig:vcloop}(c) can be written as
\begin{equation}
\delta V_{\textrm{(c)}}^c = -f^{cae} f^{beg} T^g R^{ab}, \label{eq:vc1c}
\end{equation}
where
\begin{equation}
R^{ab} = \frac12 \left[ I_c(m_\pi,m_K,\mu;0) + I_c(m_K,m_\eta,\mu;0) \right] \delta^{ab},
\end{equation}
where the loop integral $I_c(m_1,m_2,\mu;q^2)$ in the $q^2\to 0$ limit is given in Eq.~(\ref{eq:ic}) of Appendix \ref{app:integrals}. This contribution breaks $SU(3)$ as a flavor singlet.

\subsection{Figure \ref{fig:vcloop}(d)}

The Feynman diagram of Fig.~\ref{fig:vcloop}(d) is given by
\begin{equation}
\delta V_{\textrm{(d)}}^c = - \frac12 \left[T^a,\left[T^b,V^c\right]\right] S^{ab}, \label{eq:vc1d}
\end{equation}
where $S^{ab}$ has the very same structure as $P_{(n)}^{ab}$ of Eq.~(\ref{eq:pisym}), namely,
\begin{eqnarray}
S^{ab} = I_{d,\mathbf{1}} \delta^{ab} + I_{d,\mathbf{8}} d^{ab8} + I_{d,\mathbf{27}} \left[ \delta^{a8} \delta^{b8} - \frac18 \delta^{ab} - \frac35 d^{ab8} d^{888}\right], \label{eq:gamsym}
\end{eqnarray}
where
\begin{subequations}
\begin{eqnarray}
I_{d,\mathbf{1}} & = & \frac18 \left[3I_d(m_\pi,\mu) + 4I_d(m_K,\mu) + I_d(m_\eta,\mu) \right], \label{eq:G1} \\
I_{d,\mathbf{8}} & = & \frac{2\sqrt 3}{5} \left[\frac32 I_d(m_\pi,\mu) - I_d(m_K,\mu) - \frac12 I_d(m_\eta,\mu) \right], \label{eq:G8} \\
I_{d,\mathbf{27}} & = & \frac13 I_d(m_\pi,\mu) - \frac43 I_d(m_K,\mu) + I_d(m_\eta,\mu). \label{eq:G27}
\end{eqnarray}
\label{eq:loopggs}
\end{subequations}

The integral over the loop is given in Appendix \ref{app:integrals}, (\ref{eq:id}). The different flavor contributions in Eq.~(\ref{eq:vc1d}) read

(1) Flavor singlet contribution
\begin{equation}
[T^a,[T^a,V^c]] = N_f V^c. \label{eq:sind}
\end{equation}

(2) Flavor octet contribution
\begin{equation}
d^{ab8} [T^a,[T^b,V^c]] = \frac{N_f}{2} d^{c8e} V^e. \label{eq:octd}
\end{equation}

(3) Flavor $\mathbf{27}$ contribution
\begin{equation}
[T^8,[T^8,V^c]] = f^{c8e} f^{8eg} V^g. \label{eq:27d}
\end{equation}

The straightforward combination of loop corrections \ref{fig:vcloop}(c) and \ref{fig:vcloop}(d), for the $\Lambda \to p$ process, yields
\begin{equation}
\left[ \frac{f_1^{\textrm{(c)}}+f_1^{\textrm{(d)}}}{f_1^{SU(3)}} \right]_{\Lambda p} = \frac38 \left[ H(m_\pi,m_K) + H(m_K,m_\eta) \right]. \label{eq:vccd}
\end{equation}
Equation (\ref{eq:vccd}) agrees with the results derived in Refs.~\cite{Krause:1990xc,Villadoro:2006nj,Lacour:2007wm} but differs in a global sign with respect to the expression presented in Ref.~\cite{Anderson:1993as}.

It is also interesting to remark that Eq.~(\ref{eq:vccd}) contributes at the same order in $N_c$ as Eq.~(\ref{eq:vcab}). This assertion can be proved numerically.

\subsection{Total one-loop correction to the baryon vector current}

The baryon vector current operator $V^c$ including one-loop corrections can be organized in a single expression as
\begin{equation}
V^c + \delta V^c = V^c + \delta V_{\textrm{(a)}}^c + \delta V_{\textrm{(b)}}^c + \delta V_{\textrm{(c)}}^c + \delta V_{\textrm{(d)}}^c,
\end{equation}
where $\delta V_{\textrm{(a)}}^c$, $\delta V_{\textrm{(b)}}^c$, $\delta V_{\textrm{(c)}}^c$, and $\delta V_{\textrm{(d)}}^c$ are given by Eqs.~(\ref{eq:vc1a}), (\ref{eq:vc1b}), (\ref{eq:vc1c}), and (\ref{eq:vc1d}). In the large-$N_c$ counting, each correction is suppressed at least by a factor $1/N_c$ with respect to the tree-level operator $V^c$. The loop contributions   expanded in ${(m_s-\hat{m})}$ satisfy the  AG  theorem.

The matrix elements of the operator $V^c + \delta V^c$ give the actual values of the vector form factors $f_1$ as defined in HSD. The full expressions for the processes observed are
\begin{eqnarray}
&  & \left[ \frac{f_1}{f_1^{SU(3)}} \right]_{\Lambda p} \nonumber \\
&  & \mbox{\hglue0.5truecm} = 1 + \left[ \frac38 + \frac{17}{32} a_1^2 + \frac{3}{16} a_1 b_2 + \frac{17}{48} a_1 b_3 + \frac{1}{32} b_2^2 + \frac{1}{16} b_2 b_3 + \frac{17}{288} b_3^2 \right] H(m_\pi,m_K) \nonumber \\
&  & \mbox{\hglue1.0truecm} + \left[ \frac38 + \frac{9}{32} a_1^2 + \frac{3}{16} a_1 b_2 + \frac{3}{16} a_1 b_3 + \frac{1}{32} b_2^2 + \frac{1}{16} b_2 b_3 + \frac{1}{32} b_3^2 \right] H(m_K,m_\eta) \nonumber \\
&  & \mbox{\hglue1.0truecm} + \left[ -\frac14 a_1^2 - \frac14 a_1 c_3 -\frac{1}{16} c_3^2 \right] K(m_\pi,m_K,\Delta),
\end{eqnarray}

\begin{eqnarray}
&  & \left[ \frac{f_1}{f_1^{SU(3)}} \right]_{\Sigma^-n} \nonumber \\
&  & \mbox{\hglue0.5truecm} = 1 + \left[ \frac38 - \frac{7}{32} a_1^2 - \frac{1}{16} a_1 b_2 - \frac{7}{48} a_1 b_3 + \frac{1}{32} b_2^2 - \frac{1}{48} b_2 b_3 - \frac{7}{288} b_3^2 \right] H(m_\pi,m_K) \nonumber \\
&  & \mbox{\hglue1.0truecm} + \left[ \frac38 + \frac{1}{32} a_1^2 - \frac{1}{16} a_1 b_2 + \frac{1}{48} a_1 b_3 + \frac{1}{32} b_2^2 - \frac{1}{48} b_2 b_3 + \frac{1}{288} b_3^2 \right] H(m_K,m_\eta) \nonumber \\
&  & \mbox{\hglue1.0truecm} + \left[ \frac12 a_1^2 + \frac12 a_1 c_3 +\frac18 c_3^2 \right] K(m_\pi,m_K,\Delta) + \left[ \frac14 a_1^2 + \frac14 a_1 c_3 + \frac{1}{16} c_3^2 \right] K(m_K,m_\eta,\Delta)\;\, \; , \nonumber \\
\end{eqnarray}

\begin{eqnarray}
&  & \left[ \frac{f_1}{f_1^{SU(3)}} \right]_{\Xi^-\Lambda} \nonumber \\
&  & \mbox{\hglue0.5truecm} = 1 + \left[ \frac38 + \frac{9}{32} a_1^2 + \frac{1}{16} a_1 b_2 + \frac{3}{16} a_1 b_3 + \frac{1}{32} b_2^2 + \frac{1}{48} b_2 b_3 + \frac{1}{32} b_3^2 \right] H(m_\pi,m_K) \nonumber \\
&  & \mbox{\hglue1.0truecm} + \left[ \frac38 + \frac{1}{32} a_1^2 + \frac{1}{16} a_1 b_2 + \frac{1}{48} a_1 b_3 + \frac{1}{32} b_2^2 + \frac{1}{48} b_2 b_3 + \frac{1}{288} b_3^2 \right] H(m_K,m_\eta) \nonumber \\
&  & \mbox{\hglue1.0truecm} + \left[ \frac14 a_1^2 + \frac14 a_1 c_3 +\frac{1}{16} c_3^2\right] K(m_K,m_\eta,\Delta),
\end{eqnarray}

\begin{eqnarray}
&  & \left[ \frac{f_1}{f_1^{SU(3)}} \right]_{\Xi^-\Sigma^0} \nonumber \\
&  & \mbox{\hglue0.5truecm} = 1 + \left[ \frac38 + \frac{17}{32} a_1^2 + \frac{5}{16} a_1 b_2 + \frac{17}{48} a_1 b_3 + \frac{1}{32} b_2^2 + \frac{5}{48} b_2 b_3 + \frac{17}{288} b_3^2 \right] H(m_\pi,m_K) \nonumber \\
&  & \mbox{\hglue1.0truecm} + \left[ \frac38 + \frac{25}{32} a_1^2 + \frac{5}{16} a_1 b_2 + \frac{25}{48} a_1 b_3 + \frac{1}{32} b_2^2 + \frac{5}{48} b_2 b_3 + \frac{25 }{288} b_3^2 \right] H(m_K,m_\eta) \nonumber \\
&  & \mbox{\hglue1.0truecm} + \left[-\frac14 a_1^2 - \frac14 a_1 c_3 - \frac{1}{16} c_3^2 \right] K(m_\pi,m_K,\Delta) + \left[ -\frac12 a_1^2 - \frac12 a_1 c_3 - \frac18 c_3^2 \right] K(m_K,m_\eta,\Delta)\; , \nonumber \\
\end{eqnarray}

\begin{eqnarray}
&  & \left[ \frac{f_1}{f_1^{SU(3)}} \right]_{\Xi^0\Sigma^+} \nonumber \\
&  & \mbox{\hglue0.5truecm} = 1 + \left[ \frac38 + \frac{17}{32} a_1^2 + \frac{5}{16} a_1 b_2 + \frac{17}{48} a_1 b_3 + \frac{1}{32} b_2^2 + \frac{5}{48} b_2 b_3 + \frac{17}{288} b_3^2 \right] H(m_\pi,m_K) \nonumber \\
&  & \mbox{\hglue1.0truecm} +  \left[ \frac38 + \frac{25}{32} a_1^2 + \frac{5}{16} a_1 b_2 + \frac{25}{48} a_1 b_3 + \frac{1}{32} b_2^2 + \frac{5}{48} b_2 b_3 + \frac{25}{288} b_3^2 \right] H(m_K,m_\eta) \nonumber \\
&  & \mbox{\hglue1.0truecm} + \left[-\frac14 a_1^2 - \frac14 a_1 c_3 - \frac{1}{16} c_3^2 \right] K(m_\pi,m_K,\Delta) + \left[ -\frac12 a_1^2 - \frac12 a_1 c_3 - \frac18 c_3^2 \right] K(m_K,m_\eta,\Delta). \nonumber \\
\end{eqnarray}

A full crosscheck of the above expressions has been performed with their counterparts obtained within (heavy) baryon chiral perturbation theory \cite{Krause:1990xc,Villadoro:2006nj,Lacour:2007wm,Anderson:1993as}, according to the guidelines described above. The results agree order by order.

\section{\label{sec:numbers}Numerical analysis}

An analysis of the available experimental data \cite{Beringer:1900zz} can be performed by using the results obtained here. In previous works \cite{FloresMendieta:2006ei,FloresMendieta:2012dn} a number of fits have been carried out to determine the baryon axial couplings, which are given by the matrix elements of the baryon axial current operator $A^{kc}+\delta A^{kc}$.  For octet baryons, the axial vector couplings are $g_1$ normalized in such a way that $g_1 \sim 1.27$ for neutron $\beta$ decay. For decuplet baryons, the axial vector couplings are denoted by $g$, which are extracted via  Goldberger-Treiman relations  from the widths of the strong decays of decuplet to octet baryons and pions \cite{Dai:1995zg}.

The effects related to $SU(3)$ SB are contained in $\delta A^{kc}$ in two ways: On the one hand, at tree level, all relevant operators which explicitly break $SU(3)$ at leading order are included; this contribution is loosely referred to as perturbative SB. On the other hand, in the one-loop corrections, $SU(3)$ SB is accounted for implicitly, since the loop integrals depend on the $\pi$, $K$ and $\eta$ masses.

The operator $\delta A^{kc}$ has been built up in a systematic way. In Ref.~\cite{FloresMendieta:2006ei}, $\delta A^{kc}$ was constituted only by one-loop corrections within the combined approach, while in Ref.~\cite{FloresMendieta:2012dn}, a more refined calculation was performed to include the effects of perturbative $SU(3)$ SB corrections and the effects of the baryon decuplet--octet mass splitting. The corrected axial vector current operator actually used in the numerical analysis reads
\begin{eqnarray}
A^{kc}+\delta A_{\mathrm{SB}}^{kc}+\delta A_{\mathrm{1L}}^{kc} & = & a_1 G^{kc} + b_2 \frac{1}{N_c} \mathcal{D}_2^{kc} + b_3 \frac{1}{N_c^2} \mathcal{D}_3^{kc} + c_3 \frac{1}{N_c^2} \mathcal{O}_3^{kc} + \Bigg[ d_1 d^{c8e} G^{ke} + d_2 \frac{1}{N_c} d^{c8e} \mathcal{D}_2^{ke} \nonumber \\
&  & \mbox{} + d_3 \frac{1}{N_c} \left( \{G^{kc},T^8\}-\{G^{k8},T^c\} \right) + d_4 \frac{1}{N_c} \left( \{G^{kc},T^8\}+\{G^{k8},T^c\} \right) \Bigg] \nonumber \\
&  & \mbox{} + \delta A_{\mathrm{1L}}^{kc}, \label{eq:akcsb}
\end{eqnarray}
where $\delta A_{\mathrm{SB}}^{kc}$ is the correction that arises from perturbative SB and $\delta A_{\mathrm{1L}}^{kc}$ is the one-loop correction. Note that the loop corrections are renormalized by the counter terms corresponding to the coefficients $a_i$, $b_i$, and $c_i$. Minimal subtraction is used { with renormalization scale $\mu$}. Equation (\ref{eq:akcsb}) was parametrized in Ref.~\cite{FloresMendieta:2012dn} in such a way that flavor SB took place entirely in the non-zero strangeness sector only. This involves however a bias, namely that $g_1=F+D$ for neutron $\beta$ decay \textit{even in the presence of} $SU(3)$ SB, which corresponds to a constraint on the counter term coefficients. That bias is avoided here by instead taking into account $SU(3)$ SB in the axial couplings throughout.

The scope of the numerical analyses performed in Refs.~\cite{FloresMendieta:2006ei,FloresMendieta:2012dn} within these two scenarios was somewhat limited to determining only $g_1$ and $g$, because $f_1$'s were given at their $SU(3)$ symmetric values, $f_1^{SU(3)}$, in view of the  AG  theorem. By that time, the main aim of those analyses was not to be definitive about the determination of the form factors, but rather to explore the working assumptions. The present analysis, however, is uniquely positioned in the sense that, on the same footing as $g_1$, the one-loop corrections to $f_1$ within large-$N_c$ chiral perturbation theory have been computed, including the effects of a non-zero baryon decuplet--octet mass splitting. Thus, the pattern of $SU(3)$ SB for $f_1$, which will be referred to as $f_1/f_1^{SU(3)}$ hereafter, can be evaluated.

A very important lesson learned from previous analyses is that different working assumptions yield rather different outputs, so it is hard to assess the success of a particular set of assumptions. In the present numerical analysis a more cautious approach will be adopted. A satisfactory fit will be considered as such when not only the best-fit parameters yield acceptable values of the $SU(3)$ invariant couplings $D$, $F$, $\mathcal{C}$, and $\mathcal{H}$ [c.f.\ Eq.~(\ref{eq:rel1})], but also when the predicted values of the several integrated observables are in good agreement with their experimental counterparts, which necessarily will be reflected in the goodness of the fit itself. For instance, the final fit of Ref.~\cite{FloresMendieta:2012dn} yielded best-fit parameters which fairly reproduced the $SU(3)$ invariant couplings, but some observables were not well reproduced. Based on the above two criteria, such a fit is no longer satisfactory.

The available experimental information for octet baryons is given in terms of the decay rates $R$, the ratios $g_1/f_1$, the angular correlation coefficients $\alpha_{e\nu}$, and the spin-asymmetry coefficients $\alpha_e$, $\alpha_\nu$, $\alpha_B$, $A$, and $B$. All eight decay rates and all six possible $g_1/f_1$ ratios have been measured (the ratios $g_1/f_1$ for $\Sigma^\pm\to\Lambda$ semileptonic decays are undefined). A summary of this experimental information can be found in Table II of Ref.~\cite{FloresMendieta:2012dn}, along with a detailed discussion about how this information can be matched with the one listed in Ref.~\cite{Beringer:1900zz}. That discussion is not repeated here. For decuplet baryons, the axial couplings $g$ for the processes $\Delta \to N\pi$, $\Sigma^*\to\Lambda\pi$, $\Sigma^*\to\Sigma\pi$, and $\Xi^*\to\Xi\pi$ are given in Table IX of that reference as well. For the purposes of the present work, the experimental information is arranged into three sets: $R$ and $g_1/f_1$ constitute set 1; $R$, $g_1/f_1$ and $g$ constitute set 2; and $g_1/f_1$ and $g$ constitute set 3. The latter can be enriched by adding two more pieces of information: the $g_1$ couplings for the $\Sigma^\pm\to\Lambda$ semileptonic processes, which can be obtained from their respective decay rates through a standard procedure.\footnote{Radiative corrections and a dipole parametrization of the axial vector form factors are two key considerations \cite{Garcia:1985xz}.} The values found are $g_1 =0.619\pm 0.077$ and $g_1=0.597\pm 0.014$ for $\Sigma^+\to\Lambda e^+ \nu$ and $\Sigma^-\to\Lambda e^- \overline{\nu}$, respectively. In passing, it is worth mentioning that set 3 is also particularly interesting because $g_1$ and $g$ are related in the large-$N_c$ limit; in actual numerical analyses, the fits that include $g$ yield more stable solutions \cite{FloresMendieta:2012dn}. 

From the theoretical bent, the analysis of BSD is a rather complex problem. Unlike $K_{\ell 3}$ decays, which are described in terms of two vector form factors, BSD are governed by six form factors due to the participation of both vector and axial vector weak currents, as it was discussed in the introductory section. Although the form factors $f_3(q^2)$ and $g_3(q^2)$ can be ignored for electron modes, there are four relevant form factors left to be determined.
The limit of exact flavor $SU(3)$ symmetry can be first exploited to predict $f_1$ and $f_2$ from the electromagnetic form factors of the proton and the neutron and to set $g_2$ to zero, but still $g_1$ is given in terms of the two  couplings $F$ and $D$.		

There are eight parameters to be determined in the analysis, all of them affecting directly the $g_1$'s. Four of them, $a_1$, $b_2$, $b_3$, and $c_3$ arise from the $1/N_c$ expansion of $A^{kc}$ alone, Eq.~(\ref{eq:akc}), and the remaining four, $d_1,\ldots,d_4$, come from perturbative SB, according to the discussion provided in Sec.~V.B.\ of Ref.~\cite{FloresMendieta:2012dn}. {The fits performed range from the simplest $SU(3)$ symmetric fit to the inclusion of flavor SB effects in $g_1$, $g$ and $f_1$ using data sets 1, 2 and 3. For definiteness, the physical masses of the mesons and baryons listed in Ref.~\cite{Beringer:1900zz} are used, along with $\Delta=0.231$ GeV, $F_\pi=93$ MeV, and $\mu= 1$ GeV. Also, the suggested values of the CKM matrix elements $V_{ud}$ and $V_{us}$ are used as inputs.}

For concreteness, the $SU(3)$ symmetric fit is equivalent to removing all SB effects. As a preliminary analysis, the effects of the leading order parameter $a_1$ are considered with data set 1. The output is listed under the label fit A in Table \ref{tab:summary}. As expected, a one-parameter fit straightforwardly fulfills the $SU(6)$ symmetry relations $F/D=2/3$, $\mathcal{C}=-2D$, and $\mathcal{H}=-3D$, but the corresponding $\chi^2$ is too large to consider this fit A as satisfactory. If the exercise in redone using data set 3, there are two more free parameters, namely $b_2$ and $c_3$ (in this particular case $b_3$ becomes a redundant parameter so it must be removed). The best-fit parameters are listed under the label fit B in Table \ref{tab:summary}. The high $\chi^2$ obtained in the previous cases would be an indicator that SB effects should be present. Before drawing any conclusions, a further numerical analysis will be performed by including $SU(3)$ SB effects in stages. In other words, one-loop corrections are first considered, leaving aside  perturbative SB effects and then both effects are simultaneously introduced. This way their impact on a global fit to data can be better appreciated. A note of caution is in order here: The particular choice of $\mu$ leaves in effect an ambiguity when only one-loop corrections are taken into account which is lifted once the counter terms are included.

Next, and also as exploratory, a fit by retaining only the leading order effects in the $1/N_c$ expansion using data set 1 is performed. First the limit $f_1=f_1^{SU(3)}$ is used (case a) to subsequently add SB effects in $f_1$ (case b). The fit is labeled as fit 1 in Table \ref{tab:summary}. Although the $SU(3)$ invariant couplings are also well reproduced, the total $\chi^2$ increased its value considerably with respect to fit A, to the point that it is hard to connect the output with physics. Thus, leading-order SB effects yield a poor quality fit. The inclusion of $SU(3)$ SB effects in $f_1$ (fit 1b) does not improve the situation, indeed, $\chi^2$ becomes slightly higher.

For the following fit, subleading $1/N_c$ corrections to $g_1$ are added with the inclusion of $b_2$ for the very same data set 1. This fit is listed under fit 2 in Table \ref{tab:summary}. There is a perceptible decrease in the total $\chi^2$, but it still remains too high for the fit to be considered satisfactory (notice that the ratio $F/D$ has increased its value as compared to fit 1). Again, the inclusion of $SU(3)$ SB effects in $f_1$ (fit 2b) does not have a significant effect.

A step further requires the introduction of the four free parameters of the $1/N_c$ expansion of $A^{kc}$ into the analysis, keeping data set 1. This case is listed under fit 3 in Table \ref{tab:summary}. The remarkably small $\chi^2$ obtained should be highlighted in fit 3a as compared with the previous ones. The parameters $a_1$ and $b_2$ are roughly well determined according to the $1/N_c$ expansion expectations. However, what gives some concern are the high values of $b_3$ and $c_3$, which in turn affect the $SU(3)$ couplings $\mathcal{C}$ and $\mathcal{H}$ to the point that they are beyond any reasonable physical value. The addition of $SU(3)$ SB effects in $f_1$ yields best-fit parameters in good accordance with the $1/N_c$ predictions, at the expense of increasing the value of $\chi^2$. Also, the $SU(3)$ invariant couplings are ill reproduced. The fit is however not satisfactory: the numerical analysis performed in Ref.~\cite{FloresMendieta:2006ei} showed highly unstable fits under these working assumptions. The highly discrepant values in the best-fit parameters in cases 3a and 3b should be lifted by adding the counter terms, parametrized here by the perturbative SB terms contained in $\delta A_{\mathrm{SB}}^{kc}$. A variant of the above fit is the inclusion of the decuplet baryon data, which means using data set 2. This constitutes fit 4. Although the $\chi^2$ explodes again, there are slight improvements in the $SU(3)$ couplings $\mathcal{C}$ and $\mathcal{H}$, but still it is not enough to consider the fit as satisfactory.

Finally, a fit where all $SU(3)$ SB corrections enter into play can be performed using data set 3, which is equivalent to using the data about $g_1/f_1$ and $g$.  An analysis under this circumstance has some implications. First, it has been pointed out that both $g_1$ and   $g$ are related in the large-$N_c$ limit, so for a consistent analysis they should be present simultaneously. Also, the new output can be contrasted with the equivalent one obtained in Ref.~\cite{FloresMendieta:2012dn}. But most importantly, the use of the axial couplings only will allow one to check whether the predicted decay rates and asymmetry and spin-angular correlation coefficients agree with the experimental ones. This may be a crucial test of this approach. This time it is an eight-parameter fit for 12 pieces of information. The results are labeled as fit 5 in Table \ref{tab:summary}. Some interesting features emerge in this case. First, the $a_1,\ldots,c_3$ parameters are order one, which completely agrees with expectations. Besides, the SB parameters $d_1,\ldots,d_4$ are roughly suppressed by a factor of $\epsilon\sim 0.3$ with respect to the leading ones, which is consistent with first-order SB. However, what is also worth mentioning is that the $SU(3)$ invariant couplings $D$, $F$, $\mathcal C$, and $\mathcal H$ reach values which are in good agreement with expectations (the coupling $\mathcal{H}$ still remains a little high, but possesses the correct sign). On the other hand, fit 5b deserves special attention because it is where the effects of SB in $f_1$ are evaluated. With the corresponding best-fit parameters, the different flavor contributions of the form factor $g_1$ are listed in Table \ref{tab:AFF} whereas the corresponding $SU(3)$ SB pattern of $f_1$ is displayed in Table \ref{tab:VFF}.

Armed with the vector and axial couplings from fit 5b, the integrated observables for BSD can be estimated. These values are displayed in Table \ref{tab:INTO} for the sake of completeness. The overall behavior of fit 5b is excellent in the sense that the predicted observables are in very good agreement with their experimental counterparts. This was achieved by introducing $SU(3)$ SB to first order in $g_1$ and $g$. For $f_1$, the pattern of SB systematically decreases their $SU(3)$ symmetric values in all the decays considered. The decrease ranges between 3.4 and 4.8\%, which is in perfect agreement with the expectation from second-order SB dictated by the AG theorem.

A variant of fit 5b consists of removing all subleading corrections from $f_1$ and keeping the $a_1$ contribution only. There are no significant changes in the best-fit parameters. The pattern of SB in $f_1$ varies between $\pm 1$\% but the total $\chi^2$ remains practically unaltered.

There are  other  fits which could lead to some interesting results. The ultimate aim of a SB analysis in HSD is to determine $|V_{us}|$. This problem can be addressed here. Using data set 2 with $|V_{ud}|$ as an input and $|V_{us}|$ as a free parameter, the fit yields $|V_{us}| = 0.2262\pm 0.0009$, where the error comes from the fit only. In this case $\chi^2=26.9$ for 11 degrees of freedom. The values of the best-fit parameters are very close to the ones listed for fit 5b and the SB pattern of $f_1$ remains practically unaltered. There is no need to enlarge Table \ref{tab:summary} with such small changes. However, if now $|V_{us}|$ is not restricted to predate around the vicinity of the $|V_{us}|$ from $K_{l3}$ decays but allowed to be an unconstrained parameter, the analysis yields $V_{us} = 0.2357 \pm 0.0028$, with $\chi^2 = 13.4$ for 10 degrees of freedom. The best-fit parameters change by small amounts with respect to the previous case and the pattern of SB reduces accordingly in a fraction of a percent. The $|V_{us}|$ obtained this way fails to fulfill unitarity, however.

Finally, a fit using data set 1, dropping all the decuplet effects, can be performed. This falls in the context of BChPT without dynamical degrees of freedom. In this case, the only $SU(3)$ couplings that enter are $a_1$ and $b_2$, along with the four $b_i$'s. The analysis yields $a_1=0.93\pm 0.01$, $b_2=-0.01\pm 0.07$, $d_1= 0.14\pm 0.05$, $d_2=1.01\pm 0.31$, $d_3=0.31\pm 0.06$, and $d_4=-0.15\pm 0.07$, with $\chi^2=16/8\, \mathrm{dof}$. The leading vector form factors reduce their $SU(3)$ symmetric values by $5\%$ for $\Lambda \to p$, $\Xi^-\to \Sigma^0$ and $\Xi^0\to\Sigma^+$ processes, and by $1.1\%$ and $3.6\%$ for $\Sigma^-\to n$ and $\Xi^-\to\Lambda$ processes, respectively. This output is consistent with other analysis \cite{Lacour:2007wm}.

To close this section, it should be pointed out that the SB pattern of $f_1$ observed here opposes the one observed in Refs.~\cite{FloresMendieta:1998ii,FloresMendieta:2004sk}, obtained within the $1/N_c$ expansion alone. The parent discrepancy is due to a few factors. The experimental information used by then has been partially updated (the data on the $\Xi^0\to \Sigma^+$ semileptonic decay was not available by that time). Besides, $|V_{us}|=0.2196\pm 0.0023$ was used, which is lower than the current determination. Thus, in order for the product $|V_{us}f_1|$ to remain fixed, an increase in $V_{us}$ has to be followed by a decrease in $f_1$ and vice versa. Actually, the analysis of Ref.~\cite{FloresMendieta:1998ii} can be repeated with the updated experimental information. The SB pattern of $f_1$ indeed reduces by an amount equivalent to the increase of $|V_{us}|$ in such a way that $f_1/f_1^{SU(3)}$ for $\Lambda \to p$ semileptonic decay is now slightly lower than one. This last remark leads to a final comment. One cannot yet consider the theoretical issues as closed. It is most important that within the same combined approach used to calculate
the $f_1$'s to $\ord{p^2}$, higher order corrections be also computed. The values displayed for these form factors in Tables \ref{tab:VFF} may provide useful guidance for this non-trivial enterprise.

\begin{turnpage}

\begingroup
\squeezetable
\begin{table}
\caption{\label{tab:summary}Best-fit parameters for the various fits performed. The pertinent values of the equivalent $SU(3)$ couplings $D$, $F$, $\mathcal{C}$, and $\mathcal{H}$ are also listed. The quoted errors come from the fits only.}
\begin{center}
\begin{tabular}{lrrrrrrrrrrrr}
\hline\hline
                                    &        Fit A &         Fit B &       Fit 1a &       Fit 1b &        Fit 2a &       Fit 2b &         Fit 3a &         Fit 3b &         Fit 4a &         Fit 4b &        Fit 5a &        Fit 5b \\ \hline
Data set                            &            1 &             3 &            1 &            1 &             1 &            1 &              1 &              1 &              2 &              2 &             3 &             3 \\
SB in $f_1$                         &    \ding{53} &     \ding{53} &    \ding{53} &    \ding{51} &     \ding{53} &    \ding{51} &      \ding{53} &      \ding{51} &      \ding{53} &      \ding{51} &     \ding{53} &     \ding{51} \\
SB in $g_1$                         &    \ding{53} &     \ding{53} &    \ding{51} &    \ding{51} &     \ding{51} &    \ding{51} &      \ding{51} &      \ding{51} &      \ding{51} &      \ding{51} &     \ding{51} &     \ding{51} \\
$a_1$                               & $1.53(0.01)$ & $ 1.58(0.01)$ & $1.25(0.01)$ & $1.25(0.01)$ & $ 0.99(0.01)$ & $0.99(0.01)$ & $ -0.35(0.02)$ &  $-0.36(0.02)$ & $ -0.24(0.01)$ & $ -0.24(0.01)$ & $ 0.89(0.15)$ & $ 0.95(0.14)$ \\
$b_2$                               &              & $-0.31(0.07)$ &              &              & $ 0.40(0.02)$ & $0.41(0.02)$ & $ -2.40(0.16)$ &  $-2.41(0.13)$ & $ -1.03(0.05)$ & $ -1.03(0.05)$ & $-1.03(0.19)$ & $-1.10(0.19)$ \\
$b_3$                               &              &               &              &              &               &              & $  6.54(0.16)$ &  $ 6.51(0.14)$ & $  4.97(0.04)$ & $  4.97(0.04)$ & $ 1.18(0.15)$ & $ 1.10(0.09)$ \\
$c_3$                               &              & $ 0.72(0.03)$ &              &              &               &              & $  5.86(0.29)$ &  $ 5.65(0.21)$ & $  3.46(0.02)$ & $  3.46(0.02)$ & $ 1.18(0.17)$ & $ 1.07(0.15)$ \\
$d_1$                               &              &               &              &              &               &              &                &                &                &                & $ 0.52(0.12)$ & $ 0.62(0.13)$ \\
$d_2$                               &              &               &              &              &               &              &                &                &                &                & $-0.56(0.25)$ & $-0.57(0.24)$ \\
$d_3$                               &              &               &              &              &               &              &                &                &                &                & $ 0.38(0.05)$ & $ 0.39(0.05)$ \\
$d_4$                               &              &               &              &              &               &              &                &                &                &                & $-0.05(0.08)$ & $-0.06(0.08)$ \\
$D$                                 &              & $ 0.79(0.01)$ &              &              & $ 0.49(0.01)$ & $0.49(0.01)$ & $  0.91(0.02)$ & $  0.91(0.02)$ & $  0.71(0.01)$ & $  0.71(0.01)$ & $ 0.64(0.05)$ & $ 0.66(0.05)$ \\
$F$                                 &              & $ 0.48(0.01)$ &              &              & $ 0.40(0.01)$ & $0.40(0.01)$ & $  0.21(0.02)$ & $  0.20(0.01)$ & $  0.30(0.01)$ & $  0.30(0.01)$ & $ 0.26(0.01)$ & $ 0.25(0.01)$ \\
$\mathcal{C}$                       &              & $-1.94(0.01)$ &              &              &               &              & $ -2.58(0.14)$ & $ -2.47(0.11)$ & $ -1.49(0.02)$ & $ -1.49(0.02)$ & $-1.48(0.07)$ & $-1.48(0.07)$ \\
$\mathcal{H}$                       &              &               &              &              &               &              & $-12.21(0.15)$ & $-12.13(0.15)$ & $-10.52(0.01)$ & $-10.52(0.01)$ & $-2.74(0.27)$ & $-2.50(0.17)$ \\
$F/D$                               & $0.67(0.01)$ & $ 0.60(0.01)$ & $0.67(0.01)$ & $0.67(0.01)$ & $ 0.80(0.01)$ & $0.80(0.01)$ & $  0.23(0.02)$ & $  0.22(0.02)$ & $  0.42(0.01)$ & $  0.42(0.01)$ & $ 0.40(0.03)$ & $ 0.39(0.02)$ \\
$3F-D$                              &              & $ 0.64(0.03)$ &              &              & $ 0.70(0.01)$ & $0.70(0.01)$ & $ -0.29(0.06)$ & $ -0.30(0.05)$ & $  0.19(0.02)$ & $  0.19(0.02)$ & $ 0.13(0.04)$ & $ 0.10(0.04)$ \\
$[f_1/f_1^{SU(3)}]_{\Lambda p}$     &              &               &              &      $0.939$ &               &      $0.949$ &                &        $0.950$ &                &        $0.944$ &               &       $0.952$ \\
$[f_1/f_1^{SU(3)}]_{\Sigma^-n}$     &              &               &              &      $0.979$ &               &      $0.980$ &                &        $0.919$ &                &        $0.969$ &               &       $0.966$ \\
$[f_1/f_1^{SU(3)}]_{\Xi^-\Lambda}$  &              &               &              &      $0.953$ &               &      $0.960$ &                &        $0.929$ &                &        $0.949$ &               &       $0.953$ \\
$[f_1/f_1^{SU(3)}]_{\Xi^-\Sigma^0}$ &              &               &              &      $0.939$ &               &      $0.946$ &                &        $0.982$ &                &        $0.954$ &               &       $0.962$ \\
$[f_1/f_1^{SU(3)}]_{\Xi^0\Sigma^+}$ &              &               &              &      $0.939$ &               &      $0.946$ &                &        $0.982$ &                &        $0.954$ &               &       $0.962$ \\
$\chi^2/\mathrm{dof}$               &     $105/13$ &       $432/9$ &    $2541/13$ &    $2586/13$ &     $1911/12$ &    $1937/12$ &      $17.9/10$ &      $46.8/10$ &       $558/14$ &       $579/14$ &       $5.6/4$ &       $5.5/4$ \\
\hline\hline
\end{tabular}
\end{center}
\end{table}
\endgroup

\end{turnpage}

\begingroup
\squeezetable
\begin{table}
\caption{\label{tab:AFF}Predicted axial vector couplings for non-vanishing $\Delta$. The output of fit 5b is used in the evaluation. The experimental information about $g_1$ and $g$ are used in the fit. Note that $SU(3)$ flavor symmetry breaking is taken into account in two ways: explicitly through perturbative symmetry breaking and implicitly through the integrals occurring in the one-loop corrections. The figure labels refer to Figure 1 of Ref.~\cite{FloresMendieta:2012dn}. The flavor contributions $\mathbf{1}$, $\mathbf{8}$ and $\mathbf{27}$ to $g_1$ arise from Eqs.~(26)-(28) of that reference.}
\begin{center}
\begin{tabular}{lrrrrrrrrrrrrrrr}
\hline\hline
&  & &  & \multicolumn{3}{c}{Figures 1(a)--(c), $\mathcal{O}(\Delta^0)$} & \multicolumn{3}{c}{Figures 1(a)--(c), $\mathcal{O}(\Delta)$} & \multicolumn{3}{c}{Figures 1(a)--(c), $\mathcal{O}(\Delta^2)$} & \multicolumn{3}{c}{Figure 1(d)} \\
Process & Total & Tree & SB & $\mathbf{1}$ & $\mathbf{8}$ & $\mathbf{27}$ & $\mathbf{1}$ & $\mathbf{8}$ & $\mathbf{27}$ & $\mathbf{1}$ & $\mathbf{8}$ & $\mathbf{27}$ & $\mathbf{1}$ & $\mathbf{8}$ & $\mathbf{27}$ \\ \hline
$np$                 & $ 1.270$ & $ 0.909$ & $ 0.356$ & $ 0.041$ & $-0.012$ & $ 0.003$ & $ 0.212$ & $-0.106$ & $-0.002$ & $-0.198$ & $-0.097$ & $-0.001$ & $ 0.245$ & $-0.082$ & $ 0.002$ \\
$\Sigma^\pm \Lambda$ & $ 0.597$ & $ 0.535$ & $ 0.146$ & $ 0.169$ & $-0.059$ & $-0.002$ & $-0.123$ & $-0.030$ & $ 0.001$ & $-0.127$ & $-0.012$ & $ 0.001$ & $ 0.145$ & $-0.048$ & $ 0.001$ \\
$\Lambda p$          & $-0.837$ & $-0.578$ & $-0.008$ & $ 0.119$ & $-0.076$ & $ 0.002$ & $-0.384$ & $ 0.114$ & $-0.002$ & $ 0.116$ & $ 0.039$ & $-0.001$ & $-0.156$ & $-0.026$ & $ 0.003$ \\
$\Sigma^-n$          & $ 0.328$ & $ 0.403$ & $ 0.024$ & $ 0.373$ & $-0.021$ & $-0.002$ & $-0.515$ & $ 0.060$ & $-0.001$ & $-0.112$ & $-0.004$ & $ 0.000$ & $ 0.109$ & $ 0.018$ & $-0.002$ \\
$\Xi^- \Lambda$      & $ 0.296$ & $ 0.042$ & $ 0.111$ & $-0.288$ & $-0.059$ & $ 0.001$ & $ 0.507$ & $-0.036$ & $ 0.003$ & $ 0.011$ & $-0.009$ & $ 0.001$ & $ 0.011$ & $ 0.002$ & $ 0.000$ \\
$\Xi^-\Sigma^0$      & $ 0.828$ & $ 0.643$ & $-0.126$ & $ 0.029$ & $ 0.004$ & $ 0.000$ & $ 0.150$ & $ 0.038$ & $-0.002$ & $-0.140$ & $ 0.034$ & $-0.001$ & $ 0.174$ & $ 0.029$ & $-0.004$ \\
$\Xi^0\Sigma^+$      & $ 1.170$ & $ 0.909$ & $-0.178$ & $ 0.041$ & $ 0.006$ & $-0.001$ & $ 0.212$ & $ 0.053$ & $-0.002$ & $-0.198$ & $ 0.048$ & $-0.002$ & $ 0.245$ & $ 0.041$ & $-0.005$ \\
$\Delta N$           & $-2.037$ & $-1.483$ & $-0.545$ & $-0.479$ & $ 0.200$ & $-0.004$ & $ 0.378$ & $-0.231$ & $-0.019$ & $ 0.291$ & $ 0.122$ & $ 0.002$ & $-0.401$ & $ 0.133$ & $-0.003$ \\
$\Sigma^*\Lambda$    & $-1.709$ & $-1.483$ & $-0.358$ & $-0.479$ & $ 0.206$ & $ 0.011$ & $ 0.378$ & $-0.085$ & $ 0.009$ & $ 0.291$ & $ 0.071$ & $-0.001$ & $-0.401$ & $ 0.133$ & $-0.003$ \\
$\Sigma^*\Sigma$     & $-1.751$ & $-1.483$ & $ 0.163$ & $-0.479$ & $-0.225$ & $-0.011$ & $ 0.378$ & $-0.055$ & $ 0.013$ & $ 0.291$ & $-0.068$ & $-0.006$ & $-0.401$ & $ 0.133$ & $-0.003$ \\
$\Xi^*\Xi$           & $-1.409$ & $-1.483$ & $ 0.089$ & $-0.479$ & $-0.003$ & $ 0.000$ & $ 0.378$ & $ 0.076$ & $ 0.042$ & $ 0.291$ & $-0.050$ & $-0.002$ & $-0.401$ & $ 0.133$ & $-0.003$ \\
\hline\hline
\end{tabular}
\end{center}
\end{table}
\endgroup

\begingroup
\begin{table}
\caption{\label{tab:VFF}Predicted leading vector form factors $f_1$ for non-vanishing $\Delta$. The experimental information about $g_1$ and $g$ are used in the fit.}
\begin{center}
\begin{tabular}{lcccc}
\hline\hline
Process ~~~&~ $\displaystyle \frac{f_1}{f_1^{SU(3)}}-1~$ &~ $\displaystyle \frac{f_1^{\mathrm{(a)}}+f_1^{\mathrm{(b)}}}{f_1^{SU(3)}}$~ & ~$\displaystyle \frac{f_1^{\mathrm{(c)}}+f_1^{\mathrm{(d)}}}{f_1^{SU(3)}}$ ~&~ $\displaystyle \frac{f_1}{f_1^{SU(3)}}$ \\
\hline
$\Lambda p$     & $-0.048$ & $-0.026$ & $-0.022$ & $0.952$ \\
$\Sigma^-n$     & $-0.034$ & $-0.013$ & $-0.022$ & $0.966$ \\
$\Xi^- \Lambda$ & $-0.047$ & $-0.025$ & $-0.022$ & $0.953$ \\
$\Xi^-\Sigma^0$ & $-0.038$ & $-0.016$ & $-0.022$ & $0.962$ \\
$\Xi^0\Sigma^+$ & $-0.038$ & $-0.016$ & $-0.022$ & $0.962$ \\
\hline\hline
\end{tabular}
\end{center}
\end{table}
\endgroup

\begingroup
\squeezetable
\begin{table}
\caption{\label{tab:INTO} Values of predicted observables for eight observed baryon semileptonic decays. The output of fit 5b is used in the evaluation. The units of $R$ are $10^{-3}\, \textrm{s}^{-1}$ for neutron decay and $10^6 \, \textrm{s}^{-1}$ for the others.}
\begin{center}
\begin{tabular}{lrrrrrrrr}
 \hline \hline
&
$n \to p e^- \overline \nu_e$ &
$~~~\Sigma^+ \to \Lambda e^+ \nu_e$ &
$~~~\Sigma^- \to \Lambda e^- \overline \nu_e$ &
$~~~\Lambda \to p e^- \overline \nu_e$ &
$~~~\Sigma^- \to n e^-\overline \nu_e$ &
$~~~\Xi^- \to \Lambda e^- \overline \nu_e$ &
$~~~\Xi^-\to \Sigma^0 e^- \overline \nu_e$ &
$~~~\Xi^0\to \Sigma^+ e^- \overline \nu_e$ \\
\hline
$R$
& $1.128$ & $0.232$ & $0.387$ & $2.949$ & $6.284$ & $2.844$ & $0.454$ & $0.820$ \\
$\alpha_{e\nu}$
& $-0.079$ & $-0.406$ & $-0.414$ & $-0.026$ & $0.340$ & $0.559$ &  & \\
$\alpha_e$
& $-0.087$ &  & & $0.014$ & $-0.627$ &  & & \\
$\alpha_\nu$
& $0.987$ &  & & $0.977$ & $-0.357$ &  & & \\
$\alpha_B$
&  & &  & $-0.586$ & $0.668$ &  & & \\
$A$
&  & & $0.049$ &  & & $0.583$ &  & \\
$B$
&  & & $0.888$ &  & &  & & \\
$g_1/f_1$
& $1.270$ &  & & $0.718$ & $-0.340$ & $0.254$ & $1.217$ & $1.217$ \\
\hline \hline
\end{tabular}
\end{center}
\end{table}
\endgroup

\section{\label{sec:summary}Summary and conclusions}

The leading $SU(3)$ SB corrections to the form factors $f_1$ of the vector currents were calculated in a framework consistent with chiral symmetry and the $1/N_c$ expansion. Those corrections are $\ord{p^2}$ in the chiral expansion, and included higher orders of the $1/N_c$ expansion. The results were compared with previous calculations and also were confronted with the experimental observables for BSD and the strong decays of the decuplet baryons. Several different fits were carried out in order to elucidate the relative importance of the various effects. A summary of those fits is presented in Table \ref{tab:summary}. The following conclusions can be derived from those results:

a)  The $1/N_c$ corrections to the axial vector currents are very important. These are reflected in the deviation of the relations between the couplings  $F$,  $D$, ${\cal{ C}}$  and ${\cal{ H}}$  which hold in the $SU(6)$ limit. Both, the strong decuplet to octet strong transitions as well as the weak decays are sensitive to those sub-leading corrections.

b) The effects of $SU(3)$ SB in $f_1$ are calculable at the order considered here and turn out to be about $-5\%$.   The hyperon weak decay observables at the current degree of accuracy  are not sensitive to those effects, endorsing the same claim made by Cabibbo, Swallow and Winston  \cite{Cabibbo:2003cu}. It is noted that the $1/N_c$ corrections to the axial currents which determine the vertices in the loop-diagrams  do not affect significantly the correction to $f_1$.

c) The effects of $SU(3)$ SB on the axial vector couplings  $g_1$ are on the other hand very important, as shown in Table V. The octet pieces of the SB are the dominant ones with magnitudes up to 0.3, while the 27-plet pieces are much smaller, at most 20\% of the octet ones and in most cases much smaller than that. Because of the small tree level value of the axial coupling of the transition $\Xi^-\Lambda$, the subleading corrections, which include the SB effects, turn out to be larger than the leading term.  For the other cases the subleading corrections do not exceed the expected 30\% of the leading order value.

d) In the calculation of the $SU(3)$ SB corrections to $f_1$ it is noted that the inclusion of the subleading in $1/N_c$ corrections to the meson-baryon couplings produce small deviations, and to the current level of accuracies they are unnoticeable. Similarly, any $SU(3)$ breaking effects on those couplings turn out to be insignificant: they are of higher order in the chiral expansion, but they were evaluated in order to check their insignificance.

e) Perhaps the most important reason for accurate calculations of HSD is to provide an additional accurate extraction of $|V_{us}|$. At present the ratio of $K_{\ell 2}$ to $\pi_{\ell 2}$ decay together with the ratio $F_K/F_\pi$ from lattice QCD and $|V_{ud}|$ from super-allowed $\beta$ decay, and the $K_{\ell 3}$  decays give the most accurate determinations. The smallness of $|V_{ub}|$ means that $|V_{us}|$ is very close to the unitarity limit. A test of unitarity the CKM matrix requires as accurate as possible results for $|V_{us}|$, for which the increase in precision from HSD would be welcome. This however will require further experimental progress in the determination of the various HSD  parameters.

The natural next step in the study of the BSD in the present framework is the calculation in the combined framework of the $\xi$ expansion to $\ord{\xi^3}$. This is the next order beyond the one presented here. While such a complete calculation for the axial currents is already available  for two flavors, it needs to be implemented for three flavors and also for the vector currents. This will be the objective of future work.
 
To close this article, it is worth quoting a sentence found in Ref.~\cite{Dashen:1993jt}: \lq\lq It will take a lot more work to see whether the $1/N_c$ expansion can be combined with baryon chiral perturbation theory to analyze
baryon properties in a systematic and controlled expansion." Two decades later, one can claim that this task is indeed possible.

\begin{acknowledgments}
This work was supported in part by DOE Contract No.\ DE-AC05-06OR23177 under which JSA operates the Thomas Jefferson National Accelerator Facility (RFM and JLG), and by the National Science Foundation (USA) through grant PHY-1307413 (JLG), and by Consejo Nacional de Ciencia y Tecnolog{\'\i}a and Fondo de Apoyo a la Investigaci\'on (Universidad Aut\'onoma de San Luis Potos{\'\i}), Mexico (RFM).
\end{acknowledgments}

\appendix

\section{\label{app:integrals}Loop integrals}

The integrals over the loops displayed in Fig.~\ref{fig:vcloop} are fully discussed in this section and the most general results needed in the present analysis are provided for the sake of completeness.

First, for the Feynman diagram displayed in Fig.~\ref{fig:vcloop}(a), the loop integral can be written in the most general way as
\bea
&  & J_{ij}^\mu(m_1,m_2,\Delta,\mu;q^2) \nonumber \\
&  & \mbox{\hglue0.5truecm} = \frac{i}{F_\pi^2} (\mu^2)^{\frac{4-d}{2}} \int \frac{d^dk}{(2\pi)^d} \frac{(2k+q)^\mu k_i(k+q)_j}{(k^2-m_1^2+i\varepsilon)[(k+q)^2-m_2^2+i\varepsilon](p^0-k^0-\Delta+i\varepsilon)}, \label{eq:iadef}
\eea
where $m_1$ and $k$ and $m_2$ and $k+q$ denote respectively the masses and four-momenta of the mesons in the loop, $q$ is the four-momentum transfer, $\Delta$ is the baryon decuplet-octet mass difference, and $d=4-\epsilon$ to use dimensional regularization with scale $\mu$. Due to the Lorentz structure of $J_{ij}^\mu(m_1,m_2,\Delta,\mu;q^2)$, it can be separated into temporal $J_{ij}^0(m_1,m_2,\Delta,\mu;q^2)$ and spatial $J_{ij}^k(m_1,m_2,\Delta,\mu;q^2)$ components. The former, which is the one needed here, can be decomposed as $J_A(m_1,m_2,\Delta,\mu;q^2)\delta_{ij}+J_B(m_1,m_2,\Delta,\mu;q^2)q_iq_j$. In the $q^2\to 0$ limit,
\begin{equation}
I_a(m_1,m_2,\Delta,\mu;0) \equiv \lim_{q^2\to 0}J_A(m_1,m_2,\Delta,\mu;q^2),
\end{equation}
where $I_a$ is the integral associated to the one-loop correction to the baryon vector current of Fig.~\ref{fig:vcloop}(a) at zero recoil. Without further ado, the resultant expression reads,
\begin{eqnarray}
&  & 32\pi^2f^2I_a(m_1,m_2,\Delta,\mu;0) \nonumber \\
&  & \mbox{\hglue0.5truecm} = -(m_1^2+m_2^2-4\Delta^2)\lambda_\epsilon - \frac32 (m_1^2+m_2^2) + \frac{28}{3}\Delta^2 \nonumber \\
&  & \mbox{\hglue1.00truecm} + \frac{1}{3(m_1^2-m_2^2)} \left[(3m_1^4-12m_1^2\Delta^2+8\Delta^4) \ln\frac{m_1^2}{\mu^2} - (3m_2^4-12m_2^2\Delta^2+8\Delta^4) \ln\frac{m_2^2}{\mu^2} \right] \nonumber \\
&  & \mbox{\hglue1.0truecm} + \frac83 \frac{\Delta}{m_1^2-m_2^2} \times \left\{ \begin{array}{ll} 
\displaystyle 2(m_1^2-\Delta^2)^{3/2} \left[ \frac{\pi}{2} -\tan^{-1} \left[ \frac{\Delta}{\sqrt{m_1^2-\Delta^2}} \right] \right] \\[6mm]
\displaystyle - 2(m_2^2-\Delta^2)^{3/2} \left[ \frac{\pi}{2} - \tan^{-1} \left[ \frac{\Delta}{\sqrt{m_2^2-\Delta^2}} \right] \right], & |\Delta| < m_1 < m_2 \\[6mm]
\displaystyle -(\Delta^2-m_1^2)^{3/2} \ln \left[\frac{\Delta-\sqrt{\Delta^2-m_1^2}}{\Delta+\sqrt{\Delta^2-m_1^2}} \right] \\[6mm]
\displaystyle - 2(m_2^2-\Delta^2)^{3/2} \left[ \frac{\pi}{2} -\tan^{-1} \left[ \frac{\Delta}{\sqrt{m_2^2-\Delta^2}} \right] \right], & m_1 < |\Delta| < m_2 \\[6mm]
\displaystyle -(\Delta^2-m_1^2)^{3/2} \ln \left[\frac{\Delta-\sqrt{\Delta^2-m_1^2}}{\Delta+\sqrt{\Delta^2-m_1^2}}\right] \\[6mm]
\displaystyle + (\Delta^2-m_2^2)^{3/2} \ln \left[ \frac{\Delta-\sqrt{\Delta^2-m_2^2}}{\Delta+\sqrt{\Delta^2-m_2^2}}\right], & m_1 < m_2 < |\Delta|
\end{array}
\right. \label{eq:ia}
\end{eqnarray}
where
\begin{equation}
\lambda_\epsilon = \frac{2}{\epsilon} - \gamma + \ln(4\pi),
\end{equation}
with $\gamma \simeq 0.577216$ the Euler constant. Without loss of generality, the condition $m_1<m_2$ has been assumed in order to get the above result.

Now, the correction arising from the Feynman diagram displayed in Fig.~\ref{fig:vcloop}(b) is given in terms of the derivatives of the basic loop integral \cite{FloresMendieta:2000mz}
\begin{equation}
\delta^{ij} I_b(m,\Delta,\mu) = \frac{i}{F_\pi^2}(\mu^2)^\frac{4-d}{2} \int \frac{d^dk}{(2\pi)^d} \frac{(k^i)(-k^j)}{(k^0-\Delta+i\varepsilon)(k^2-m^2+i\varepsilon)}.
\end{equation}
An explicit calculation yields\footnote{Here the sign in front of the term $\frac72 \Delta m^2$ in the function $I_b(m,\Delta,\mu)$ has been fixed. The opposite sign, which is incorrect, was used in Refs.~\cite{FloresMendieta:2000mz,FloresMendieta:2006ei,FloresMendieta:2009rq,FloresMendieta:2012dn,Ahuatzin:2010ef}.}.
\begin{eqnarray}
24\pi^2 f^2 I_b(m,\Delta,\mu) & = & -\Delta\left[\Delta^2-\frac32 m^2\right] \lambda_\epsilon + \Delta \left[\Delta^2-\frac32 m^2\right] \ln{\frac{m^2}{\mu^2}} - \frac83 \Delta^3 + \frac72 \Delta m^2 \nonumber \\
&  & \mbox{} + \left\{ \begin{array}{ll}
\displaystyle 2(m^2-\Delta^2)^{3/2} \left[\frac{\pi}{2} - \tan^{-1} \left[\frac{\Delta}{\sqrt{m^2-\Delta ^2}}\right] \right], & |\Delta|< m \\[3mm]
\displaystyle - (\Delta^2-m^2)^{3/2} \left[-2i \pi +\ln \left[\frac{\Delta - \sqrt{\Delta^2-m^2}}{\Delta + \sqrt{\Delta^2-m^2}} \right] \right], & |\Delta| > m. \end{array} \right. \label{eq:ib}
\end{eqnarray}

From this function it follows that
\bea
16\pi^2f^2 I_b^{(1)}(m,\Delta,\mu) & = & (m^2-2\Delta^2)\left[ \lambda_\epsilon + 1 - \ln \frac{m^2}{\mu^2} \right] - 2\Delta^2 \nonumber\\
&  & \mbox{} -\left\{ \begin{array}{ll} \displaystyle 4\Delta\sqrt{m^2-\Delta^2} \left[ \frac{\pi}{2}-\tan^{-1} \left[ \frac{\Delta}{\sqrt{m^2-\Delta^2}}\right]\right], & |\Delta|< m \\[6mm]
\displaystyle 2\Delta \sqrt{\Delta^2-m^2} \left[ -2i\pi + \ln \left[ \frac{\Delta-\sqrt{\Delta^2-m2}}{\Delta+\sqrt{\Delta^2-m^2}} \right]\right], & |\Delta| > m. \end{array} \right. \label{eq:ibp}
\eea
\bea
4\pi^2f^2 I_b^{(2)}(m,\Delta,\mu) & = & -\Delta \left[ \lambda_\epsilon + 1 - \ln \frac{m^2}{\mu^2} \right] \nonumber \\
&  & \mbox{} + \left\{ \begin{array}{ll} \displaystyle -\frac{m^2-2\Delta^2}{\sqrt{m^2-\Delta^2}} \left[ \frac{\pi}{2}-\tan^{-1} \left[ \frac{\Delta}{\sqrt{m^2-\Delta^2}}\right]\right], & |\Delta|< m \\[6mm] 
\displaystyle \frac{m^2-2\Delta^2}{2\sqrt{\Delta^2-m^2}} \left[-2i\pi + \ln \left[ \frac{\Delta-\sqrt{\Delta^2-m^2}}{\Delta+\sqrt{\Delta^2-m2}}\right]\right],  & |\Delta| > m. \end{array} \right. \label{eq:ibpp}
\eea
\bea
4\pi^2f^2 I_b^{(3)}(m,\Delta,\mu) & = & -\lambda_\epsilon -\frac{\Delta^2}{m^2-\Delta^2} + \ln \frac{m^2}{\mu^2} \nonumber \\
&  & \mbox{} + \left\{ \begin{array}{ll} \displaystyle \frac{\Delta(3m^2-2\Delta^2)}{(m^2-\Delta^2)^{3/2}} \left[ \frac{\pi}{2} -\tan^{-1} \left[ \frac{\Delta}{\sqrt{m^2-\Delta^2}}\right] \right], & |\Delta|< m \\[6mm] 
\displaystyle \frac{\Delta(3m^2-2\Delta^2)}{2(\Delta^2-m^2)^{3/2}} \left[-2i\pi + \ln \left[ \frac{\Delta-\sqrt{\Delta^2-m^2}}{\Delta+\sqrt{\Delta^2-m^2}}\right]\right], & |\Delta| > m.
\end{array} \right. \label{eq:ibppp}
\eea

Therefore, the function $I_b(m,\Delta,\mu)$ and its derivatives in the $\Delta \to 0$ limit follow accordingly; they read
\begin{equation}
I_b(m,0,\mu) = \frac{m^3}{24\pi^2 f^2},
\end{equation}
\begin{equation}
I_b^{(1)}(m,0,\mu) = \frac{m^2}{16\pi^2 f^2} \left[ \lambda_\epsilon + 1 - \ln{\frac{m^2}{\mu^2}} \right], \label{eq:ib1}
\end{equation}
\begin{equation}
I_b^{(2)}(m,0,\mu) = -\frac{m}{8\pi f^2}, \label{eq:ib2}
\end{equation}
and
\begin{equation}
I_b^{(3)}(m,0,\mu) = \frac{1}{4\pi^2f^2} \left[-\lambda_\epsilon + \ln{\frac{m^2}{\mu^2}} \right]. \label{eq:ib3}
\end{equation}

Next, for the Feynman diagrams displayed in Figs.~\ref{fig:vcloop}(c) and \ref{fig:vcloop}(d), it is useful to introduce the scalar function
\begin{equation}
L_{r,m} = i(\mu^2)^{\frac{4-d}{2}} \int\frac{d^d\ell}{(2\pi)^d}\frac{(\ell^2)^r}{(\ell^2-\beta+i\varepsilon)^m},
\end{equation}
where $r$ and $m$ are integers and $\beta$ is an independent function of $\ell^2$. An explicit calculation yields
\begin{equation}
L_{r,m} = \frac{(-1)^{r-m+1}}{16\pi^2} \frac{\Gamma(r+2-\frac{\epsilon}{2}) \Gamma(-r+m-2+\frac{\epsilon}{2})}{\Gamma(m) \Gamma(2-\frac{\epsilon}{2})} (4\pi \mu^2)^{\frac{\epsilon}{2}} \beta^{r-m+2-\frac{\epsilon}{2}}, \label{eq:lrm}
\end{equation}

Now, the loop integral of Fig.~\ref{fig:vcloop}(c) is given in chiral perturbation theory by
\begin{equation}
I^\alpha(m_1,m_2,\mu;q^2) = \frac{i}{F_\pi^2}(\mu^2)^{\frac{4-d}{2}} \int \frac{d^dk}{(2\pi)^d}\frac{(2\slash{k}-\slash{q})q^\alpha}{[(k-q)^2-m_2^2+i\varepsilon](k^2-m_1^2+i\varepsilon)}. \label{eq:icdef}
\end{equation}
$I^\alpha$ will have a piece proportional to $\gamma^\alpha$ and another one proportional to $q^\alpha$. The former is the one related to the vector form factor $f_1(q^2)$.

By using the conventional Feynman method to combine denominators, it is easy to see that the contribution of $I^\alpha$ proportional to $\gamma^\alpha$ can be written as
\begin{equation}
I_c(m_1,m_2,\mu;q^2) = \frac{1}{F_\pi^2} \int_0^1dx \frac{2}{4-\epsilon} L_{1,2},
\end{equation}
where $L_{1,2}$ can be obtained from Eq.~(\ref{eq:lrm}) and
\begin{equation}
\beta=-q^2x(1-x) + m_2^2x + m_1^2(1-x)
\end{equation}
accordingly.

A standard calculation yields
\begin{eqnarray}
16\pi^2f^2 I_c(m_1,m_2,\mu;q^2) & = & \frac16 (q^2-3m_1^2-3m_2^2) \lambda_\epsilon - \frac{1}{12} (q^2-3m_1^2-3m_2^2) \left[ \ln{\frac{m_1^2}{\mu ^2}} + \ln{\frac{m_2^2}{\mu^2}} \right] \nonumber \\
&  & \mbox{} + \frac49 q^2 - \frac76 (m_1^2+m_2^2) + \frac{(m_1^2-m_2^2)^2}{6q^2} - \frac{m_1^4-m_2^4}{4q^2} \ln{\frac{m_2^2}{m_1^2}} \nonumber \\
&  & \mbox{} + \frac{(m_1^2-m_2^2)^3 + \left[q^2(q^2-2m_1^2-2m_2^2)+(m_1^2-m_2^2)^2\right]^{3/2}}{12(q^2)^2} \ln{\frac{m_2^2}{m_1^2}} \nonumber \\
&  & \mbox{} + \frac{\left[q^2(q^2-2m_1^2-2m_2^2)+(m_1^2-m_2^2)^2\right]^{3/2}}{6(q^2)^2} \nonumber \\
&  & \times \ln{\left[\frac{-q^2-m_1^2+m_2^2+\sqrt{q^2(q^2-2m_1^2-2m_2^2)+(m_1^2-m_2^2)^2}}{q^2-m_1^2+m_2^2+\sqrt{q^2(q^2-2m_1^2-2m_2^2)+(m_1^2-m_2^2)^2}}\right]}.\nonumber \\
\end{eqnarray}

In the $q^2\to 0$ limit, $I_c(m_1,m_2,\mu;q^2)$ reduces to
\begin{equation}
32\pi^2f^2 I_c(m_1,m_2,\mu;0) = -(m_1^2+m_2^2)\lambda_\epsilon - \frac32 (m_1^2+m_2^2) + \frac{1}{m_1^2-m_2^2}\left[ m_1^4 \ln{\frac{m_1^2}{\mu ^2}} - m_2^4 \ln{\frac{m_2^2}{\mu^2}} \right].
\label{eq:ic}
\end{equation}

Finally, for the Feynman diagram displayed in Fig.~\ref{fig:vcloop}(d), the integral over the loop is
\begin{eqnarray}
I_d(m,\mu) & = & \frac{i}{F_\pi^2}(\mu^2)^\frac{4-d}{2} \int \frac{d^dk}{(2\pi)^d} \frac{1}{k^2-m^2+i\varepsilon} \nonumber \\
& = & \frac{1}{F_\pi^2} L_{0,1},
\end{eqnarray}
where $\beta=m^2$ in this case. A straightforward calculation yields
\begin{equation}
I_d(m,\mu) = \frac{m^2}{16\pi^2\,F_\pi^2}\left[ -\lambda_\epsilon - 1 + \ln{\frac{m^2}{\mu^2}} \right]. \label{eq:id}
\end{equation}

\section{\label{app:reduc1}Reduction of baryon operators}

The full list of operator reductions performed in the current analysis is presented in this appendix. For $N_c=3$, there appeared operator products containing up to eight-body operators for which the reductions turned out to be quite involved. Due to the fact that for any $SU(6)$ representation polynomials in the spin-flavor generators $J^i$, $T^a$ and $G^{ia}$ form a complete set of operators, the reductions were always possible. Apart from using well-known decompositions among operators, a particularly useful identity was also used, namely,
\begin{equation}
[T^a,X^b] = if^{abc} X^c\; , \nonumber
\end{equation}
where $X^b$ stands for \textit{any} spin-0 or spin-1 flavor octet. For instance,
\begin{equation}
[T^a,A^{ib}]= if^{abc} A^{ic}\; , \nonumber
\end{equation}
where $A^{ib}$ is the axial-vector current operator (\ref{eq:akc}), or
\begin{equation}
[T^a,[J^2,A^{ib}]] = if^{abc}[J^2,A^{ic}]\; , \nonumber
\end{equation}
or
\begin{equation}
d^{abe}[T^c,\{J^2,\{T^a,T^b\}\}] = if^{ceg}d^{gde}\{J^2,\{T^d,T^e\}\}\; , \nonumber
\end{equation}
to name but a few.

For computational ease, the second and third summands of Eq.~(\ref{eq:vc1b}) can be respectively rewritten as
\begin{equation}
\{A^{ja},[T^c,[J^2,A^{jb}]]\} = if^{cbe} \{A^{ja},[J^2,A^{je}]\} \nonumber
\end{equation}
and
\bea
&  & [A^{ja},[[J^2,[J^2,A^{jb}]],T^c]] - \frac12 [[J^2,A^{ja}],[[J^2,A^{jb}],T^c]] \nonumber \\
&  & \mbox{\hglue0.5truecm} = \frac32 if^{bce} [[J^2,A^{je}],[J^2,A^{ja}]] - if^{bce} [J^2,[[J^2,A^{je}],A^{ja}]]\; , \nonumber
\eea
where
\begin{equation}
if^{bce} [J^2,[[J^2,A^{je}],A^{ja}]] F^{ab} = 0\; , \nonumber
\end{equation}
for $F^{ab} = \delta^{ab}$, $d^{ab8}$, or $\delta^{a8}\delta^{b8}$.

The operator reductions performed, for arbitrary $N_c$ and $N_f$, are listed below. These expressions are to be evaluated at the physical values $N_f=N_c=3$.

\subsection{$i f^{acb} A^{ia}A^{ib}$}

\begin{equation}
i f^{acb} G^{ia}G^{ib} = \frac38 N_f T^c,
\end{equation}

\begin{equation}
i f^{acb}(G^{ia}\mathcal{D}_2^{ib} + \mathcal{D}_2^{ia}G^{ib}) = \frac12 N_f\{J^r,G^{rc}\},
\end{equation}

\begin{equation}
if^{acb}(G^{ia}\mathcal{D}_3^{ib} + \mathcal{D}_3^{ia}G^{ib}) = (N_c+N_f) \{J^r,G^{rc}\} + \frac12 (N_f-2) \{J^2,T^c\},
\end{equation}

\begin{equation}
if^{acb}(G^{ia}\mathcal{O}_3^{ib} + \mathcal{O}_3^{ia}G^{ib}) = \frac32 N_f T^c - \frac32 (N_c+N_f) \{J^r,G^{rc}\} + \frac12 (N_f+3) \{J^2,T^c\}, 
\end{equation}

\begin{equation}
if^{acb}\mathcal{D}_2^{ia}\mathcal{D}_2^{ib} = \frac14 N_f\{J^2,T^c\},
\end{equation}

\begin{equation}
if^{acb}(\mathcal{D}_2^{ia}\mathcal{D}_3^{ib} + \mathcal{D}_3^{ia}\mathcal{D}_2^{ib}) = N_f \{J^2,\{J^r,G^{rc}\}\}, 
\end{equation}

\begin{equation}
 if^{acb}(\mathcal{D}_2^{ia}\mathcal{O}_3^{ib} + \mathcal{O}_3^{ia}\mathcal{D}_2^{ib}) = 0, 
\end{equation}

\begin{equation}
if^{acb}\mathcal{D}_3^{ia}\mathcal{D}_3^{ib} = (N_c+N_f) \{J^2,\{J^r,G^{rc}\}\} + \frac12 (N_f-2) \{J^2,\{J^2,T^c\}\}, 
\end{equation}

\begin{equation}
 if^{acb}(\mathcal{D}_3^{ia}\mathcal{O}_3^{ib} + \mathcal{O}_3^{ia}\mathcal{D}_3^{ib}) = 0,
\end{equation}

\begin{eqnarray}
if^{acb}\mathcal{O}_3^{ia}\mathcal{O}_3^{ib} & = & \frac32 N_f T^c - \frac32 (N_c+N_f) \{J^r,G^{rc}\} + \frac12 (4N_f+3) \{J^2,T^c\} \nonumber \\
&  & \mbox{} - \frac54 (N_c+N_f) \{J^2,\{J^r,G^{rc}\}\} + \frac14 (N_f+5) \{J^2,\{J^2,T^c\}\}.
\end{eqnarray}

\subsection{$i f^{acb} A^{ia} J^2 A^{ib}$}

\begin{equation}
if^{acb}G^{ia} J^2 G^{ib} = \frac34 N_f T^c - \frac12 (N_c+N_f) \{J^r,G^{rc}\} + \frac{1}{16} (3N_f+8) \{J^2,T^c\}, 
\end{equation}

\begin{equation}
if^{acb}(G^{ia} J^2 \mathcal{D}_2^{ib} + \mathcal{D}_2^{ia} J^2 G^{ib}) = \frac14 N_f \{J^2,\{J^r,G^{rc}\}\},
\end{equation}

\begin{equation}
if^{acb}(G^{ia} J^2 \mathcal{D}_3^{ib} + \mathcal{D}_3^{ia} J^2 G^{ib}) = \frac12 (N_c+N_f) \{J^2,\{J^r,G^{rc}\}\} + \frac14 (N_f-2) \{J^2,\{J^2,T^c\}\}, 
\end{equation}

\begin{eqnarray}
if^{acb}(G^{ia} J^2 \mathcal{O}_3^{ib} + \mathcal{O}_3^{ia} J^2 G^{ib}) & = & 3 N_f T^c - 3 (N_c+N_f) \{J^r,G^{rc}\} + \frac14 (13N_f+12) \{J^2,T^c\} \nonumber \\
&  & \mbox{} -\frac74 (N_c+N_f) \{J^2,\{J^r,G^{rc}\}\} + \frac14 (N_f+7) \{J^2,\{J^2,T^c\}\}\; , \nonumber \\
\end{eqnarray}

\begin{equation}
if^{acb}\mathcal{D}_2^{ia} J^2 \mathcal{D}_2^{ib} = \frac18 N_f\{J^2,\{J^2,T^c\}\},
\end{equation}

\begin{equation}
if^{acb}(\mathcal{D}_2^{ia} J^2 \mathcal{D}_3^{ib} + \mathcal{D}_3^{ia} J^2 \mathcal{D}_2^{ib}) = \frac12 N_f\{J^2,\{J^2,\{J^r,G^{rc}\}\}\}, 
\end{equation}

\begin{equation}
if^{acb}(\mathcal{D}_2^{ia} J^2 \mathcal{O}_3^{ib} + \mathcal{O}_3^{ia} J^2 \mathcal{D}_2^{ib}) = 0,
\end{equation}

\begin{equation}
if^{acb}\mathcal{D}_3^{ia} J^2 \mathcal{D}_3^{ib} = \frac12 (N_c+N_f) \{J^2,\{J^2,\{J^r,G^{rc}\}\}\}+ \frac14 (N_f-2) \{J^2,\{J^2,\{J^2,T^c\}\}\},
\end{equation}

\begin{equation}
if^{acb}(\mathcal{D}_3^{ia} J^2 \mathcal{O}_3^{ib} + \mathcal{O}_3^{ia} J^2 \mathcal{D}_3^{ib}) = 0, 
\end{equation}

\begin{eqnarray}
if^{acb}\mathcal{O}_3^{ia} J^2 \mathcal{O}_3^{ib} & = & 3 N_f T^c - 3 (N_c+N_f) \{J^r,G^{rc}\} + \frac14 (25N_f+12) \{J^2,T^c\} \nonumber\\
&  & -\frac{19}{4} (N_c+N_f) \{J^2,\{J^r,G^{rc}\}\} + \frac14 (11N_f+19) \{J^2,\{J^2,T^c\}\} \nonumber \\
&  & -\frac98 (N_c+N_f) \{J^2,\{J^2,\{J^r,G^{rc}\}\}\} + \frac18 (N_f+9) \{J^2,\{J^2,\{J^2,T^c\}\}\}. \nonumber \\
\end{eqnarray}

\subsection{$i(f^{aec}d^{be8}-f^{bec}d^{ae8}-f^{abe}d^{ec8}) A^{ia} A^{ib}$}

\begin{equation}
i(f^{aec}d^{be8}-f^{bec}d^{ae8}-f^{abe}d^{ec8}) G^{ia} G^{ib} = 0,
\end{equation}

\begin{equation}
i(f^{aec}d^{be8}-f^{bec}d^{ae8}-f^{abe}d^{ec8}) (G^{ia} \mathcal{D}_2^{ib} + \mathcal{D}_2^{ia}G^{ib}) = 0,
\end{equation}

\begin{equation}
i(f^{aec}d^{be8}-f^{bec}d^{ae8}-f^{abe}d^{ec8}) (G^{ia} \mathcal{D}_3^{ib} + \mathcal{D}_3^{ia}G^{ib}) = \{T^c,\{J^r,G^{r8}\}\} - \{T^8,\{J^r,G^{rc}\}\}, 
\end{equation}

\begin{equation}
i(f^{aec}d^{be8}-f^{bec}d^{ae8}-f^{abe}d^{ec8}) (G^{ia} \mathcal{O}_3^{ib} + \mathcal{O}_3^{ia} G^{ib}) = -\frac32 \{T^c,\{J^r,G^{r8}\}\} + \frac32 \{T^8,\{J^r,G^{rc}\}\},
\end{equation}

\begin{equation}
i(f^{aec}d^{be8}-f^{bec}d^{ae8}-f^{abe}d^{ec8}) \mathcal{D}_2^{ia} \mathcal{D}_2^{ib} = 0, 
\end{equation}

\begin{equation}
i(f^{aec}d^{be8}-f^{bec}d^{ae8}-f^{abe}d^{ec8}) (\mathcal{D}_2^{ia} \mathcal{D}_3^{ib} + \mathcal{D}_3^{ia} \mathcal{D}_2^{ib}) = 0, 
\end{equation}

\begin{equation}
i(f^{aec}d^{be8}-f^{bec}d^{ae8}-f^{abe}d^{ec8}) (\mathcal{D}_2^{ia} \mathcal{O}_3^{ib} + \mathcal{O}_3^{ia} \mathcal{D}_2^{ib}) = 0,
\end{equation}

\begin{equation}
i(f^{aec}d^{be8}-f^{bec}d^{ae8}-f^{abe}d^{ec8}) \mathcal{D}_3^{ia} \mathcal{D}_3^{ib} = \{J^2,\{T^c,\{J^r,G^{r8}\}\}\} - \{J^2,\{T^8,\{J^r,G^{rc}\}\}\},
\end{equation}

\begin{equation}
i(f^{aec}d^{be8}-f^{bec}d^{ae8}-f^{abe}d^{ec8}) (\mathcal{D}_3^{ia} \mathcal{O}_3^{ib} + \mathcal{O}_3^{ia}\mathcal{D}_3^{ib}) = 0, 
\end{equation}

\begin{eqnarray}
i(f^{aec}d^{be8}-f^{bec}d^{ae8}-f^{abe}d^{ec8}) \mathcal{O}_3^{ia} \mathcal{O}_3^{ib} & = & - \frac32 \{T^c,\{J^r,G^{r8}\}\} + \frac32 \{T^8,\{J^r,G^{rc}\}\} \nonumber \\ 
&  & - \frac54 \{J^2,\{T^c,\{J^r,G^{r8}\}\}\} + \frac54 \{J^2,\{T^8,\{J^r,G^{rc}\}\}\}. \nonumber \\
\end{eqnarray}

\subsection{$i(f^{aec}d^{be8}-f^{bec}d^{ae8}-f^{abe}d^{ec8}) A^{ia} J^2 A^{ib}$}

\begin{equation}
i(f^{aec}d^{be8}-f^{bec}d^{ae8}-f^{abe}d^{ec8}) G^{ia} J^2 G^{ib} = - \frac12 \{T^c,\{J^r,G^{r8}\}\} + \frac12 \{T^8,\{J^r,G^{rc}\}\},
\end{equation}

\begin{equation}
i(f^{aec}d^{be8}-f^{bec}d^{ae8}-f^{abe}d^{ec8}) (G^{ia} J^2 \mathcal{D}_2^{ib} + \mathcal{D}_2^{ia} J^2 G^{ib}) = 0,
\end{equation}

\bea
&  & i(f^{aec}d^{be8}-f^{bec}d^{ae8}-f^{abe}d^{ec8})(G^{ia} J^2 \mathcal{D}_3^{ib} + \mathcal{D}_3^{ia} J^2 G^{ib}) \nonumber\\
&  & \mbox{\hglue0.5truecm} = \frac12 \{J^2,\{T^c,\{J^r,G^{r8}\}\}\} - \frac12 \{J^2,\{T^8,\{J^r,G^{rc}\}\}\},
\eea

\begin{eqnarray}
&  & i(f^{aec}d^{be8}-f^{bec}d^{ae8}-f^{abe}d^{ec8})(G^{ia} J^2 \mathcal{O}_3^{ib} + \mathcal{O}_3^{ia} J^2 G^{ib}) \nonumber \\
&  & \mbox{\hglue0.5truecm} = - 3 \{T^c,\{J^r,G^{r8}\}\} + 3 \{T^8,\{J^r,G^{rc}\}\} - \frac74 \{J^2,\{T^c,\{J^r,G^{r8}\}\}\} \nonumber \\
&  & \mbox{\hglue1.0truecm} + \frac74 \{J^2,\{T^8,\{J^r,G^{rc}\}\}\},
\end{eqnarray}

\begin{equation}
i(f^{aec}d^{be8}-f^{bec}d^{ae8}-f^{abe}d^{ec8}) \mathcal{D}_2^{ia} J^2 \mathcal{D}_2^{ib} = 0,
\end{equation}

\begin{equation}
i(f^{aec}d^{be8}-f^{bec}d^{ae8}-f^{abe}d^{ec8}) (\mathcal{D}_2^{ia} J^2 \mathcal{D}_3^{ib} + \mathcal{D}_3^{ia} J^2 \mathcal{D}_2^{ib}) = 0,
\end{equation}

\begin{equation}
i(f^{aec}d^{be8}-f^{bec}d^{ae8}-f^{abe}d^{ec8}) (\mathcal{D}_2^{ia} J^2 \mathcal{O}_3^{ib} + \mathcal{O}_3^{ia} J^2 \mathcal{D}_2^{ib}) = 0,
\end{equation}

\begin{eqnarray}
&  & i(f^{aec}d^{be8}-f^{bec}d^{ae8}-f^{abe}d^{ec8}) \mathcal{D}_3^{ia} J^2 \mathcal{D}_3^{ib} \nonumber \\
&  & \mbox{\hglue0.5truecm} = \frac12 \{J^2,\{J^2,\{T^c,\{J^r,G^{r8}\}\}\}\} - \frac12 \{J^2,\{J^2,\{T^8,\{J^r,G^{rc}\}\}\}\}, 
\end{eqnarray}

\begin{equation}
i(f^{aec}d^{be8}-f^{bec}d^{ae8}-f^{abe}d^{ec8}) (\mathcal{D}_3^{ia} J^2 \mathcal{O}_3^{ib} + \mathcal{O}_3^{ia} J^2 \mathcal{D}_3^{ib}) = 0,
\end{equation}

\begin{eqnarray}
&  & i(f^{aec}d^{be8}-f^{bec}d^{ae8}-f^{abe}d^{ec8}) \mathcal{O}_3^{ia} J^2 \mathcal{O}_3^{ib} \nonumber \\
&  & \mbox{\hglue0.5truecm} = - 3 \{T^c,\{J^r,G^{r8}\}\} + 3 \{T^8,\{J^r,G^{rc}\}\} - \frac{19}{4} \{J^2,\{T^c,\{J^r,G^{r8}\}\}\} \nonumber \\
&  & \mbox{\hglue1.0truecm} + \frac{19}{4} \{J^2,\{T^8,\{J^r,G^{rc}\}\}\} - \frac98 \{J^2,\{J^2,\{T^c,\{J^r,G^{r8}\}\}\}\} \nonumber \\
&  & \mbox{\hglue1.0truecm} + \frac98 \{J^2,\{J^2,\{T^8,\{J^r,G^{rc}\}\}\}\}.
\end{eqnarray}

\subsection{$[A^{ia},[A^{ia},T^c]]$}

\begin{equation}
[G^{ia},[G^{ia},T^c]] = \frac{3}{4} N_f T^c,
\end{equation}

\begin{equation}
[G^{ia},[\mathcal{D}_2^{ia},T^c]] + [\mathcal{D}_2^{ia},[G^{ia},T^c]] = N_f\{J^r,G^{rc}\},
\end{equation}

\begin{equation}
[G^{ia},[\mathcal{D}_3^{ia},T^c]]+[\mathcal{D}_3^{ia},[G^{ia},T^c]] = 2(N_c+N_f)\{J^r,G^{rc}\} + (N_f-2)\{J^2,T^c\},
\end{equation}

\begin{equation}
[G^{ia},[\mathcal{O}_3^{ia},T^c]]+[\mathcal{O}_3^{ia},[G^{ia},T^c]] = 3 N_f T^c - 3(N_c+N_f)\{J^r,G^{rc}\} + (N_f+3)\{J^2,T^c\},
\end{equation}

\begin{equation}
[\mathcal{D}_2^{ia},[\mathcal{D}_2^{ia},T^c]] = \frac12 N_f\{J^2,T^c\},
\end{equation}

\begin{equation}
[\mathcal{D}_2^{ia},[\mathcal{D}_3^{ia},T^c]]+[\mathcal{D}_3^{ia},[\mathcal{D}_2^{ia},T^c]] = 2 N_f\{J^2,\{J^r,G^{rc}\}\},
\end{equation}

\begin{equation}
[\mathcal{D}_2^{ia},[\mathcal{O}_3^{ia},T^c]]+[\mathcal{O}_3^{ia},[\mathcal{D}_2^{ia},T^c]] = 0,
\end{equation}

\begin{equation}
[\mathcal{D}_3^{ia},[\mathcal{D}_3^{ia},T^c]] = 2 (N_c+N_f)\{J^2,\{J^r,G^{rc}\}\} + (N_f-2)\{J^2,\{J^2,T^c\}\},
\end{equation}

\begin{equation}
[\mathcal{D}_3^{ia},[\mathcal{O}_3^{ia},T^c]]+[\mathcal{O}_3^{ia},[\mathcal{D}_3^{ia},T^c]] = 0,
\end{equation}

\begin{eqnarray}
[\mathcal{O}_3^{ia},[\mathcal{O}_3^{ia},T^c]] & = & 3 N_f T^c - 3 (N_c+N_f)\{J^r,G^{rc}\} + (4N_f+3)\{J^2,T^c\} \nonumber \\
&  & -\frac52 (N_c+N_f)\{J^2,\{J^r,G^{rc}\}\} + \frac12 (N_f+5)\{J^2,\{J^2,T^c\}\}.
\end{eqnarray}

\subsection{$d^{ab8} [A^{ia},[A^{ib},T^c]]$}

\begin{equation}
d^{ab8} [G^{ia},[G^{ib},T^c]] = \frac{3}{8} N_f d^{c8e}T^e,
\end{equation}

\begin{equation}
d^{ab8} ([G^{ia},[\mathcal{D}_2^{ib},T^c]]+[\mathcal{D}_2^{ia},[G^{ib},T^c]]) = \frac12 N_f d^{c8e} \{J^r,G^{re}\},
\end{equation}

\begin{eqnarray}
d^{ab8} ([G^{ia},[\mathcal{D}_3^{ib},T^c]]+[\mathcal{D}_3^{ia},[G^{ib},T^c]]) & = & (N_c+N_f) d^{c8e}\{J^r,G^{re}\} - \{T^c,\{J^r,G^{r8}\}\} \nonumber \\
&  & + \{T^8,\{J^r,G^{rc}\}\} + \frac12 (N_f-2) d^{c8e}\{J^2,T^e\}\; , \nonumber \\
\end{eqnarray}

\begin{eqnarray}
d^{ab8} ([G^{ia},[\mathcal{O}_3^{ib},T^c]]+[\mathcal{O}_3^{ia},[G^{ib},T^c]]) & = & \frac32 N_f d^{c8e} T^e + \frac32\{T^c,\{J^r,G^{r8}\}\} - \frac32\{T^8,\{J^r,G^{rc}\}\} \nonumber \\
&  & + \frac12 (N_f+3) d^{c8e} \{J^2,T^e\} - \frac32 (N_c+N_f) d^{c8e} \{J^r,G^{re}\}\; , \nonumber \\
\end{eqnarray}

\begin{equation}
d^{ab8} [\mathcal{D}_2^{ia},[\mathcal{D}_2^{ib},T^c]] = \frac14 N_f d^{c8e} \{J^2,T^e\},
\end{equation}

\begin{equation}
d^{ab8} ([\mathcal{D}_2^{ia},[\mathcal{D}_3^{ib},T^c]]+[\mathcal{D}_3^{ia},[\mathcal{D}_2^{ib},T^c]]) = N_f d^{c8e} \{J^2,\{J^r,G^{re}\}\},
\end{equation}

\begin{equation}
d^{ab8} ([\mathcal{D}_2^{ia},[\mathcal{O}_3^{ib},T^c]]+[\mathcal{O}_3^{ia},[\mathcal{D}_2^{ib},T^c]]) = 0,
\end{equation}

\begin{eqnarray}
d^{ab8} [\mathcal{D}_3^{ia},[\mathcal{D}_3^{ib},T^c]] & = & (N_c+N_f) d^{c8e}\{J^2,\{J^r,G^{re}\}\} - \{J^2,\{T^c,\{J^r,G^{r8}\}\}\} \nonumber \\
&  & + \{J^2,\{T^8,\{J^r,G^{rc}\}\}\} + \frac12 (N_f-2) d^{c8e}\{J^2,\{J^2,T^e\}\},
\end{eqnarray}

\begin{equation}
d^{ab8} ([\mathcal{D}_3^{ia},[\mathcal{O}_3^{ib},T^c]]+[\mathcal{O}_3^{ia},[\mathcal{D}_3^{ib},T^c]]) = 0,
\end{equation}

\begin{eqnarray}
d^{ab8} [\mathcal{O}_3^{ia},[\mathcal{O}_3^{ib},T^c]] & = & \frac32 N_f d^{c8e} T^e - \frac32 (N_c+N_f) d^{c8e}\{J^r,G^{re}\} + \frac32 \{T^c,\{J^r,G^{r8}\}\} \nonumber \\
&  & - \frac32 \{T^8,\{J^r,G^{rc}\}\} + \frac12 (4N_f+3) d^{c8e}\{J^2,T^e\} + \frac54 \{J^2,\{T^c,\{J^r,G^{r8}\}\}\} \nonumber \\
&  & - \frac54 \{J^2,\{T^8,\{J^r,G^{rc}\}\}\} + \frac14 (N_f+5) d^{c8e}\{J^2,\{J^2,T^e\}\} \nonumber \\
&  & - \frac54 (N_c+N_f) d^{c8e}\{J^2,\{J^r,G^{re}\}\}.
\end{eqnarray}

\subsection{$[A^{i8},[A^{i8},T^c]]$}

\begin{equation}
[G^{i8},[G^{i8},T^c]] = \frac34 f^{c8e} f^{8eg} T^g,
\end{equation}

\begin{equation}
[G^{i8},[\mathcal{D}_2^{i8},T^c]]+[\mathcal{D}_2^{i8},[G^{i8},T^c]] = f^{c8e} f^{8eg} \{J^r,G^{rg}\},
\end{equation}

\begin{equation}
[G^{i8},[\mathcal{D}_3^{i8},T^c]]+[\mathcal{D}_3^{i8},[G^{i8},T^c]] = 3 f^{c8e} f^{8eg} T^g + f^{c8e} f^{8eg} \{J^2,T^g\} - 2 \epsilon^{ijk} f^{c8e} \{G^{ke},\{J^i,G^{j8}\}\},
\end{equation}

\begin{equation}
[G^{i8},[\mathcal{O}_3^{i8},T^c]]+[\mathcal{O}_3^{i8},[G^{i8},T^c]] = - \frac32 f^{c8e} f^{8eg} T^g + f^{c8e} f^{8eg} \{J^2,T^g\} + 3 \epsilon^{ijk} f^{c8e} \{G^{ke},\{J^i,G^{j8}\}\},
\end{equation}

\begin{equation}
[\mathcal{D}_2^{i8},[\mathcal{D}_2^{i8},T^c]] = \frac12 f^{c8e} f^{8eg} \{J^2,T^g\},
\end{equation}

\begin{equation}
[\mathcal{D}_2^{i8},[\mathcal{D}_3^{i8},T^c]]+[\mathcal{D}_3^{i8},[\mathcal{D}_2^{i8},T^c]] = 2 f^{c8e} f^{8eg} \{J^2,\{J^r,G^{rg}\}\},
\end{equation}

\begin{equation}
[\mathcal{D}_2^{i8},[\mathcal{O}_3^{i8},T^c]]+[\mathcal{O}_3^{i8},[\mathcal{D}_2^{i8},T^c]] = 0,
\end{equation}

\begin{equation}
[\mathcal{D}_3^{i8},[\mathcal{D}_3^{i8},T^c]] = 3 f^{c8e} f^{8eg} \{J^2,T^g\} + f^{c8e} f^{8eg} \{J^2,\{J^2,T^g\}\} -2 \epsilon^{ijk} f^{c8e} \{J^2,\{G^{ke},\{J^i,G^{j8}\}\}\},
\end{equation}

\begin{equation}
[\mathcal{D}_3^{i8},[\mathcal{O}_3^{i8},T^c]]+[\mathcal{O}_3^{i8},[\mathcal{D}_3^{i8},T^c]] = 0,
\end{equation}

\begin{eqnarray}
[\mathcal{O}_3^{i8},[\mathcal{O}_3^{i8},T^c]] & = & - \frac32 f^{c8e} f^{8eg} T^g + \frac14 f^{c8e} f^{8eg} \{J^2,T^g\} + 3 \epsilon^{ijk} f^{c8e} \{G^{ke},\{J^i,G^{j8}\}\} \nonumber \\
&  & \mbox{} + \frac12 f^{c8e} f^{8eg} \{J^2,\{J^2,T^g\}\} + \frac52 \epsilon^{ijk} f^{c8e} \{J^2,\{G^{ke},\{J^i,G^{j8}\}\}\}.
\end{eqnarray}

\subsection{$if^{cae} \{A^{ia},[J^2,A^{ie}]\}$}

\begin{equation}
i f^{cae} \{G^{ia},[J^2,G^{ie}]\} = - \frac32 N_f T^c + (N_c+N_f) \{J^r,G^{rc}\} - \{J^2,T^c\},
\end{equation}

\begin{equation}
i f^{cae} \{\mathcal{D}_2^{ia},[J^2,G^{ie}]\} = 0,
\end{equation}

\begin{equation}
i f^{cae} \{\mathcal{D}_3^{ia},[J^2,G^{ie}]\} = 0,
\end{equation}

\begin{eqnarray}
&  & if^{cae} (\{\mathcal{O}_3^{ia},[J^2,G^{ie}]\} + \{G^{ia},[J^2,\mathcal{O}_3^{ie}]\}) \nonumber \\
&  & \mbox{\hglue0.5truecm} = - 6 N_f T^c + 6 (N_c+N_f)\{J^r,G^{rc}\} - (5N_f+6) \{J^2,T^c\} + 2 (N_c+N_f)\{J^2,\{J^r,G^{rc}\}\} \nonumber \\
&  & \mbox{\hglue1.0truecm} - 2\{J^2,\{J^2,T^c\}\},
\end{eqnarray}

\begin{equation}
i f^{cae} \{\mathcal{D}_2^{ia},[J^2,\mathcal{O}_3^{ie}]\} = 0,
\end{equation}

\begin{equation}
i f^{cae} \{\mathcal{D}_3^{ia},[J^2,\mathcal{O}_3^{ie}]\} = 0,
\end{equation}

\begin{eqnarray}
if^{cae} \{\mathcal{O}_3^{ia},[J^2,\mathcal{O}_3^{ie}]\} & = & -6 N_f T^c + 6 (N_c+N_f)\{J^r,G^{rc}\} - (11N_f+6) \{J^2,T^c\} \nonumber \\
&  & \mbox{} + 8(N_c+N_f)\{J^2,\{J^r,G^{rc}\}\} - \frac12 (7N_f+16)\{J^2,\{J^2,T^c\}\} \nonumber \\
&  & \mbox{} + (N_c+N_f) \{J^2,\{J^2,\{J^r,G^{rc}\}\}\} - \{J^2,\{J^2,\{J^2,T^c\}\}\}.
\end{eqnarray}

\subsection{$id^{ab8} f^{cbe} \{A^{ia},[J^2,A^{ie}]\}$}

\begin{eqnarray}
id^{ab8} f^{cbe} \{G^{ia},[J^2,G^{ie}]\} & = & - \frac34 N_f d^{c8e} T^e + \frac12 (N_c+N_f) d^{c8e}\{J^r,G^{re}\} - \frac12\{T^c,\{J^r,G^{r8}\}\} \nonumber \\
&  & \mbox{} + \frac12\{T^8,\{J^r,G^{rc}\}\} - \frac12 d^{c8e}\{J^2,T^e\},
\end{eqnarray}

\begin{equation}
id^{ab8} f^{cbe} \{\mathcal{D}_2^{ia},[J^2,G^{ie}]\} = 0,
\end{equation}

\begin{equation}
id^{ab8} f^{cbe} \{\mathcal{D}_3^{ia},[J^2,G^{ie}]\} = 0,
\end{equation}

\begin{eqnarray}
&  & id^{ab8} f^{cbe} (\{\mathcal{O}_3^{ia},[J^2,G^{ie}]\} + \{G^{ia},[J^2,\mathcal{O}_3^{ie}]\}) \nonumber \\
&  & \mbox{\hglue0.5truecm} = - 3 N_f d^{c8e} T^e + 3 (N_c+N_f) d^{c8e}\{J^r,G^{re}\} - 3 \{T^c,\{J^r,G^{r8}\}\} + 3\{T^8,\{J^r,G^{rc}\}\} \nonumber \\
&  & \mbox{\hglue1.0truecm} - \frac12 (5N_f+6) d^{c8e} \{J^2,T^e\} + (N_c+N_f) d^{c8e}\{J^2,\{J^r,G^{re}\}\} - \{J^2,\{T^c,\{J^r,G^{r8}\}\}\} \nonumber \\
&  & \mbox{\hglue1.0truecm} + \{J^2,\{T^8,\{J^r,G^{rc}\}\}\} - d^{c8e} \{J^2,\{J^2,T^e\}\},
\end{eqnarray}

\begin{equation}
id^{ab8} f^{cbe} \{\mathcal{D}_2^{ia},[J^2,\mathcal{O}_3^{ie}]\} = 0,
\end{equation}

\begin{equation}
id^{ab8} f^{cbe} \{\mathcal{D}_3^{ia},[J^2,\mathcal{O}_3^{ie}]\} = 0, 
\end{equation}

\begin{eqnarray}
&  & i d^{ab8} f^{cbe} \{\mathcal{O}_3^{ia},[J^2,\mathcal{O}_3^{ie}]\} \nonumber \\
&  & \mbox{\hglue0.5truecm} = - 3 N_f d^{c8e} T^e + 3 (N_c+N_f) d^{c8e}\{J^r,G^{re}\} - 3 \{T^c,\{J^r,G^{r8}\}\} + 3 \{T^8,\{J^r,G^{rc}\}\} \nonumber \\
&  & \mbox{\hglue1.0truecm} - \frac12 (11N_f-6) d^{c8e}\{J^2,T^e\} + 4 (N_c+N_f) d^{c8e}\{J^2,\{J^r,G^{re}\}\} - 4 \{J^2,\{T^c,\{J^r,G^{r8}\}\}\} \nonumber \\
&  & \mbox{\hglue1.0truecm} + 4 \{J^2,\{T^8,\{J^r,G^{rc}\}\}\} - \frac14 (7N_f+16) d^{c8e}\{J^2,\{J^2,T^e\}\} \nonumber \\
&  & \mbox{\hglue1.0truecm} + \frac12 (N_c+N_f) d^{c8e}\{J^2,\{J^2,\{J^r,G^{re}\}\}\} - \frac12\{J^2,\{J^2,\{T^c,\{J^r,G^{r8}\}\}\}\} \nonumber \\
&  & \mbox{\hglue1.0truecm} + \frac12\{J^2,\{J^2,\{T^8,\{J^r,G^{rc}\}\}\}\} - \frac12 d^{c8e}\{J^2,\{J^2,\{J^2,T^e\}\}\},
\end{eqnarray}

\subsection{$if^{c8e} \{A^{i8},[J^2,A^{ie}]\}$}

\begin{equation}
if^{c8e} \{G^{i8},[J^2,G^{ie}]\} = - \epsilon^{ijk} f^{c8e} \{G^{ke},\{J^i,G^{j8}\}\},
\end{equation}

\begin{equation}
if^{c8e} \{\mathcal{D}_2^{i8},[J^2,G^{ie}]\} = 0,
\end{equation}

\begin{equation}
if^{c8e} \{\mathcal{D}_3^{i8},[J^2,G^{ie}]\} = 0,
\end{equation}

\begin{eqnarray}
&  & if^{c8e} (\{\mathcal{O}_3^{i8},[J^2,G^{ie}]\} + \{G^{i8},[J^2,\mathcal{O}_3^{ie}]\}) \nonumber \\
&  & \mbox{\hglue0.5truecm} = 3 f^{c8e} f^{8eg} T^g- 2 f^{c8e} f^{8eg} \{J^2,T^g\} - 6 \epsilon^{ijk} f^{c8e} \{G^{ke},\{J^i,G^{j8}\}\} \nonumber \\
&  & \mbox{\hglue1.0truecm} - 2 \epsilon^{ijk} f^{c8e} \{J^2,\{G^{ke},\{J^i,G^{j8}\}\}\},
\end{eqnarray}

\begin{equation}
if^{c8e} \{\mathcal{D}_2^{i8},[J^2,\mathcal{O}_3^{ie}]\} = 0,
\end{equation}

\begin{equation}
if^{c8e} \{\mathcal{D}_3^{i8},[J^2,\mathcal{O}_3^{ie}]\} = 0,
\end{equation}

\begin{eqnarray}
&  & if^{c8e} \{\mathcal{O}_3^{i8},[J^2,\mathcal{O}_3^{ie}]\} \nonumber \\
&  & \mbox{\hglue0.5truecm} = 3 f^{c8e} f^{8eg} T^g + f^{c8e} f^{8eg} \{J^2,T^g\} - 6 \epsilon^{ijk} f^{c8e} \{G^{ke},\{J^i,G^{j8}\}\} - 2 f^{c8e} f^{8eg} \{J^2,\{J^2,T^g\}\} \nonumber \\
&  & \mbox{\hglue1.0truecm} - 8\epsilon^{ijk} f^{c8e} \{J^2,\{G^{ke},\{J^i,G^{j8}\}\}\} - \epsilon^{ijk} f^{c8e} \{J^2,\{J^2,\{G^{ke},\{J^i,G^{j8}\}\}\}\}.
\end{eqnarray}

\subsection{$i f^{ace}[[J^2,A^{ie}],[J^2,A^{ia}]]$}

\begin{equation}
if^{ace}[[J^2,G^{ie}],[J^2,G^{ia}]] = 3 N_f T^c - 3 (N_c+N_f)\{J^r,G^{rc}\} + (N_f+3)\{J^2,T^c\},
\end{equation}

\begin{eqnarray}
if^{ace}[[J^2,G^{ie}],[J^2,\mathcal{O}_3^{ia}]] & = & 6 N_f T^c - 6 (N_c+N_f) \{J^r,G^{rc}\} + (8N_f+6) \{J^2,T^c\} \nonumber \\
&  & \mbox{} - 5 (N_c+N_f) \{J^2,\{J^r,G^{rc}\}\} + (N_f+5) \{J^2,\{J^2,T^c\}\}\; , \nonumber \\
\end{eqnarray}

\begin{eqnarray}
if^{ace}[[J^2,\mathcal{O}_3^{ie}],[J^2,G^{ia}]] & = & 6 N_f T^c - 6 (N_c+N_f) \{J^r,G^{rc}\} + (8N_f+6) \{J^2,T^c\} \nonumber \\
&  & \mbox{} - 5 (N_c+N_f) \{J^2,\{J^r,G^{rc}\}\} + (N_f+5) \{J^2,\{J^2,T^c\}\}\; , \nonumber \\
\end{eqnarray}

\begin{eqnarray}
&  & if^{ace}[[J^2,\mathcal{O}_3^{ie}],[J^2,\mathcal{O}_3^{ia}]] \nonumber \\
&  & \mbox{\hglue0.2truecm} = 12 N_f T^c - 12 (N_c+N_f) \{J^r,G^{rc}\} + 4 (7N_f+3) \{J^2,T^c\} - 22 (N_c+N_f) \{J^2,\{J^r,G^{rc}\}\} \nonumber \\
&  & \mbox{\hglue0.7truecm} + (15N_f+22) \{J^2,\{J^2,T^c\}\} - 7 (N_c+N_f) \{J^2,\{J^2,\{J^r,G^{rc}\}\}\} \nonumber \\
&  & \mbox{\hglue0.7truecm} + (N_f+7) \{J^2,\{J^2,\{J^2,T^c\}\}\},
\end{eqnarray}

\subsection{$id^{ab8} f^{bce}[[J^2,A^{ie}],[J^2,A^{ia}]]$}

\begin{eqnarray}
id^{ab8} f^{bce}[[J^2,G^{ie}],[J^2,G^{ia}]] & = & \frac32 N_f d^{c8e} T^e - \frac32 (N_c+N_f) d^{c8e} \{J^r,G^{re}\} + \frac32 \{T^c,\{J^r,G^{r8}\}\} \nonumber \\
&  & \mbox{} - \frac32 \{T^8,\{J^r,G^{rc}\}\} + \frac12 (N_f+3) d^{c8e} \{J^2,T^e\},
\end{eqnarray}

\begin{eqnarray}
&  & id^{ab8} f^{bce}[[J^2,G^{ie}],[J^2,\mathcal{O}_3^{ia}]] \nonumber \\
&  & \mbox{\hglue0.5truecm} = 3 N_f d^{c8e} T^e - 3 (N_c+N_f) d^{c8e} \{J^r,G^{re}\} + 3 \{T^c,\{J^r,G^{r8}\}\} - 3 \{T^8,\{J^r,G^{rc}\}\} \nonumber\\
&  & \mbox{\hglue1.0truecm} + (4N_f+3) d^{c8e} \{J^2,T^e\} + \frac52 \{J^2,\{T^c,\{J^r,G^{r8}\}\}\} - \frac52 \{J^2,\{T^8,\{J^r,G^{rc}\}\}\} \nonumber\\
&  & \mbox{\hglue1.0truecm} + \frac12 (N_f+5) d^{c8e} \{J^2,\{J^2,T^e\}\} - \frac52 (N_c+N_f) d^{c8e} \{J^2,\{J^r,G^{re}\}\},
\end{eqnarray}

\begin{eqnarray}
&  & id^{ab8} f^{bce}[[J^2,\mathcal{O}_3^{ie}],[J^2,G^{ia}]] \nonumber \\
&  & \mbox{\hglue0.5truecm} = 3 N_f d^{c8e} T^e - 3 (N_c+N_f) d^{c8e} \{J^r,G^{re}\} + 3 \{T^c,\{J^r,G^{r8}\}\} - 3 \{T^8,\{J^r,G^{rc}\}\} \nonumber\\
&  & \mbox{\hglue1.0truecm} + (4N_f+3) d^{c8e} \{J^2,T^e\} + \frac52 \{J^2,\{T^c,\{J^r,G^{r8}\}\}\} - \frac52 \{J^2,\{T^8,\{J^r,G^{rc}\}\}\} \nonumber\\
&  & \mbox{\hglue1.0truecm} + \frac12 (N_f+5) d^{c8e} \{J^2,\{J^2,T^e\}\} - \frac52 (N_c+N_f) d^{c8e} \{J^2,\{J^r,G^{re}\}\},
\end{eqnarray}

\begin{eqnarray}
&  & id^{ab8} f^{bce}[[J^2,\mathcal{O}_3^{ie}],[J^2,\mathcal{O}_3^{ia}]] \nonumber \\
&  & \mbox{\hglue0.5truecm} = - 6 f^{c8e} f^{8eg} T^g - 5 f^{c8e} f^{8eg} \{J^2,T^g\} + 12 \epsilon^{ijk} f^{c8e} \{G^{ke},\{J^i,G^{j8}\}\} \nonumber \\
&  & \mbox{\hglue1.0truecm} + \frac92 f^{c8e} f^{8eg} \{J^2,\{J^2,T^g\}\} + 22 \epsilon^{ijk} f^{c8e} \{J^2,\{G^{ke},\{J^i,G^{j8}\}\}\} \nonumber \\
&  & \mbox{\hglue1.0truecm} + f^{c8e} f^{8eg} \{J^2,\{J^2,\{J^2,T^g\}\}\} + 7 \epsilon^{ijk} f^{c8e} \{J^2,\{J^2,\{G^{ke},\{J^i,G^{j8}\}\}\}\},
\end{eqnarray}

\subsection{$i f^{8ce}[[J^2,A^{ie}],[J^2,A^{i8}]]$}

\begin{equation}
if^{8ce}[[J^2,G^{ie}],[J^2,G^{i8}]] = - \frac32 f^{c8e} f^{8eg} T^g + f^{c8e} f^{8eg} \{J^2,T^g\} + 3 \epsilon^{ijk} f^{c8e} \{G^{ke},\{J^i,G^{j8}\}\},
\end{equation}

\begin{eqnarray}
if^{8ce}[[J^2,G^{ie}],[J^2,\mathcal{O}_3^{i8}]] & = & - 3 f^{c8e} f^{8eg} T^g + \frac12 f^{c8e} f^{8eg} \{J^2,T^g\} + f^{c8e} f^{8eg} \{J^2,\{J^2,T^g\}\} \nonumber \\
&  & \mbox{} + 6 \epsilon^{ijk} f^{c8e} \{G^{ke},\{J^i,G^{j8}\}\} + 5 \epsilon^{ijk} f^{c8e}\{J^2,\{G^{ke},\{J^i,G^{j8}\}\}\}\; , \nonumber \\
\end{eqnarray}

\begin{eqnarray}
if^{8ce}[[J^2,\mathcal{O}_3^{ie}],[J^2,G^{i8}]] & = & - 3 f^{c8e} f^{8eg} T^g + \frac12 f^{c8e} f^{8eg} \{J^2,T^g\} + f^{c8e} f^{8eg} \{J^2,\{J^2,T^g\}\} \nonumber\\
&  & \mbox{} + 6 \epsilon^{ijk} f^{c8e} \{G^{ke},\{J^i,G^{j8}\}\} + 5 \epsilon^{ijk} f^{c8e} \{J^2,\{G^{ke},\{J^i,G^{j8}\}\}\}\; , \nonumber \\ 
\end{eqnarray}

\begin{eqnarray}
&  & if^{8ce}[[J^2,\mathcal{O}_3^{ie}],[J^2,\mathcal{O}_3^{i8}]] \nonumber \\
&  & \mbox{\hglue0.5truecm} = - 6 f^{c8e} f^{8eg} T^g - 5 f^{c8e} f^{8eg} \{J^2,T^g\} + 12 \epsilon^{ijk} f^{c8e} \{G^{ke},\{J^i,G^{j8}\}\} \nonumber \\
&  & \mbox{\hglue1.0truecm} + \frac92 f^{c8e} f^{8eg} \{J^2,\{J^2,T^g\}\} + 22 \epsilon^{ijk} f^{c8e} \{J^2,\{G^{ke},\{J^i,G^{j8}\}\}\} \nonumber \\
&  & \mbox{\hglue1.0truecm} + f^{c8e} f^{8eg} \{J^2,\{J^2,\{J^2,T^g\}\}\} + 7 \epsilon^{ijk} f^{c8e} \{J^2,\{J^2,\{G^{ke},\{J^i,G^{j8}\}\}\}\}.
\end{eqnarray}

\section{\label{app:coeff}Operator coefficients}

The several operator products involved in the analysis can be cast into rather compact forms. They can be written as summations involving an operator coefficient times a corresponding operator belonging to the $SU(3)$ flavor representations ${\mathbf 1}$, ${\mathbf 8}$ and ${\mathbf 27}$, listed in Eqs.~(\ref{eq:1op}), (\ref{eq:8op}) and (\ref{eq:27op}) respectively.

The compact expressions are listed as follows.

\begin{equation}
if^{acb}A^{ia}A^{ib} = \sum_{n=1}^7 a_{n}^{\mathbf{8}} S_n^c,
\end{equation}
where
\bea
%\begin{equation}
a_{1}^{\mathbf{8}} & = & \frac{3N_f}{8} a_1^2 + \frac{3N_f}{2N_c^2} a_1 c_3 + \frac{3N_f}{2N_c^4} c_3^2\; , \nonumber \\
%\end{equation}
%\begin{equation}
a_{2}^{\mathbf{8}} & = & \frac{N_f}{2N_c} a_1 b_2 + \frac{N_c+N_f}{N_c^2} a_1 b_3 - \frac{3(N_c+N_f)}{2N_c^2} a_1 c_3 - \frac{3(N_c+N_f)}{2N_c^4}c_3^2\; , \nonumber \\
%\end{equation}
%\begin{equation}
a_{3}^{\mathbf{8}} & = & \frac{N_f-2}{2N_c^2} a_1 b_3 + \frac{N_f+3}{2N_c^2} a_1 c_3 + \frac{N_f}{4N_c^2} b_2^2 + \frac{4N_f+3}{2N_c^4} c_3^2\; , \nonumber \\
%\end{equation}
%\begin{equation}
a_{4}^{\mathbf{8}} & = & \frac{N_f}{N_c^3} b_2 b_3 + \frac{N_c+N_f}{N_c^4} b_3^2 - \frac{5(N_c+N_f)}{4N_c^4} c_3^2\; , \nonumber \\
%\end{equation}
%\begin{equation}
a_{5}^{\mathbf{8}} & = & \frac{N_f-2}{2N_c^4} b_3^2 + \frac{N_f+5}{4N_c^4} c_3^2\; , \nonumber \\
%\end{equation}
%\begin{equation}
a_{6}^{\mathbf{8}} & = & 0\; , \nonumber \\
%\end{equation}
%\begin{equation}
a_{7}^{\mathbf{8}} & = & 0\; . \nonumber
%\end{equation}
\eea

\begin{equation}
if^{acb}A^{ia} J^2 A^{ib} = \sum_{n=1}^7 \overline{a}_{n}^{\mathbf{8}} S_n^c,
\end{equation}
where
\bea
%\begin{equation}
\overline{a}_{1}^{\mathbf{8}} & = & \frac{3N_f}{4} a_1^2 + \frac{3N_f}{N_c^2} a_1 c_3 + \frac{3N_f}{N_c^4} c_3^2\; , \nonumber \\
%\end{equation}
%\begin{equation}
\overline{a}_{2}^{\mathbf{8}} & = & -\frac{N_c+N_f}{2} a_1^2 - \frac{3(N_c+N_f)}{N_c^2} a_1 c_3 - \frac{3(N_c+N_f)}{N_c^4} c_3^2\; , \nonumber \\
%\end{equation}
%\begin{equation}
\overline{a}_{3}^{\mathbf{8}} & = & \frac{3N_f+8}{16} a_1^2 + \frac{13N_f+12}{4N_c^2} a_1 c_3 + \frac{25N_f+12}{4N_c^4} c_3^2\; , \nonumber \\
%\end{equation}
%\begin{equation}
\overline{a}_{4}^{\mathbf{8}} & = & \frac{N_f}{4N_c} a_1 b_2 + \frac{N_c+N_f}{2N_c^2} a_1 b_3 - \frac{7(N_c+N_f)}{4N_c^2} a_1 c_3 - \frac{19(N_c+N_f)}{4N_c^4} c_3^2\; , \nonumber \\
%\end{equation}
%\begin{equation}
\overline{a}_{5}^{\mathbf{8}} & = & \frac{N_f-2}{4N_c^2} a_1 b_3 + \frac{N_f+7}{4N_c^2} a_1 c_3 + \frac{N_f}{8N_c^2} b_2^2 + \frac{11N_f+19}{4N_c^4} c_3^2\; , \nonumber \\
%\end{equation}
%\begin{equation}
\overline{a}_{6}^{\mathbf{8}} & = & \frac{N_f}{2N_c^3} b_2 b_3 + \frac{N_c+N_f}{2N_c^4} b_3^2 - \frac{9(N_c+N_f)}{8N_c^4} c_3^2\; , \nonumber \\
%\end{equation}
%\begin{equation}
\overline{a}_{7}^{\mathbf{8}} & = & \frac{N_f-2}{4N_c^4} b_3^2 + \frac{N_f+9}{8N_c^4} c_3^2\; . \nonumber
%\end{equation}
\eea

\begin{equation}
i(f^{aec}d^{be8}-f^{bec}d^{ae8}-f^{abe}d^{ec8}) A^{ia}A^{ib} = \sum_{n=1}^{13} b_{n}^{\mathbf{10}+\overline{\mathbf{10}}} O_n^c,
\end{equation}
where
\bea
%\begin{equation}
b_{1}^{\mathbf{10}+\overline{\mathbf{10}}} & = & 0\; , \nonumber \\
%\end{equation}
%\begin{equation}
b_{2}^{\mathbf{10}+\overline{\mathbf{10}}} & = & 0\; , \nonumber \\
%\end{equation}
%\begin{equation}
b_{3}^{\mathbf{10}+\overline{\mathbf{10}}} & = & 0\; , \nonumber \\
%\end{equation}
%\begin{equation}
b_{4}^{\mathbf{10}+\overline{\mathbf{10}}} & = & \frac{1}{N_c^2}a_1 b_3 - \frac{3}{2N_c^2} a_1 c_3 - \frac{3}{2N_c^4} c_3^2\; , \nonumber \\
%\end{equation}
%\begin{equation}
b_{5}^{\mathbf{10}+\overline{\mathbf{10}}} & = & -\frac{1}{N_c^2}a_1 b_3 + \frac{3}{2N_c^2} a_1 c_3 + \frac{3}{2N_c^4} c_3^2\; , \nonumber \\
%\end{equation}
%\begin{equation}
b_{6}^{\mathbf{10}+\overline{\mathbf{10}}} & = & 0\; , \nonumber \\
%\end{equation}
%\begin{equation}
b_{7}^{\mathbf{10}+\overline{\mathbf{10}}} & = & 0\; , \nonumber \\
%\end{equation}
%\begin{equation}
b_{8}^{\mathbf{10}+\overline{\mathbf{10}}} & = & \frac{1}{N_c^4}b_3^2 - \frac{5}{4N_c^4} c_3^2\; , \nonumber \\
%\end{equation}
%\begin{equation}
b_{9}^{\mathbf{10}+\overline{\mathbf{10}}} & = & -\frac{1}{N_c^4} b_3^2 + \frac{5}{4N_c^4} c_3^2\; , \nonumber \\
%\end{equation}
%\begin{equation}
b_{10}^{\mathbf{10}+\overline{\mathbf{10}}} & = & 0\; , \nonumber \\
%\end{equation}
%\begin{equation}
b_{11}^{\mathbf{10}+\overline{\mathbf{10}}} & = & 0\; , \nonumber \\
%\end{equation}
%\begin{equation}
b_{12}^{\mathbf{10}+\overline{\mathbf{10}}} & = & 0\; , \nonumber \\
%\end{equation}
%\begin{equation}
b_{13}^{\mathbf{10}+\overline{\mathbf{10}}} & = & 0\; . \nonumber
%\end{equation}
\eea

\begin{equation}
i(f^{aec}d^{be8}-f^{bec}d^{ae8}-f^{abe}d^{ec8}) A^{ia} J^2 A^{ib} = \sum_{n=1}^{13} \overline{b}_{n}^{\mathbf{10}+\overline{\mathbf{10}}} O_n^c,
\end{equation}
where
\bea
%\begin{equation}
\overline{b}_{1}^{\mathbf{10}+\overline{\mathbf{10}}} & = & 0\; , \nonumber \\
%\end{equation}
%\begin{equation}
\overline{b}_{2}^{\mathbf{10}+\overline{\mathbf{10}}} & = & 0\; , \nonumber \\
%\end{equation}
%\begin{equation}
\overline{b}_{3}^{\mathbf{10}+\overline{\mathbf{10}}} & = & 0\; , \nonumber \\
%\end{equation}
%\begin{equation}
\overline{b}_{4}^{\mathbf{10}+\overline{\mathbf{10}}} & = & -\frac12 a_1^2 - \frac{3}{N_c^2} a_1 c_3 - \frac{3}{N_c^4} c_3^2\; , \nonumber \\
%\end{equation}
%\begin{equation}
\overline{b}_{5}^{\mathbf{10}+\overline{\mathbf{10}}} & = & \frac12 a_1^2 + \frac{3}{N_c^2} a_1 c_3 + \frac{3}{N_c^4} c_3^2\; , \nonumber \\
%\end{equation}
%\begin{equation}
\overline{b}_{6}^{\mathbf{10}+\overline{\mathbf{10}}} & = & 0\; , \nonumber \\
%\end{equation}
%\begin{equation}
\overline{b}_{7}^{\mathbf{10}+\overline{\mathbf{10}}} & = & 0\; , \nonumber \\
%\end{equation}
%\begin{equation}
\overline{b}_{8}^{\mathbf{10}+\overline{\mathbf{10}}} & = & \frac{1}{2N_c^2}a_1 b_3 - \frac{7}{4N_c^2} a_1 c_3 - \frac{19}{4N_c^4} c_3^2\; , \nonumber \\
%\end{equation}
%\begin{equation}
\overline{b}_{9}^{\mathbf{10}+\overline{\mathbf{10}}} & = & -\frac{1}{2N_c^2}a_1 b_3 + \frac{7}{4N_c^2} a_1 c_3 + \frac{19}{4N_c^4} c_3^2\; , \nonumber \\
%\end{equation}
%\begin{equation}
\overline{b}_{10}^{\mathbf{10}+\overline{\mathbf{10}}} & = & 0\; , \nonumber \\
%\end{equation}
%\begin{equation}
\overline{b}_{11}^{\mathbf{10}+\overline{\mathbf{10}}} & = & 0\; , \nonumber \\
%\end{equation}
%\begin{equation}
\overline{b}_{12}^{\mathbf{10}+\overline{\mathbf{10}}} & = & \frac{1}{2N_c^4}b_3^2 - \frac{9}{8N_c^4} c_3^2\; , \nonumber \\
%\end{equation}
%\begin{equation}
\overline{b}_{13}^{\mathbf{10}+\overline{\mathbf{10}}} & = & -\frac{1}{2N_c^4}b_3^2 + \frac{9}{8N_c^4} c_3^2\; . \nonumber
%\end{equation}
\eea

\begin{equation}
\frac12 [A^{ia},[A^{ia},T^c]] = \sum_{n=1}^7 c_{n}^{\mathbf{1}} S_n^c,
\end{equation}
where
\bea
%\begin{equation}
c_{1}^{\mathbf{1}} & = & \frac{3N_f}{8} a_1^2 + \frac{3N_f}{2N_c^2} a_1 c_3 + \frac{3N_f}{2N_c^4} c_3^2\; , \nonumber \\
%\end{equation}
%\begin{equation}
c_{2}^{\mathbf{1}} & = & \frac{N_f}{2N_c} a_1 b_2 + \frac{N_c+N_f}{N_c^2} a_1 b_3 - \frac{3(N_c+N_f)}{2N_c^2} a_1 c_3 - \frac{3(N_c+N_f)}{2N_c^4}c_3^2\; , \nonumber \\
%\end{equation}
%\begin{equation}
c_{3}^{\mathbf{1}} & = & \frac{N_f-2}{2N_c^2} a_1 b_3 + \frac{N_f+3}{2N_c^2} a_1 c_3 + \frac{N_f}{4N_c^2} b_2^2 + \frac{4N_f+3}{2N_c^4} c_3^2\; , \nonumber \\
%\end{equation}
%\begin{equation}
c_{4}^{\mathbf{1}} & = & \frac{N_f}{N_c^3} b_2 b_3 + \frac{N_c+N_f}{N_c^4} b_3^2 - \frac{5(N_c+N_f)}{4N_c^4} c_3^2\; , \nonumber \\
%\end{equation}
%\begin{equation}
c_{5}^{\mathbf{1}} & = & \frac{N_f-2}{2N_c^4} b_3^2 + \frac{N_f+5}{4N_c^4} c_3^2\; , \nonumber \\
%\end{equation}
%\begin{equation}
c_{6}^{\mathbf{1}} & = & 0\; , \nonumber \\
%\end{equation}
%\begin{equation}
c_{7}^{\mathbf{1}} & = & 0\; . \nonumber
%\end{equation}
\eea

\begin{equation}
-\frac12 \{A^{ja},[V^c,[\mathcal{M},A^{ja}]]\} = \sum_{n=1}^7 d_n^{\mathbf{1}} S_n^c,
\end{equation}
where
\bea
%\begin{equation}
d_{1}^{\mathbf{1}} & = & \left( \frac{3N_f}{4} a_1^2 + \frac{3N_f}{N_c^2} a_1 c_3 + \frac{3 N_f}{N_c^4} c_3^2 \right) \frac{\Delta}{N_c}\; , \nonumber \\
%\end{equation}
%\begin{equation}
d_{2}^{\mathbf{1}} & = & \left( -\frac12(N_c+N_f) a_1^2 - \frac{3(N_c+N_f)}{N_c^2} a_1 c_3 - \frac{3(N_c+N_f)}{N_c^4} c_3^2 \right) \frac{\Delta}{N_c}\; , \nonumber \\
%\end{equation}
%\begin{equation}
d_{3}^{\mathbf{1}} & = & \left( \frac12 a_1^2 + \frac{5N_f+6}{2N_c^2} a_1 c_3 + \frac{11N_f+6}{2N_c^4} c_3^2 \right) \frac{\Delta}{N_c}\; , \nonumber \\
%\end{equation}
%\begin{equation}
d_{4}^{\mathbf{1}} & = & \left( -\frac{N_c+N_f}{N_c^2} a_1 c_3 - \frac{4(N_c+N_f)}{N_c^4} c_3^2 \right) \frac{\Delta}{N_c}\; , \nonumber \\
%\end{equation}
%\begin{equation}
d_{5}^{\mathbf{1}} & = & \left( \frac{1}{N_c^2} a_1 c_3 + \frac{7N_f+16}{4N_c^4} c_3^2 \right) \frac{\Delta}{N_c}\; , \nonumber \\
%\end{equation}
%\begin{equation}
d_{6}^{\mathbf{1}} & = & \left( -\frac{N_c+N_f}{2N_c^4} c_3^2 \right) \frac{\Delta}{N_c}\; , \nonumber \\
%\end{equation}
%\begin{equation}
d_{7}^{\mathbf{1}} & = & \left( \frac{1}{2N_c^4} c_3^2 \right) \frac{\Delta}{N_c}\; . \nonumber
%\end{equation}
\eea

\begin{equation}
\frac16 \left([A^{ja},[[\mathcal{M},[\mathcal{M},A^{ja}]],V^c]] - \frac12 [[\mathcal{M},A^{ja}],[[\mathcal{M},A^{ja}],V^c]]\right) = \sum_{n=1}^7 e_n^{\mathbf{1}} S_n^c, 
\end{equation}
where
\bea
%\begin{equation}
e_{1}^{\mathbf{1}} & = & \left( \frac{3N_f}{4} a_1^2 + \frac{3 N_f}{N_c^2} a_1 c_3 + \frac{3N_f}{N_c^4} c_3^2\right) \frac{\Delta^2}{N_c^2}\; , \nonumber \\
%\end{equation}
%\begin{equation}
e_{2}^{\mathbf{1}} & = & \left( -\frac34(N_c+N_f) a_1^2 - \frac{3(N_c+N_f)}{N_c^2} a_1 c_3 - \frac{3(N_c+N_f)}{N_c^4} c_3^2\right) \frac{\Delta^2}{N_c^2}\; , \nonumber \\
%\end{equation}
%\begin{equation}
e_{3}^{\mathbf{1}} & = & \left( \frac14(N_f+3) a_1^2 + \frac{4N_f+3}{N_c^2} a_1 c_3 + \frac{7N_f+3}{N_c^4} c_3^2\right) \frac{\Delta^2}{N_c^2}\; , \nonumber \\
%\end{equation}
%\begin{equation}
e_{4}^{\mathbf{1}} & = & \left( -\frac{5(N_c+N_f)}{2N_c^2} a_1 c_3 - \frac{11(N_c+N_f)}{2N_c^4} c_3^2\right) \frac{\Delta^2}{N_c^2}\; , \nonumber \\
%\end{equation}
%\begin{equation}
e_{5}^{\mathbf{1}} & = & \left( \frac{N_f+5}{2N_c^2} a_1 c_3 + \frac{15N_f+22}{4N_c^4} c_3^2\right) \frac{\Delta^2}{N_c^2}\; , \nonumber \\
%\end{equation}
%\begin{equation}
e_{6}^{\mathbf{1}} & = & \left( -\frac{7(N_c+N_f)}{4 N_c^4} c_3^2\right) \frac{\Delta^2}{N_c^2}\; , \nonumber \\
%\end{equation}
%\begin{equation}
e_{7}^{\mathbf{1}} & = & \left( \frac{N_f+7}{4N_c^4} c_3^2\right) \frac{\Delta^2}{N_c^2}\; . \nonumber
%\end{equation}
\eea

\begin{equation}
\frac12 d^{ab8} [A^{ja},[A^{jb},V^c]] = \sum_{n=1}^{13} c_{n}^{\mathbf{8}} O_n^c,
\end{equation}
where
\bea
%\begin{equation}
c_{1}^{\mathbf{8}} & = & \frac{3N_f}{16} a_1^2 + \frac{3N_f}{4N_c^2} a_1 c_3 + \frac{3N_f}{4N_c^4} c_3^2\; , \nonumber \\
%\end{equation}
%\begin{equation}
c_{2}^{\mathbf{8}} & = & \frac{N_f}{4N_c} a_1 b_2 + \frac{N_c+N_f}{2N_c^2} a_1 b_3 - \frac{3(N_c+N_f)}{4N_c^2} a_1 c_3 - \frac{3(N_c+N_f)}{4N_c^4} c_3^2\; , \nonumber \\
%\end{equation}
%\begin{equation}
c_{3}^{\mathbf{8}} & = & \frac{N_f-2}{4N_c^2} a_1 b_3 + \frac{N_f+3}{4N_c^2} a_1 c_3 + \frac{N_f}{8N_c^2} b_2^2 + \frac{4N_f+3}{4N_c^4} c_3^2\; , \nonumber \\
%\end{equation}
%\begin{equation}
c_{4}^{\mathbf{8}} & = & -\frac{1}{2N_c^2}a_1 b_3 + \frac{3}{4N_c^2} a_1 c_3 + \frac{3}{4N_c^4} c_3^2\; , \nonumber \\
%\end{equation}
%\begin{equation}
c_{5}^{\mathbf{8}} & = & \frac{1}{2N_c^2}a_1 b_3 - \frac{3}{4N_c^2} a_1 c_3 - \frac{3}{4N_c^4} c_3^2\; , \nonumber \\
%\end{equation}
%\begin{equation}
c_{6}^{\mathbf{8}} & = & \frac{N_f}{2N_c^3} b_2 b_3 + \frac{N_c+N_f}{2N_c^4} b_3^2 - \frac{5(N_c+N_f)}{8N_c^4} c_3^2\; , \nonumber \\
%\end{equation}
%\begin{equation}
c_{7}^{\mathbf{8}} & = & \frac{N_f-2}{4N_c^4} b_3^2 + \frac{N_f+5}{8N_c^4} c_3^2\; , \nonumber \\
%\end{equation}
%\begin{equation}
c_{8}^{\mathbf{8}} & = & -\frac{1}{2N_c^4}b_3^2 + \frac{5}{8N_c^4} c_3^2\; , \nonumber \\
%\end{equation}
%\begin{equation}
c_{9}^{\mathbf{8}} & = & \frac{1}{2N_c^4}b_3^2 - \frac{5}{8N_c^4} c_3^2\; , \nonumber \\
%\end{equation}
%\begin{equation}
c_{10}^{\mathbf{8}} & = & 0\; , \nonumber \\
%\end{equation}
%\begin{equation}
c_{11}^{\mathbf{8}} & = & 0\; , \nonumber \\
%\end{equation}
%\begin{equation}
c_{12}^{\mathbf{8}} & = & 0\; , \nonumber \\
%\end{equation}
%\begin{equation}
c_{13}^{\mathbf{8}} & = & 0\; . \nonumber
%\end{equation}
\eea

\begin{equation}
- \frac12 d^{ab8} \{A^{ja},[V^c,[\mathcal{M},A^{jb}]]\} = \sum_{n=1}^{13} d_{n}^{\mathbf{8}} O_n^c,
\end{equation}
where
\bea
%\begin{equation}
d_{1}^{\mathbf{8}} & = & \left( \frac{3N_f}{8} a_1^2 + \frac{3N_f}{2N_c^2} a_1 c_3 + \frac{3N_f}{2N_c^4} c_3^2 \right) \frac{\Delta}{N_c}\; , \nonumber \\
%\end{equation}
%\begin{equation}
d_{2}^{\mathbf{8}} & = & \left( -\frac14(N_c+N_f) a_1^2 - \frac{3(N_c+N_f)}{2N_c^2} a_1 c_3 - \frac{3(N_c+N_f)}{2N_c^4} c_3^2 \right) \frac{\Delta}{N_c}\; , \nonumber \\
%\end{equation}
%\begin{equation}
d_{3}^{\mathbf{8}} & = & \left( \frac14 a_1^2 + \frac{5N_f+6}{4N_c^2} a_1 c_3 + \frac{11 N_f+6}{4N_c^4} c_3^2 \right) \frac{\Delta}{N_c}\; , \nonumber \\
%\end{equation}
%\begin{equation}
d_{4}^{\mathbf{8}} & = & \left( \frac14 a_1^2 + \frac{3}{2N_c^2} a_1 c_3 + \frac{3}{2N_c^4} c_3^2 \right) \frac{\Delta}{N_c}\; , \nonumber \\
%\end{equation}
%\begin{equation}
d_{5}^{\mathbf{8}} & = & \left( -\frac14 a_1^2 - \frac{3}{2N_c^2} a_1 c_3 - \frac{3}{2N_c^4} c_3^2 \right) \frac{\Delta}{N_c}\; , \nonumber \\
%\end{equation}
%\begin{equation}
d_{6}^{\mathbf{8}} & = & \left( -\frac{N_c+N_f}{2N_c^2} a_1 c_3 - \frac{2(N_c+N_f)}{N_c^4} c_3^2 \right) \frac{\Delta}{N_c}\; , \nonumber \\
%\end{equation}
%\begin{equation}
d_{7}^{\mathbf{8}} & = & \left( \frac{1}{2N_c^2}a_1 c_3 + \frac{7N_f+16}{8N_c^4} c_3^2 \right) \frac{\Delta}{N_c}\; , \nonumber \\
%\end{equation}
%\begin{equation}
d_{8}^{\mathbf{8}} & = & \left( \frac{1}{2N_c^2}a_1 c_3 + \frac{2}{N_c^4} c_3^2 \right) \frac{\Delta}{N_c}\; , \nonumber \\
%\end{equation}
%\begin{equation}
d_{9}^{\mathbf{8}} & = & \left( -\frac{1}{2N_c^2}a_1 c_3 - \frac{2}{N_c^4} c_3^2 \right) \frac{\Delta}{N_c}\; , \nonumber \\
%\end{equation}
%\begin{equation}
d_{10}^{\mathbf{8}} & = & \left( -\frac{N_c+N_f}{4N_c^4} c_3^2 \right) \frac{\Delta}{N_c}\; , \nonumber \\
%\end{equation}
%\begin{equation}
d_{11}^{\mathbf{8}} & = & \left( \frac{1}{4N_c^4}c_3^2 \right) \frac{\Delta}{N_c}\; , \nonumber \\
%\end{equation}
%\begin{equation}
d_{12}^{\mathbf{8}} & = & \left( \frac{1}{4N_c^4}c_3^2 \right) \frac{\Delta}{N_c}\; , \nonumber \\
%\end{equation}
%\begin{equation}
d_{13}^{\mathbf{8}} & = & \left( -\frac{1}{4N_c^4} c_3^2 \right) \frac{\Delta}{N_c}\; . \nonumber
%\end{equation}
\eea

\begin{equation}
\frac16 d^{ab8} \left([A^{ja},[[\mathcal{M},[\mathcal{M},A^{jb}]],V^c]] - \frac12 [[\mathcal{M},A^{ja}],[[\mathcal{M},A^{jb}],V^c]]\right) = \sum_{n=1}^{13} e_{n}^{\mathbf{8}} O_n^c,
\end{equation}
where
\bea
%\begin{equation}
e_{1}^{\mathbf{8}} & = & \left( \frac{3N_f}{8} a_1^2 + \frac{3N_f}{2N_c^2} a_1 c_3 + \frac{3N_f}{2N_c^4} c_3^2\right) \frac{\Delta^2}{N_c^2}\; , \nonumber \\
%\end{equation}
%\begin{equation}
e_{2}^{\mathbf{8}} & = & \left( -\frac38(N_c+N_f) a_1^2 - \frac{3(N_c+N_f)}{2N_c^2} a_1 c_3 - \frac{3(N_c+N_f)}{2N_c^4} c_3^2\right) \frac{\Delta^2}{N_c^2}\; , \nonumber \\
%\end{equation}
%\begin{equation}
e_{3}^{\mathbf{8}} & = & \left( \frac18(N_f+3) a_1^2 + \frac{4N_f+3}{2N_c^2} a_1 c_3 + \frac{7N_f+3}{2N_c^4} c_3^2\right) \frac{\Delta^2}{N_c^2}\; , \nonumber \\
%\end{equation}
%\begin{equation}
e_{4}^{\mathbf{8}} & = & \left( \frac38 a_1^2 + \frac{3}{2N_c^2} a_1 c_3 + \frac{3}{2N_c^4} c_3^2\right) \frac{\Delta^2}{N_c^2}\; , \nonumber \\
%\end{equation}
%\begin{equation}
e_{5}^{\mathbf{8}} & = & \left( -\frac38 a_1^2 - \frac{3}{2N_c^2} a_1 c_3 - \frac{3}{2N_c^4} c_3^2\right) \frac{\Delta^2}{N_c^2}\; , \nonumber \\
%\end{equation}
%\begin{equation}
e_{6}^{\mathbf{8}} & = & \left( -\frac{5(N_c+N_f)}{4N_c^2} a_1 c_3 - \frac{11(N_c+N_f)}{4N_c^4} c_3^2\right) \frac{\Delta^2}{N_c^2}\; , \nonumber \\
%\end{equation}
%\begin{equation}
e_{7}^{\mathbf{8}} & = & \left( \frac{N_f+5}{4N_c^2} a_1 c_3 + \frac{15N_f+22}{8N_c^4} c_3^2\right) \frac{\Delta^2}{N_c^2}\; , \nonumber \\
%\end{equation}
%\begin{equation}
e_{8}^{\mathbf{8}} & = & \left( \frac{5}{4N_c^2} a_1 c_3 + \frac{11}{4N_c^4} c_3^2\right) \frac{\Delta^2}{N_c^2}\; , \nonumber \\
%\end{equation}
%\begin{equation}
e_{9}^{\mathbf{8}} & = & \left( -\frac{5}{4N_c^2} a_1 c_3 - \frac{11}{4N_c^4} c_3^2\right) \frac{\Delta^2}{N_c^2}\; , \nonumber \\
%\end{equation}
%\begin{equation}
e_{10}^{\mathbf{8}} & = & \left( -\frac{7(N_c+N_f)}{8N_c^4} c_3^2\right) \frac{\Delta^2}{N_c^2}\; , \nonumber \\
%\end{equation}
%\begin{equation}
e_{11}^{\mathbf{8}} & = & \left( \frac{N_f+7}{8N_c^4} c_3^2\right) \frac{\Delta^2}{N_c^2}\; , \nonumber \\
%\end{equation}
%\begin{equation}
e_{12}^{\mathbf{8}} & = & \left( \frac{7}{8N_c^4} c_3^2\right) \frac{\Delta^2}{N_c^2}\; , \nonumber \\
%\end{equation}
%\begin{equation}
e_{13}^{\mathbf{8}} & = & \left( -\frac{7}{8N_c^4} c_3^2\right) \frac{\Delta^2}{N_c^2}\; . \nonumber
%\end{equation}
\eea

\begin{equation}
\frac12 [A^{j8},[A^{j8},V^c]] = \sum_{n=1}^{9} c_{n}^{\mathbf{27}} T_n^c,
\end{equation}
where
\bea
%\begin{equation}
c_{1}^{\mathbf{27}} & = & \frac38 a_1^2 + \frac{3}{2N_c^2} a_1 b_3 - \frac{3}{4N_c^2} a_1 c_3 - \frac{3}{4N_c^4} c_3^2\; , \nonumber \\
%\end{equation}
%\begin{equation}
c_{2}^{\mathbf{27}} & = & \frac{1}{2N_c}a_1 b_2\; , \nonumber \\
%\end{equation}
%\begin{equation}
c_{3}^{\mathbf{27}} & = & \frac{1}{2N_c^2}a_1 b_3 + \frac{1}{2N_c^2} a_1 c_3 + \frac{1}{4N_c^2}b_2^2 + \frac{3}{2N_c^4} b_3^2 + \frac{1}{8N_c^4} c_3^2\; , \nonumber \\
%\end{equation}
%\begin{equation}
c_{4}^{\mathbf{27}} & = & -\frac{1}{N_c^2} a_1 b_3 + \frac{3}{2N_c^2} a_1 c_3 + \frac{3}{2N_c^4} c_3^2\; , \nonumber \\
%\end{equation}
%\begin{equation}
c_{5}^{\mathbf{27}} & = & \frac{1}{N_c^3} b_2 b_3\; , \nonumber \\
%\end{equation}
%\begin{equation}
c_{6}^{\mathbf{27}} & = & \frac{1}{2N_c^4} b_3^2 + \frac{1}{4N_c^4} c_3^2\; , \nonumber \\
%\end{equation}
%\begin{equation}
c_{7}^{\mathbf{27}} & = & -\frac{1}{N_c^4} b_3^2 + \frac{5}{4N_c^4} c_3^2\; , \nonumber \\
%\end{equation}
%\begin{equation}
c_{8}^{\mathbf{27}} & = & 0\; , \nonumber \\
%\end{equation}
%\begin{equation}
c_{9}^{\mathbf{27}} & = & 0\; . \nonumber
%\end{equation}
\eea

\begin{equation}
- \frac12 \{A^{j8},[V^c,[\mathcal{M},A^{j8}]]\} = \sum_{n=1}^{9} d_{n}^{\mathbf{27}} T_n^c,
\end{equation}
where
\bea
%\begin{equation}
d_{1}^{\mathbf{27}} & = & \left( -\frac{3}{2N_c^2} a_1 c_3 - \frac{3}{2N_c^4} c_3^2 \right) \frac{\Delta}{N_c}\; , \nonumber \\
%\end{equation}
%\begin{equation}
d_{2}^{\mathbf{27}} & = & 0\; , \nonumber \\
%\end{equation}
%\begin{equation}
d_{3}^{\mathbf{27}} & = & \left( \frac{1}{N_c^2} a_1 c_3 - \frac{1}{2N_c^4} c_3^2 \right) \frac{\Delta}{N_c}\; , \nonumber \\
%\end{equation}
%\begin{equation}
d_{4}^{\mathbf{27}} & = & \left( \frac12 a_1^2 + \frac{3}{N_c^2} a_1 c_3 + \frac{3}{N_c^4} c_3^2 \right) \frac{\Delta}{N_c}\; , \nonumber \\
%\end{equation}
%\begin{equation}
d_{5}^{\mathbf{27}} & = & 0\; , \nonumber \\
%\end{equation}
%\begin{equation}
d_{6}^{\mathbf{27}} & = & \left( \frac{1}{N_c^4} c_3^2 \right) \frac{\Delta}{N_c}\; , \nonumber \\
%\end{equation}
%\begin{equation}
d_{7}^{\mathbf{27}} & = & \left( \frac{1}{N_c^2} a_1 c_3 + \frac{4}{N_c^4} c_3^2 \right) \frac{\Delta}{N_c}\; , \nonumber \\
%\end{equation}
%\begin{equation}
d_{8}^{\mathbf{27}} & = & 0\; , \nonumber \\
%\end{equation}
%\begin{equation}
d_{9}^{\mathbf{27}} & = & \left( \frac{1}{2N_c^4} c_3^2 \right) \frac{\Delta}{N_c}\; . \nonumber
%\end{equation}
\eea

\begin{equation}
\frac16 \left([A^{j8},[[\mathcal{M},[\mathcal{M},A^{j8}]],V^c]] - \frac12 [[\mathcal{M},A^{j8}],[[\mathcal{M},A^{j8}],V^c]]\right) = \sum_{n=1}^{9} e_{n}^{\mathbf{27}} T_n^c,
\end{equation}
where
\bea
%\begin{equation}
e_{1}^{\mathbf{27}} & = & \left( -\frac38 a_1^2 - \frac{3}{2N_c^2} a_1 c_3 - \frac{3}{2N_c^4} c_3^2\right) \frac{\Delta^2}{N_c^2}\; , \nonumber \\
%\end{equation}
%\begin{equation}
e_{2}^{\mathbf{27}} & = & 0\; , \nonumber \\
%\end{equation}
%\begin{equation}
e_{3}^{\mathbf{27}} & = & \left( \frac14 a_1^2 + \frac{1}{4N_c^2}a_1 c_3 - \frac{5}{4N_c^4} c_3^2\right) \frac{\Delta^2}{N_c^2}\; , \nonumber \\
%\end{equation}
%\begin{equation}
e_{4}^{\mathbf{27}} & = & \left( \frac34 a_1^2 + \frac{3}{N_c^2} a_1 c_3 + \frac{3}{N_c^4} c_3^2\right) \frac{\Delta^2}{N_c^2}\; , \nonumber \\
%\end{equation}
%\begin{equation}
e_{5}^{\mathbf{27}} & = & 0\; , \nonumber \\
%\end{equation}
%\begin{equation}
e_{6}^{\mathbf{27}} & = & \left( \frac{1}{2N_c^2} a_1 c_3 + \frac{9}{8N_c^4} c_3^2\right) \frac{\Delta^2}{N_c^2}\; , \nonumber \\
%\end{equation}
%\begin{equation}
e_{7}^{\mathbf{27}} & = & \left( \frac{5}{2N_c^2} a_1 c_3 + \frac{11}{2N_c^4} c_3^2\right) \frac{\Delta^2}{N_c^2}\; , \nonumber \\
%\end{equation}
%\begin{equation}
e_{8}^{\mathbf{27}} & = & \left( \frac{1}{4N_c^4} c_3^2\right) \frac{\Delta^2}{N_c^2}\; , \nonumber \\
%\end{equation}
%\begin{equation}
e_{9}^{\mathbf{27}} & = & \left( \frac{7}{4N_c^4} c_3^2\right) \frac{\Delta^2}{N_c^2}\; . \nonumber
\eea

\bibliography{Refs}

\begin{thebibliography}{29}
\expandafter\ifx\csname natexlab\endcsname\relax\def\natexlab#1{#1}\fi
\expandafter\ifx\csname bibnamefont\endcsname\relax
  \def\bibnamefont#1{#1}\fi
\expandafter\ifx\csname bibfnamefont\endcsname\relax
  \def\bibfnamefont#1{#1}\fi
\expandafter\ifx\csname citenamefont\endcsname\relax
  \def\citenamefont#1{#1}\fi
\expandafter\ifx\csname url\endcsname\relax
  \def\url#1{\texttt{#1}}\fi
\expandafter\ifx\csname urlprefix\endcsname\relax\def\urlprefix{URL }\fi
\providecommand{\bibinfo}[2]{#2}
\providecommand{\eprint}[2][]{\url{#2}}

\bibitem[{\citenamefont{Garcia et~al.}(1985)\citenamefont{Garcia, Kielanowski,
  and Bohm}}]{Garcia:1985xz}
\bibinfo{author}{\bibfnamefont{A.}~\bibnamefont{Garcia}},
  \bibinfo{author}{\bibfnamefont{P.}~\bibnamefont{Kielanowski}},
  \bibnamefont{and} \bibinfo{author}{\bibfnamefont{A.}~\bibnamefont{Bohm}},
  \bibinfo{journal}{Lect.Notes Phys.} \textbf{\bibinfo{volume}{222}},
  \bibinfo{pages}{1} (\bibinfo{year}{1985}).

\bibitem[{\citenamefont{Flores-Mendieta}(2004)}]{FloresMendieta:2004sk}
\bibinfo{author}{\bibfnamefont{R.}~\bibnamefont{Flores-Mendieta}},
  \bibinfo{journal}{Phys.Rev.} \textbf{\bibinfo{volume}{D70}},
  \bibinfo{pages}{114036} (\bibinfo{year}{2004}), \eprint{hep-ph/0410171}.

\bibitem[{\citenamefont{Leutwyler and Roos}(1984)}]{Leutwyler:1984je}
\bibinfo{author}{\bibfnamefont{H.}~\bibnamefont{Leutwyler}} \bibnamefont{and}
  \bibinfo{author}{\bibfnamefont{M.}~\bibnamefont{Roos}},
  \bibinfo{journal}{Z.Phys.} \textbf{\bibinfo{volume}{C25}},
  \bibinfo{pages}{91} (\bibinfo{year}{1984}).

\bibitem[{\citenamefont{Gasser and Leutwyler}(1985)}]{Gasser:1984ux}
\bibinfo{author}{\bibfnamefont{J.}~\bibnamefont{Gasser}} \bibnamefont{and}
  \bibinfo{author}{\bibfnamefont{H.}~\bibnamefont{Leutwyler}},
  \bibinfo{journal}{Nucl.Phys.} \textbf{\bibinfo{volume}{B250}},
  \bibinfo{pages}{517} (\bibinfo{year}{1985}).

\bibitem[{\citenamefont{Jenkins}(1996)}]{Jenkins:1995gc}
\bibinfo{author}{\bibfnamefont{E.~E.} \bibnamefont{Jenkins}},
  \bibinfo{journal}{Phys.Rev.} \textbf{\bibinfo{volume}{D53}},
  \bibinfo{pages}{2625} (\bibinfo{year}{1996}), \eprint{hep-ph/9509433}.

\bibitem[{\citenamefont{Flores-Mendieta and
  Hofmann}(2006)}]{FloresMendieta:2006ei}
\bibinfo{author}{\bibfnamefont{R.}~\bibnamefont{Flores-Mendieta}}
  \bibnamefont{and} \bibinfo{author}{\bibfnamefont{C.~P.}
  \bibnamefont{Hofmann}}, \bibinfo{journal}{Phys.Rev.}
  \textbf{\bibinfo{volume}{D74}}, \bibinfo{pages}{094001}
  (\bibinfo{year}{2006}), \eprint{hep-ph/0609120}.

\bibitem[{\citenamefont{Flores-Mendieta
  et~al.}(2012)\citenamefont{Flores-Mendieta, Hernandez-Ruiz, and
  Hofmann}}]{FloresMendieta:2012dn}
\bibinfo{author}{\bibfnamefont{R.}~\bibnamefont{Flores-Mendieta}},
  \bibinfo{author}{\bibfnamefont{M.~A.} \bibnamefont{Hernandez-Ruiz}},
  \bibnamefont{and} \bibinfo{author}{\bibfnamefont{C.~P.}
  \bibnamefont{Hofmann}}, \bibinfo{journal}{Phys.Rev.}
  \textbf{\bibinfo{volume}{D86}}, \bibinfo{pages}{094041}
  (\bibinfo{year}{2012}), \eprint{1210.8445}.

\bibitem[{\citenamefont{Flores-Mendieta}(2009)}]{FloresMendieta:2009rq}
\bibinfo{author}{\bibfnamefont{R.}~\bibnamefont{Flores-Mendieta}},
  \bibinfo{journal}{Phys.Rev.} \textbf{\bibinfo{volume}{D80}},
  \bibinfo{pages}{094014} (\bibinfo{year}{2009}), \eprint{0910.1103}.

\bibitem[{\citenamefont{Ahuatzin et~al.}(2014)\citenamefont{Ahuatzin,
  Flores-Mendieta, Hernandez-Ruiz, and Hofmann}}]{Ahuatzin:2010ef}
\bibinfo{author}{\bibfnamefont{G.}~\bibnamefont{Ahuatzin}},
  \bibinfo{author}{\bibfnamefont{R.}~\bibnamefont{Flores-Mendieta}},
  \bibinfo{author}{\bibfnamefont{M.~A.} \bibnamefont{Hernandez-Ruiz}},
  \bibnamefont{and} \bibinfo{author}{\bibfnamefont{C.}~\bibnamefont{Hofmann}},
  \bibinfo{journal}{Phys.Rev.} \textbf{\bibinfo{volume}{D89}},
  \bibinfo{pages}{034012} (\bibinfo{year}{2014}), \eprint{1011.5268}.

\bibitem[{\citenamefont{Cordon and Goity}(2013)}]{CalleCordon:2012xz}
\bibinfo{author}{\bibfnamefont{A.~C.} \bibnamefont{Cordon}} \bibnamefont{and}
  \bibinfo{author}{\bibfnamefont{J.}~\bibnamefont{Goity}},
  \bibinfo{journal}{Phys.Rev.} \textbf{\bibinfo{volume}{D87}},
  \bibinfo{pages}{016019} (\bibinfo{year}{2013}), \eprint{1210.2364}.

\bibitem[{\citenamefont{Cordon et~al.}(2014)\citenamefont{Cordon, DeGrand, and
  Goity}}]{Cordon:2014sda}
\bibinfo{author}{\bibfnamefont{A.~C.} \bibnamefont{Cordon}},
  \bibinfo{author}{\bibfnamefont{T.}~\bibnamefont{DeGrand}}, \bibnamefont{and}
  \bibinfo{author}{\bibfnamefont{J.}~\bibnamefont{Goity}}
  (\bibinfo{year}{2014}), \eprint{1404.2301}.

\bibitem[{\citenamefont{Krause}(1990)}]{Krause:1990xc}
\bibinfo{author}{\bibfnamefont{A.}~\bibnamefont{Krause}},
  \bibinfo{journal}{Helv.Phys.Acta} \textbf{\bibinfo{volume}{63}},
  \bibinfo{pages}{3} (\bibinfo{year}{1990}).

\bibitem[{\citenamefont{Anderson and Luty}(1993)}]{Anderson:1993as}
\bibinfo{author}{\bibfnamefont{J.}~\bibnamefont{Anderson}} \bibnamefont{and}
  \bibinfo{author}{\bibfnamefont{M.~A.} \bibnamefont{Luty}},
  \bibinfo{journal}{Phys.Rev.} \textbf{\bibinfo{volume}{D47}},
  \bibinfo{pages}{4975} (\bibinfo{year}{1993}), \eprint{hep-ph/9301219}.

\bibitem[{\citenamefont{Villadoro}(2006)}]{Villadoro:2006nj}
\bibinfo{author}{\bibfnamefont{G.}~\bibnamefont{Villadoro}},
  \bibinfo{journal}{Phys.Rev.} \textbf{\bibinfo{volume}{D74}},
  \bibinfo{pages}{014018} (\bibinfo{year}{2006}), \eprint{hep-ph/0603226}.

\bibitem[{\citenamefont{Lacour et~al.}(2007)\citenamefont{Lacour, Kubis, and
  Meissner}}]{Lacour:2007wm}
\bibinfo{author}{\bibfnamefont{A.}~\bibnamefont{Lacour}},
  \bibinfo{author}{\bibfnamefont{B.}~\bibnamefont{Kubis}}, \bibnamefont{and}
  \bibinfo{author}{\bibfnamefont{U.-G.} \bibnamefont{Meissner}},
  \bibinfo{journal}{JHEP} \textbf{\bibinfo{volume}{0710}}, \bibinfo{pages}{083}
  (\bibinfo{year}{2007}), \eprint{0708.3957}.

\bibitem[{\citenamefont{Geng et~al.}(2009)\citenamefont{Geng, Martin~Camalich,
  and Vicente~Vacas}}]{Geng:2009ik}
\bibinfo{author}{\bibfnamefont{L.}~\bibnamefont{Geng}},
  \bibinfo{author}{\bibfnamefont{J.}~\bibnamefont{Martin~Camalich}},
  \bibnamefont{and}
  \bibinfo{author}{\bibfnamefont{M.}~\bibnamefont{Vicente~Vacas}},
  \bibinfo{journal}{Phys.Rev.} \textbf{\bibinfo{volume}{D79}},
  \bibinfo{pages}{094022} (\bibinfo{year}{2009}), \eprint{0903.4869}.

\bibitem[{\citenamefont{Dashen and
  Manohar}(1993{\natexlab{a}})}]{Dashen:1993as}
\bibinfo{author}{\bibfnamefont{R.~F.} \bibnamefont{Dashen}} \bibnamefont{and}
  \bibinfo{author}{\bibfnamefont{A.~V.} \bibnamefont{Manohar}},
  \bibinfo{journal}{Phys.Lett.} \textbf{\bibinfo{volume}{B315}},
  \bibinfo{pages}{425} (\bibinfo{year}{1993}{\natexlab{a}}),
  \eprint{hep-ph/9307241}.

\bibitem[{\citenamefont{Dashen et~al.}(1995)\citenamefont{Dashen, Jenkins, and
  Manohar}}]{Dashen:1994qi}
\bibinfo{author}{\bibfnamefont{R.~F.} \bibnamefont{Dashen}},
  \bibinfo{author}{\bibfnamefont{E.~E.} \bibnamefont{Jenkins}},
  \bibnamefont{and} \bibinfo{author}{\bibfnamefont{A.~V.}
  \bibnamefont{Manohar}}, \bibinfo{journal}{Phys.Rev.}
  \textbf{\bibinfo{volume}{D51}}, \bibinfo{pages}{3697} (\bibinfo{year}{1995}),
  \eprint{hep-ph/9411234}.

\bibitem[{\citenamefont{Gervais and Sakita}(1984)}]{Gervais:1984rc}
\bibinfo{author}{\bibfnamefont{J.-L.} \bibnamefont{Gervais}} \bibnamefont{and}
  \bibinfo{author}{\bibfnamefont{B.}~\bibnamefont{Sakita}},
  \bibinfo{journal}{Phys.Rev.} \textbf{\bibinfo{volume}{D30}},
  \bibinfo{pages}{1795} (\bibinfo{year}{1984}).

\bibitem[{\citenamefont{Jenkins and Lebed}(1995)}]{Jenkins:1995td}
\bibinfo{author}{\bibfnamefont{E.~E.} \bibnamefont{Jenkins}} \bibnamefont{and}
  \bibinfo{author}{\bibfnamefont{R.~F.} \bibnamefont{Lebed}},
  \bibinfo{journal}{Phys.Rev.} \textbf{\bibinfo{volume}{D52}},
  \bibinfo{pages}{282} (\bibinfo{year}{1995}), \eprint{hep-ph/9502227}.

\bibitem[{\citenamefont{Dashen and
  Manohar}(1993{\natexlab{b}})}]{Dashen:1993ac}
\bibinfo{author}{\bibfnamefont{R.~F.} \bibnamefont{Dashen}} \bibnamefont{and}
  \bibinfo{author}{\bibfnamefont{A.~V.} \bibnamefont{Manohar}},
  \bibinfo{journal}{Phys.Lett.} \textbf{\bibinfo{volume}{B315}},
  \bibinfo{pages}{438} (\bibinfo{year}{1993}{\natexlab{b}}),
  \eprint{hep-ph/9307242}.

\bibitem[{\citenamefont{Dai et~al.}(1996)\citenamefont{Dai, Dashen, Jenkins,
  and Manohar}}]{Dai:1995zg}
\bibinfo{author}{\bibfnamefont{J.}~\bibnamefont{Dai}},
  \bibinfo{author}{\bibfnamefont{R.~F.} \bibnamefont{Dashen}},
  \bibinfo{author}{\bibfnamefont{E.~E.} \bibnamefont{Jenkins}},
  \bibnamefont{and} \bibinfo{author}{\bibfnamefont{A.~V.}
  \bibnamefont{Manohar}}, \bibinfo{journal}{Phys.Rev.}
  \textbf{\bibinfo{volume}{D53}}, \bibinfo{pages}{273} (\bibinfo{year}{1996}),
  \eprint{hep-ph/9506273}.

\bibitem[{\citenamefont{Flores-Mendieta
  et~al.}(1998)\citenamefont{Flores-Mendieta, Jenkins, and
  Manohar}}]{FloresMendieta:1998ii}
\bibinfo{author}{\bibfnamefont{R.}~\bibnamefont{Flores-Mendieta}},
  \bibinfo{author}{\bibfnamefont{E.~E.} \bibnamefont{Jenkins}},
  \bibnamefont{and} \bibinfo{author}{\bibfnamefont{A.~V.}
  \bibnamefont{Manohar}}, \bibinfo{journal}{Phys.Rev.}
  \textbf{\bibinfo{volume}{D58}}, \bibinfo{pages}{094028}
  (\bibinfo{year}{1998}), \eprint{hep-ph/9805416}.

\bibitem[{\citenamefont{Jenkins and
  Manohar}(1991{\natexlab{a}})}]{Jenkins:1990jv}
\bibinfo{author}{\bibfnamefont{E.~E.} \bibnamefont{Jenkins}} \bibnamefont{and}
  \bibinfo{author}{\bibfnamefont{A.~V.} \bibnamefont{Manohar}},
  \bibinfo{journal}{Phys.Lett.} \textbf{\bibinfo{volume}{B255}},
  \bibinfo{pages}{558} (\bibinfo{year}{1991}{\natexlab{a}}).

\bibitem[{\citenamefont{Flores-Mendieta
  et~al.}(2000)\citenamefont{Flores-Mendieta, Hofmann, Jenkins, and
  Manohar}}]{FloresMendieta:2000mz}
\bibinfo{author}{\bibfnamefont{R.}~\bibnamefont{Flores-Mendieta}},
  \bibinfo{author}{\bibfnamefont{C.~P.} \bibnamefont{Hofmann}},
  \bibinfo{author}{\bibfnamefont{E.~E.} \bibnamefont{Jenkins}},
  \bibnamefont{and} \bibinfo{author}{\bibfnamefont{A.~V.}
  \bibnamefont{Manohar}}, \bibinfo{journal}{Phys.Rev.}
  \textbf{\bibinfo{volume}{D62}}, \bibinfo{pages}{034001}
  (\bibinfo{year}{2000}), \eprint{hep-ph/0001218}.

\bibitem[{\citenamefont{Jenkins and
  Manohar}(1991{\natexlab{b}})}]{Jenkins:1991ne}
\bibinfo{author}{\bibfnamefont{E.~E.} \bibnamefont{Jenkins}} \bibnamefont{and}
  \bibinfo{author}{\bibfnamefont{A.~V.} \bibnamefont{Manohar}},
  \bibinfo{journal}{Phys.Lett.} \textbf{\bibinfo{volume}{B259}},
  \bibinfo{pages}{353} (\bibinfo{year}{1991}{\natexlab{b}}).

\bibitem[{\citenamefont{Beringer et~al.}(2012)}]{Beringer:1900zz}
\bibinfo{author}{\bibfnamefont{J.}~\bibnamefont{Beringer}} \bibnamefont{et~al.}
  (\bibinfo{collaboration}{Particle Data Group}), \bibinfo{journal}{Phys.Rev.}
  \textbf{\bibinfo{volume}{D86}}, \bibinfo{pages}{010001}
  (\bibinfo{year}{2012}).

\bibitem[{\citenamefont{Cabibbo et~al.}(2003)\citenamefont{Cabibbo, Swallow,
  and Winston}}]{Cabibbo:2003cu}
\bibinfo{author}{\bibfnamefont{N.}~\bibnamefont{Cabibbo}},
  \bibinfo{author}{\bibfnamefont{E.~C.} \bibnamefont{Swallow}},
  \bibnamefont{and} \bibinfo{author}{\bibfnamefont{R.}~\bibnamefont{Winston}},
  \bibinfo{journal}{Ann.Rev.Nucl.Part.Sci.} \textbf{\bibinfo{volume}{53}},
  \bibinfo{pages}{39} (\bibinfo{year}{2003}), \eprint{hep-ph/0307298}.

\bibitem[{\citenamefont{Dashen et~al.}(1994)\citenamefont{Dashen, Jenkins, and
  Manohar}}]{Dashen:1993jt}
\bibinfo{author}{\bibfnamefont{R.~F.} \bibnamefont{Dashen}},
  \bibinfo{author}{\bibfnamefont{E.~E.} \bibnamefont{Jenkins}},
  \bibnamefont{and} \bibinfo{author}{\bibfnamefont{A.~V.}
  \bibnamefont{Manohar}}, \bibinfo{journal}{Phys.Rev.}
  \textbf{\bibinfo{volume}{D49}}, \bibinfo{pages}{4713} (\bibinfo{year}{1994}),
  \eprint{hep-ph/9310379}.

\end{thebibliography}

\end{document}